\documentclass[a4paper,11pt]{article}
\pdfoutput=1 
\usepackage{jheppub} 
\usepackage[T1]{fontenc} 

\usepackage{graphicx} 
\usepackage{tensor}
\usepackage{comment}
\usepackage{appendix}

\usepackage{subfigure}
\usepackage{svg}
\usepackage{braket}

\usepackage[dvipsnames]{xcolor}
\usepackage{float}

\newcommand{\tr}{\operatorname{Tr}}

\def \bal#1\eal  {\begin{align} #1 \end{align}}
\def\({\left(}
\def\){\right)}
\newcommand{\beq} {\begin{equation}}
\newcommand{\eeq} {\end{equation}}

\newcommand{\mc} {\mathcal}

\title{Superconformal indices and black hole saddles}

 \author{Maciej Kolanowski, Donald Marolf, Zi-Yue Wang, and Wenwen Zheng}
\affiliation{
	Department of Physics, University of California, Santa Barbara, CA 93106, USA}
\emailAdd{mkolanowski@ucsb.edu}
\emailAdd{marolf@ucsb.edu}
\emailAdd{zi-yue@ucsb.edu}
\emailAdd{wenwenzheng@ucsb.edu}

\begin{abstract}
{The AdS/CFT correspondence implies that the superconformal index ${\mathcal I}$ in ${\mathcal N=4}$ SU(N) supersymmetric Yang-Mills theory can be computed using the dual bulk theory.  In particular, in the limit of large $N$, the index should be given by a sum over appropriate saddles.  However, ${\mathcal I}$ depends on potentials $\sigma, \tau, \vec \Delta$ and, at large ${\rm Im}\, \tau = {\rm Im}\, \sigma$, the CFT index ${\mathcal I}$  rapidly approaches $1$ at all values of $N$.   As a result, black hole saddles associated with exponentially large contributions in $N$ cannot contribute in this limit.  This in  particular excludes saddles that were previously suggested to be relevant in such regimes.   We thus consider an approach to the bulk path integral motivated by taking it to be defined as an integral over real Lorentz-signature spacetimes with codimension-2 singularities.  This approach leads only to saddles that satisfy the above bound, and to the enforcement of this bound via Stokes phenomena. We also find similar results for bulk AdS$_4$ calculations of the ABJM superconformal index.}
\end{abstract}

\begin{document}

\maketitle
\flushbottom

\section{Introduction}

Supersymmetric indices are of great interest as they are often invariant under continuous changes of parameters.  As a result, in appropriate contexts they can be directly computed in a weakly-coupled supersymmetric theory and then used to understand aspects of the theory even at strong coupling.  This makes them especially useful in the context  of the AdS/CFT correspondence, where quantities on the two sides are generally computable only in non-overlapping regimes.

In particular, let us consider the superconformal index of ${\mathcal N}=4$ $SU(N)$ super Yang-Mills  theory in $d=4$ spacetime dimensions (SYM$_4$) \cite{Romelsberger:2005eg,Kinney:2005ej}. We denote this index by ${\mathcal I}(\sigma, \tau, \vec \Delta)$, where $\sigma, \tau, \vec \Delta=(\Delta_1,\Delta_2, \Delta_3)$  weight the counting of the relevant angular momenta and $R$-charges and where supersymmetry requires
\begin{equation}
\label{eq:SUSYPot}
\sigma+ \tau -\sum_{i=1}^3 \Delta_i = 2m+1, \ \ \ m \in {\mathbb Z}.
\end{equation}
The weights $\sigma, \tau, \vec \Delta$ are related to familiar $U(1)$ potentials $\vec \Phi$, angular velocities $\Omega_1, \Omega_2$, and the inverse temperature $\beta$ through
\begin{equation}
\tau = \frac{\beta\left(\Omega_1-1\right)}{2\pi i}, \ \sigma = \frac{\beta\left(\Omega_2-1\right)}{2\pi i}, \ \Delta_i = \frac{\beta\left(\Phi_i-1\right)}{2\pi i}.   
\label{eq:complexpotentials}
\end{equation}
Indeed, the index turns out to be just the special case of the usual thermodynamic partition function $Z(\beta, \Omega_1, \Omega_2, \Phi_1, \Phi_2, \Phi_3)$ in which the potentials satisfy  both \eqref{eq:complexpotentials} and \eqref{eq:SUSYPot}, so that the expected $(-1)^F$ emerges from the imaginary parts of $\beta\Omega_1,\beta\Omega_2,\beta\vec\Phi$.  For later use, we note that in terms of $\beta, \tau, \sigma, \vec \Delta$, the usual partition function becomes
\begin{equation}
Z(\beta, \tau, \sigma, \vec \Delta)
=\mathrm{Tr}_{\mathcal H_{\rm phys}}
\!\left(e^{-\beta\tilde E}\zeta\right) \ \ \ {\rm with} \ \ \ 
\zeta = 
\,e^{2\pi i(\tau J_1+\sigma J_2+\frac{1}{2}\sum_i \Delta_i Q_i)},
\label{eq:index_def}
\end{equation} 
where $\tilde E = E- J_1 -J_2 -(Q_1+Q_2+Q_3)/2$ is the energy relative to the  BPS bound.  When $\tau, \sigma, \vec \Delta$ satisfy \eqref{eq:SUSYPot}, this partition function can be shown to be independent of $\beta$ at fixed $\tau, \sigma, \vec \Delta$ and to agree with the desired index ${\mathcal I}(\tau,\sigma,\vec \Delta)$.

The superconformal index of \cite{Romelsberger:2005eg,Kinney:2005ej} counts states that are $\frac{1}{16}$ BPS. Both on the CFT side and in the bulk, the history of the study of this index has been rather complicated and has been filled with various twists and turns.
 However, this process greatly progressed in \cite{Choi:2018hmj,Cabo-Bizet:2018ehj,Benini:2018ywd, Copetti:2020dil}, where with appropriate fugacities the index was argued to be dominated by vacuum, by a bulk saddle point described by a complex black hole, or by a limit of such black holes.  The bulk computation of the index was then further refined in \cite{Aharony:2021zkr},  which identified additional bulk saddles that match non-perturbative corrections suggested by Bethe ansatz computations of the index on the Yang-Mills side of the correspondence (building on e.g. \cite{Closset:2017bse,Benini:2018mlo,Benini:2018ywd}). Nevertheless, as we discuss below, the discussion in the above literature cannot yet be complete\footnote{It should be pointed out that the concerns raised below are not the only open questions when it comes to these indices. For other possible issues regarding microcanonical indices and their phases, see \cite{Choi:2025lck}}. 
 
 A similar situation can be found when one considers a superconformal index \cite{Bhattacharya:2008zy, Bhattacharya:2008bja} in the ABJM theory \cite{Aharony:2008ug} with $U(N)_1 \times U(N)_{-1}$ gauge group and $\mathcal{N}=8$ superconformal symmetry. One can write this index as a matrix integral and study saddle points in the large $N$ limit \cite{Cabo-Bizet:2019eaf}, again finding a match between their contributions and Euclidean actions of the BPS black holes in the dual theory. But there are again concerns about which of these saddles actually contribute to the index and about reproducing the behavior of the matrix model at large ${\rm Im}\, \tau$. To keep this introduction concise, we now specialize to the case of $\mathcal{N}=4$ SYM$_4$, saving further comments on the ABJM index for Sec. \ref{AdS4}. 

To describe the issue we wish to resolve, let us recall that  the SYM$_4$ index ${\mathcal I}(\sigma, \tau, \vec \Delta)$ is constructed so as to be independent of $\beta$ at fixed $\sigma, \tau, \vec \Delta$ \cite{Romelsberger:2005eg,Kinney:2005ej,Benini:2018ywd}.  This is, of course, rather different behavior than what is expected for the thermodynamic partition function $Z(\beta, \sigma, \tau, \vec \Delta)$ evaluated at real potentials $\beta, \vec \Omega, \vec \Phi$.  However, the two differ only by the inclusion in the index of appropriate phases defined by the imaginary parts of $\beta, \beta \vec \Omega, \beta \vec \Phi$ that are
required for $\sigma, \tau, \vec \Delta$ to satisfy \eqref{eq:SUSYPot}. 
(These phases give the desired factor of $(-1)^F$, where $F$ represents fermion number.)  As a result, each index $\mathcal{I}$ is bounded by a partition function $Z$ evaluated at appropriate real potentials determined by the real parts of $\beta, \beta \vec \Omega, \beta \vec \Phi$.

In addition, it is well known that (even without requiring supersymmetry) real Lorentz-signature extreme AdS black holes cannot have both $\Omega$ and $\mu$ close to zero.  Instead, real Lorentz-signature extreme black holes exist only when the pair $(\Omega, \mu)$ is sufficiently far from $(0,0)$.  
This is associated with Hawking-Page-like behavior whereby the partition function $Z$ at small (real) $\Omega, \mu$ and large positive (real) $\beta$ is dominated by a thermal AdS saddle instead of a black hole saddle.  
Furthermore, at least in the uncharged case, this phase transition is directly mirrored by the counting of gauge-singlet operators in weakly-coupled ${\mathcal N}=4$ SYM \cite{Aharony:2004ir}.
As a result, in this regime $Z$ is only of order one\footnote{Here we follow the convention common in discussing the index ${\mathcal I}$ of using normalizations for both ${\mathcal I}$ and $Z$ that remove contributions from any vacuum (Casimir) energy.} at large $N$.


We should thus expect corresponding  behavior for the supersymmetric index ${\mathcal I}$.
In particular, from \eqref{eq:complexpotentials}, we see that real $\beta, \vec \Omega$ with $\beta >0$ and $\Omega_1,\Omega_2 < 1$ yield ${\rm Im}\, \tau , {\rm Im}\, \sigma >0$.  The above considerations thus motivate us to explore the behavior of the index ${\mathcal I}$ in the limit ${\rm Im}\, \tau , {\rm Im}\, \sigma \rightarrow +\infty$ (with $\vec \Delta$ satisfying \eqref{eq:SUSYPot}), and to conjecture that we should find thermal AdS saddles to dominate the corresponding bulk path integrals.  Such a result was in fact already found numerically on the CFT side in \cite{Copetti:2020dil} for the case
$\tau =\sigma$  with $\Delta_1 = \Delta_2 = \Delta_3 =:\Delta$; see their figure 7 and note that our ${\rm Im}\, \tau \rightarrow +\infty$ corresponds to their $y\rightarrow 0$.  Furthermore, as shown in appendix \ref{sec:bound}, for such potentials, methods similar to those used in \cite{Copetti:2020dil} allow one to analytically derive the above expectation at any fixed finite $N$.

In contrast, however, a black hole saddle was discussed in \cite{Kinney:2005ej,Cabo-Bizet:2018ehj,Choi:2018vbz,Benini:2018ywd,Aharony:2021zkr} in the context of being relevant for general ${\rm Im}\, \tau >0$ (again with $\tau=\sigma$ and $\Delta_1 = \Delta_2 = \Delta_3$).  This saddle gives a contribution of the form 
\begin{equation}
\exp\!\left[-N^2\,\frac{i\pi(2\tau-1)^3}{27\tau^2}\right].
\label{eq:BH_saddle}
\end{equation}
But such contributions would be exponentially large in $N$ whenever we have both ${\rm Im} \, \tau \gg 1$ and ${\rm Im} \, \tau \gg {\rm Re} \, \tau$, so that thermal AdS would not dominate in that regime. 
This tension is the main issue we wish to explore in the work below.  \footnote{\label{foot:cancel} It is technically possible that the saddle could be relevant in a Picard-Lefschetz sense but that its contribution is canceled by that of another (yet-to-be-identified) saddle having both {\it identical} classical action and (up to sign) {\it identical} corrections at all perturbative orders.  But we consider a different resolution below.}

In particular, we wish to more directly analyze the relevance of the various black hole saddles to the appropriate bulk gravitational path integral as a function of the desired potentials.  Unfortunately,  as emphasized in e.g. \cite{Aharony:2021zkr}, there are not yet generally-accepted rules for working with intrinsically-complex gravitational saddles and for determining this relevance with certainty; see also recent discussions in \cite{Held:2026bbo,Kolanowski:2026gii}.    In particular, any such rule would presumably be equivalent to choosing a convergent contour of integration for the so-called Euclidean gravitational path integral. Recall that the specification of such a contour has long been understood to be an important issue since the well-known conformal factor problem causes the gravitational path integral to diverge when the integral is performed over the space of real Euclidean metrics \cite{Gibbons:1978ac}; see also \cite{Horowitz:2025zpx} for a more modern perspective.  

Despite the lack of broad current agreement with regard to how this issue should be resolved, there have historically been many suggestions 
\cite{Hartle:2020glw,Schleich:1987fm,Mazur:1989by,Giddings:1989ny,Giddings:1990yj,Marolf:1996gb,Gratton:1999ya,Dasgupta:2001ue,Ambjorn:2002gr,Feldbrugge:2017kzv,Feldbrugge:2017fcc,Feldbrugge:2017mbc,Brown:2017wpl}
that the contour should in fact be defined by integrating over real {\it Lorentz}-signature metrics. Indeed, 
since a version of the Lorentzian path integral can be derived from canonical quantization for spacetimes with topology $\Sigma \times {\mathbb R}$, at least in that case it should thus be free of divergences after imposing an appropriate UV cutoff.  A Lorentzian approach should thus at least greatly ameliorate the Euclidean conformal factor problem. 

Below, we will follow an approach motivated by a particular version of this idea described in \cite{Marolf:2022ybi} (building on \cite{Dong:2016hjy,Colin-Ellerin:2020mva,Marolf:2020rpm,Colin-Ellerin:2021jev}).  In that work it was argued that, by allowing the off-shell spacetimes in the path integral to contain certain codimension-2 Lorentzian analogues of conical singularities\footnote{The Einstein-Hilbert action is to be defined on such spacetimes in parallel with the treatment of \cite{Louko:1995jw}. While much remains to be understood regarding theories with higher derivative terms, in such cases it was proposed in \cite{Colin-Ellerin:2020mva} to define the action using the Legendre transform of the action described in appendix B of \cite{Dong:2019piw}. }, the resulting path integral could be approximated by a statistical-mechanics-like integral over a set of {\it smooth} real Lorentz-signature stationary black holes (or, more generally, over black-hole-like geometries; see section \ref{subsec:orbifolds} for a discussion of quotients of black hole spacetimes).  It was also suggested that an improved approximation would be obtained by expanding the domain of integration to include {\it all} smooth 
real Lorentz-signature stationary black hole horizons.\footnote{\label{foot:inner} In particular, in the prescription of \cite{Marolf:2022ybi} one should include both inner and outer horizons, though loop corrections  may be substantially different in the two cases. See \cite{Kolanowski:2026gii} for further comments. 
One should also include degenerate horizons (extremal black holes).  
Since extremal black holes have an internal infinity, for charges corresponding to such black holes there will be no saddles (or even fixed-area and fixed-charge constrained saddles) satisfying the desired boundary conditions.  Nevertheless, the associated set of off-shell configurations will include parameters that specify the depth of the would-be black hole throat.  Since the extremal black hole is given by a limit where this depth diverges, in the semiclassical approximation it can be considered to provide an endpoint contribution (or, better, a boundary contribution) which behaves much as if it were in fact a constrained saddle. }

This generalization was then derived in \cite{Chen:2025leq} for black holes with  Maxwell charges and also for black holes with angular momentum in 2+1 dimensions.  While a corresponding derivation is not yet available for angular momentum in higher dimensions,  in the present work we will simply take the above as motivation to write the corresponding formulae for the partition function in the presence of general complex chemical potentials $\Omega_1,\Omega_2, \vec \Phi$,
\begin{eqnarray}
\label{eq:ansatz}
Z(\beta, \Omega_1, \Omega_2,\vec \Phi) &=& 
\sum_{{{n_{J_1},n_{J_2},n_{Q_1},n_{Q_2},n_{Q_3} \in {\mathbb Z}} \atop {n_{J_1}+ n_{J_2}+n_{Q_1}+ n_{Q_2} +n_{Q_3} \in 2{\mathbb Z}}}} 
 Z_{n_{J_1},n_{J_2},n_{Q_1},n_{Q_2},n_{Q_3}}
\ \ \ {\rm with} \nonumber \\
 Z_{n_{J_1},n_{J_2},n_{Q_1},n_{Q_2},n_{Q_3}}
 &\approx&
\int d{\mathcal A} \, dJ_1  \, dJ_2  \, dQ_1 \, dQ_2 \, dQ_3 \, \Bigl[e^{{\mathcal A}/4G_5} 
 e^{-\beta \left(E - \Omega_1 J_1 - \Omega_2 J_2 - \frac{1}{2}\sum_{i=1}^3 \Phi_i Q_i   \right)}
\nonumber \\ 
&\times&
e^{2\pi i \left(\sum_{i=1}^2 n_{J_i} J_i +\sum_{j=1}^3 n_{Q_j} \frac{Q_j}{2}  
\right)} 
\Bigr],
\end{eqnarray}
for the AdS$_5$ case of interest, where $G_5$ is the 5-dimensional Newton constant, $E$ is the energy relative to the ground state, and  we follow the convention (see e.g. \cite{Aharony:2021zkr}) of using $\frac{1}{2}\Phi_i$ to denote the coefficient of $Q_i$.
One may of course specialize the above ansatz to potentials satisfying \eqref{eq:complexpotentials} and \eqref{eq:SUSYPot} in order to attempt to study the desired index ${\mathcal I}(\Omega_1, \Omega_2,\vec \Phi)$ and the associated ${\mathcal I}_{n_{J_1},n_{J_2},n_{Q_1},n_{Q_2},n_{Q_3}}(\Omega_1, \Omega_2,\vec \Phi)$ (though one of the arguments is now redundant due to the constraint \eqref{eq:SUSYPot}). Similar ans\"atze and Picard-Lefschetz analyses were recently employed in \cite{Mahajan:2025bzo, Singhi:2025rfy, Ailiga:2025osa} in the context of black hole thermodynamics (for real potentials) and in \cite{Barbon:2026tri} in the computation of the gravitational spectral form-factor.

Let us comment briefly on the sum over `shifts' $n_{J_1},n_{J_2},n_{Q_1},n_{Q_2},n_{Q_3}$, which is an analogue of the corresponding sum in \cite{Boruch:2022tno}.  While this sum did not appear in \cite{Marolf:2022ybi}, a version did appear in \cite{Chen:2025leq} and, as discussed in \cite{Kolanowski:2026gii}, such sums are generally associated with the freedom to include certain large gauge transformations in the bulk description. See e.g. \cite{Aharony:2021zkr} for a discussion of the requirement that $n_{J_1}+ n_{J_2}+n_{Q_1}+ n_{Q_2} +n_{Q_3}$ be even.

As noted above, the range of parameters $({\mathcal A},J_1,J_2,Q_1,Q_2,Q_3)$ over which the above integral is to be performed is just the space of allowed horizon areas (again including inner horizons and degenerate horizons), angular momenta, and $U(1)$ charges for the stationary real Lorentz-signature black hole solutions allowed by the given theory. Despite the exponentially large factor of $e^{{\mathcal A}/4G_5}$, the relation between the energy $E$ and the integration variables ${\mathcal A}, \vec J, \vec Q$ will make the integrand a bounded function and, in fact, renders the integral absolutely convergent. 
As we discuss further below, the use of the symbol $\approx$ in \eqref{eq:ansatz} indicates that the right-hand-side should capture leading-order effects in the saddle point approximation, but that it may not capture quantum corrections.  As a result, we may interpret the degenerate limit $({\mathcal A}, \vec J, \vec Q) \rightarrow (0, \vec 0, \vec 0)$ as representing thermal AdS contributions. 

We emphasize  that the potentials $\beta,\Omega_1,\Omega_2, \vec \Phi$ in \eqref{eq:ansatz}  are to be regarded as external parameters that weight the contributions of all possible stationary real Lorentz-signature black holes, where we think of the energy $E$ as the corresponding function ${\mathcal A},J_1,J_2,Q_1,Q_2,Q_3$.  In particular, the integral in \eqref{eq:ansatz} is {\it not} restricted a priori to the special set of
stationary real Lorentz-signature black holes which have Hawking temperature $1/\beta$ and/or angular velocities $\Omega_1, \Omega_2$ and/or electric potentials $\vec \Phi$.  Instead, such identifications will hold for saddle points of \eqref{eq:ansatz}, whether or not those saddles are real or of Lorentz signature.

Note that the effect of the sum over $n_{J_i}, n_{Q_j}$ and the phase factor on the final line is simply to transform the integrals over $J_i,Q_j$ into discrete sums over the allowed quantized charges; i.e., the ansatz \eqref{eq:ansatz} is equivalent to
writing
\begin{equation}
\label{eq:ansatz3}
Z (\beta, \Omega_1, \Omega_2,\vec \Phi) \approx 
(2\pi)^5{\hspace {-0.5cm}}\sum_{{{J_1},{J_2} \in \frac{1}{2}{\mathbb Z}} \atop {{{{Q_1},{Q_2},{Q_3}}\in {\mathbb Z}}\atop {2J_1\equiv 2J_2 \equiv Q_1\equiv Q_2\equiv Q_3 \, mod \, 2} }   }
 \int_{X_{{J_1},{J_2},{Q_1},{Q_2},{Q_3}}} {\hspace {-1.5cm}} d{\mathcal A} \,e^{{\mathcal A}/4G_5}  
 e^{-\beta \left(E - \Omega_1 J_1 - \Omega_2 J_2 - \frac{1}{2}\sum_{i=1}^3 \Phi_i Q_i   \right)},
\end{equation}
where the domain of integration $X_{{J_1},{J_2},{Q_1},{Q_2},{Q_3}}$ for each integral over ${\mathcal A}$ corresponds to the allowed horizon areas (including inner horizons) for classical real Lorentz-signature stationary black hole solutions with charges ${J_1},{J_2},{Q_1},{Q_2},{Q_3}$. Note that the convention of \cite{Aharony:2021zkr} to weight $Q_i$ by $\frac{1}{2} \Phi_i$ places $\Phi_i$ and $\Omega_j$ on the same status in the above sum as both multiply half-integers ($\frac{Q_i}{2}$ or $J_i$).  Here we have kept the overall factor of $(2\pi)^5$ to make \eqref{eq:ansatz3} exactly equivalent to \eqref{eq:ansatz}, though this factor is irrelevant at the leading semiclassical order at which we will work.
We refer the interested reader to \cite{Kolanowski:2026gii} for a full discussion of the status of deriving the ansatz \eqref{eq:ansatz} from a Lorentzian path integral that allows the above-mentioned singularities.   However, we also note that  one might simply choose on purely physical grounds to take \eqref{eq:ansatz3} as a starting point for a leading-order semiclassical analysis. 

Let us now take a moment to state explicitly the sense in which 
we expect \eqref{eq:ansatz} and \eqref{eq:ansatz3} to capture  ``leading-order'' semiclassical effects.  In addition to the leading saddle-point behavior, it is of course natural to expect 
 \eqref{eq:ansatz3}  to capture contributions from sub-leading complex black hole saddles as well.  However, without a better understanding of the detailed integration measure to be used in \eqref{eq:ansatz3}, there is no reason for this ansatz to capture quantum corrections to any saddle.

This would not be a significant limitation in contexts where quantum corrections are small.    However, in discussions of supersymmetric indices one often encounters fermion zero modes (see e.g. \cite{Iliesiu:2021are} in the gravitational context) which can completely remove the contribution of certain saddles.  The effect of such quantum corrections can thus be quite significant indeed.  We will discuss this further in section \ref{sec:BPSansatz} where we argue that, when the potentials satisfy \eqref{eq:SUSYPot}, we expect the main effect of such zero modes to simply be to restrict the domain of integration in \eqref{eq:ansatz} (or, equivalently, in \eqref{eq:ansatz3}) to the locus defined by real Lorentz-signature black holes saturating the BPS bound.    We then propose that other fermion zero-mode effects can be incorporated through an appropriate truncation of the sum over saddles.  These expectations can be equivalently stated in terms of the $\rm{Re} \, \beta \rightarrow +\infty$ limit of \eqref{eq:ansatz} (again perhaps with a corresponding truncation). 
Section \ref{sec:BPSansatz} also describes an important consistency check on the above proposal that will be performed in later sections for the models of interest.

We begin our discussion in section  \ref{sec:PLO} with a brief review of Picard-Lefschetz methods for determining the relevance of saddles to the evaluation of integrals in the semiclassical approximation, and also for incorporating any associated boundary contributions (e.g., the higher-dimensional generalization of the endpoint contributions that arise for integrals over contours with finite endpoints in the complex plane).  We then address the above issues associated with fermion zero modes in section \ref{sec:BPSansatz},  describing the above-mentioned proposed modifications of \eqref{eq:ansatz} (associated with restricting to real BPS black holes and truncating the sum over saddles), and discussing the advertised consistency check.   

Section \ref{sec:BHreview} provides the final bit of preparatory material by giving a brief review of black holes in the AdS$_5 \times S^5$ supergravity theory of interest.  This sets the stage for section \ref{subsec:one-dimensional-BPS-integral} to analyze the  possible contributions from bulk saddles and to perform the desired consistency checks.   For simplicity we impose 
$\tau =\sigma$ with $\Delta_1 = \Delta_2 = \Delta_3 =:\Delta$ and correspondingly assume that we may truncate \eqref{eq:ansatz} to black holes with $J_1 = J_2=:J$ and $Q_1=Q_2=Q_3:=Q$. In this context, section \ref{subsec:one-dimensional-BPS-integral} shows explicitly that the semiclassical approximation to our restricted BPS-only ansatz yields ${\mathcal I}$ of order $N^0$ at sufficiently large ${\rm Im} \, \tau$.  In particular, while  there are contributions of the form \eqref{eq:BH_saddle} at certain values of $\tau$, at most one such saddle is relevant at any given value of $\tau$. Analogous statements are also shown to hold for the orbifolds of \cite{Aharony:2021zkr}. Furthermore, a Stokes' phenomenon renders the associated saddles irrelevant above some finite ${\rm Im} \, \tau$. We then discuss similar issues regarding the ABJM indices in Sec. \ref{AdS4}. 

We close with a discussion of open questions and future directions in section \ref{sec:disc}. In particular, we emphasize that our analysis is incomplete since we study only the AdS$_5$ truncation of the bulk theory and do not consider effects (such as the potential D-brane production instabilities of \cite{Aharony:2021zkr}) associated with breaking the SO(6) symmetry of the $S^5$ factor in the bulk.  Some initial explorations of \eqref{eq:ansatz} at finite $\beta$ (relevant to computations of correlation functions with insertions that soak up fermion zero modes) are also included in appendix \ref{sec:thimbles}.

\section{Picard-Lefschetz Overview}
\label{sec:PLO}

The modern understanding of the asymptotic expansion of integrals is often described in terms of Picard-Lefschetz theory and, in particular, in terms of the corresponding Lefschetz thimbles.  For readers who may not be well-acquainted with this subject,  we now provide a brief introduction to the associated terminology and to the most useful results as described in \cite{FAs,FP,AGV,BH,BH2,H}; see e.g. \cite{Witten:2010cx} for a physicist-oriented review, and see appendix A of \cite{Held:2026huj} for an even more condensed review that nevertheless explicitly discusses contours with finite endpoints.

In particular, let us  
consider a finite-dimensional integral of the schematic form
\begin{equation}
Z=\int_{\mathcal C} \mathrm{d}^n z \; e^{-S_E(z)} ,
\label{eq:schematic}
\end{equation}
where $z\in\mathbb{C}^n$ and $S_E(z)$ is holomorphic in the domain of interest.  We assume that the integral converges absolutely. Critical points $z_\sigma$ are defined by
\begin{equation}
\partial_i S_E(z_\sigma)=0 .
\label{eq:crit}
\end{equation} 
Associated to each critical point is a Lefschetz thimble $\mathcal J_\sigma$, defined as the union of all\footnote{In particular, in $n=1$ complex dimension, and for non-degenerate saddles, the thimble is the curve defined by the union of precisely two such flows.} downward gradient-flow trajectories of $e^{-S_E}$, 
\begin{equation}
\frac{d z^i}{d\lambda}=\overline{\partial_i S_E(z)} ,
\label{eq:downflow}
\end{equation}
with boundary condition $z(\lambda)\to z_\sigma$ as $\lambda\to-\infty$. 
Thus $\mathrm{Re}\,S_E$ increases monotonically along the flow away from the critical point while $\mathrm{Im}\,S_E$ is constant. Hence the integrand has a fixed phase on each thimble and the integral of $e^{-S_E}$ generically converges exponentially when performed along each $\mathcal J_\sigma$. The exceptional cases are known as Stokes' rays, Stokes' phenomena, or Stokes' transitions, and occur when the downward flow from a saddle $\sigma$ ends at another saddle $\sigma'$ rather than flowing to $S_E = +\infty$.  We set aside such exceptional cases for now, though we will return to discuss them below.

Note that in \eqref{eq:downflow} we have implicitly introduced a flat metric $\delta^{ij}$ on the space of parameters $z^i$ and used it to relate the vector on the left-hand-side to the covector on the right.  In general, one can use any smooth positive-definite real-valued metric. For $n>1$, this choice will in general modify $\mathcal{J}_\sigma$. However, the final results for the saddle point-approximated integrals as stated below will not depend on this choice.  Furthermore, for $n=1$ complex dimension changes in this metric can  only change the choice of parameter $\lambda$ along each $\mathcal{J}_\sigma$.

For integration contours ${\mathcal C}$ with  a non-trivial boundary $\partial {\mathcal C}$ in ${\mathbb C}^n$, it is also useful to define a thimble ${\mathcal J}_{bndy}$ given by the union of downward flows away from all points of $\partial {\mathcal C}$. Here we take  $\partial {\mathcal C}$ to include asymptotic boundaries at which the decay of the integrand is slower than exponential. 
When one of the saddles lies on the boundary, we take ${\mathcal J}_{bndy}$ to include only those flows from the saddle point that are continuous limits of flows from nearby points of 
$\partial {\mathcal C}$.

When the integrand $e^{-S_E}$ is holomorphic, and away from Stokes' transitions,  the original integration cycle $\mathcal C$ can be continuously deformed until it becomes an integer linear combination of thimbles,
\begin{equation}
 \mathcal C\sim {\mathcal J}_{bndy}+ \sum_\sigma n_\sigma\, \mathcal J_\sigma ,
\label{eq:decomp}
\end{equation}
where the coefficients $n_\sigma\in\mathbb{Z}$ are intersection numbers between $\mathcal C$ and the corresponding upward-flow cycles ${\mathcal K}_\sigma$ (also called ascent contours or dual thimbles) defined by changing the sign on the right-hand side of \eqref{eq:downflow}. Here we take $\sigma$ to range only over saddles that do {\it not} lie on the boundary $\partial {\mathcal C}$; such saddles will contribute only as part of the boundary contribution associated with ${\mathcal J}_{bndy}$. Note that  $\mathcal{J}_{bndy}$ has a boundary given by $\partial \mathcal{C}$, while $\mathcal{J}_\sigma$ has no boundary except at parameter-values where the saddle $\sigma$ experiences a Stokes'  phenomenon. 

Cauchy's theorem then guarantees that the above deformation cannot change the value of our integral. 
This result can also be extended without change to the case where $e^{-S_E}$ has singularities when those singularities lie at endpoints of the integration contour; see e.g. the discussion in appendix A of \cite{Held:2026huj}. 

In the main text we will consider a slightly more general form of the integral:
\begin{equation}
    Z = \int_\mathcal{C} \mathrm{d}^n z \, g(z) e^{-N^2 S_E(z)},
\end{equation}
where we take $g(z)$ and $S_E(z)$ to be holomorphic in the relevant regime and we want to study the limit $N \to \infty$. We may use the contour decomposition with respect to $S_E$ to arrive at
\begin{subequations}
\begin{equation}
    Z = \int_{\mathcal{J}_{bndry}} \mathrm{d}^n z\ g(z) e^{-N^2 S_E(z)} +\sum_\sigma n_\sigma e^{-i N^2 {\rm Im} \, S_E(z_\sigma)} \int_{\mathcal{J}_\sigma} \mathrm{d}^n z\ g(z) e^{-N^2 \Re S_E(z)}.
\end{equation}
In the large $N$ limit, these integrals can be approximated as
\begin{equation}
  \int_{\mathcal{J}_\sigma} \mathrm{d}^n z\ g(z) e^{-N^2 S_E(z)} \approx \left(\frac{2\pi}{N^2} \right)^{\frac{n}{2}} e^{-N^2 S_E(z_\sigma)} \left( g(z_\sigma) \left(\det S_E''(z_\sigma)\right)^{-1/2}
    + O\left(N^{-2}\right)
    \right),
\end{equation}
and
\begin{equation}
\int_{\mathcal{J}_{bndry}} \mathrm{d}^n z\  g(z) e^{-N^2 S_E(z)} \approx \frac{1}{N^2}\int_{\partial C} \mathrm{d}^{n-1} y\  e^{-N^2 S_E(y)} \left(\frac{g(y)}{\partial_n S_E(y)} +O\left(N^{-2}\right)
\right),
\end{equation}
where $S_E''(z_\sigma)$ is a Hessian of $S_E$ at the critical point (and the suitable branch of the square root is understood),  $\mathrm{d}^{n-1} y$ denotes a pull-back of $\mathrm{d}^n z$ to the boundary and $\partial_n S_E(y)$ is the normal derivative in the direction towards the contour. The integral along the boundary can then be further approximated using the same techniques. 
\end{subequations}

Returning to the decomposition \eqref{eq:decomp} of the integration cycle ${\cal C}$, it is important to recall that, 
while the result of the above integral  generally depends continuously on external parameters at each finite $N$, 
the ascent thimbles ${\cal K}_\sigma$   can nevertheless jump discontinuously when the parameters are varied past certain thresholds. 
Such jumps are the Stokes' transitions (or Stokes' phenomena) mentioned above.

\begin{figure}[h!]
    \centering
    {{\includegraphics[width=0.75\linewidth]{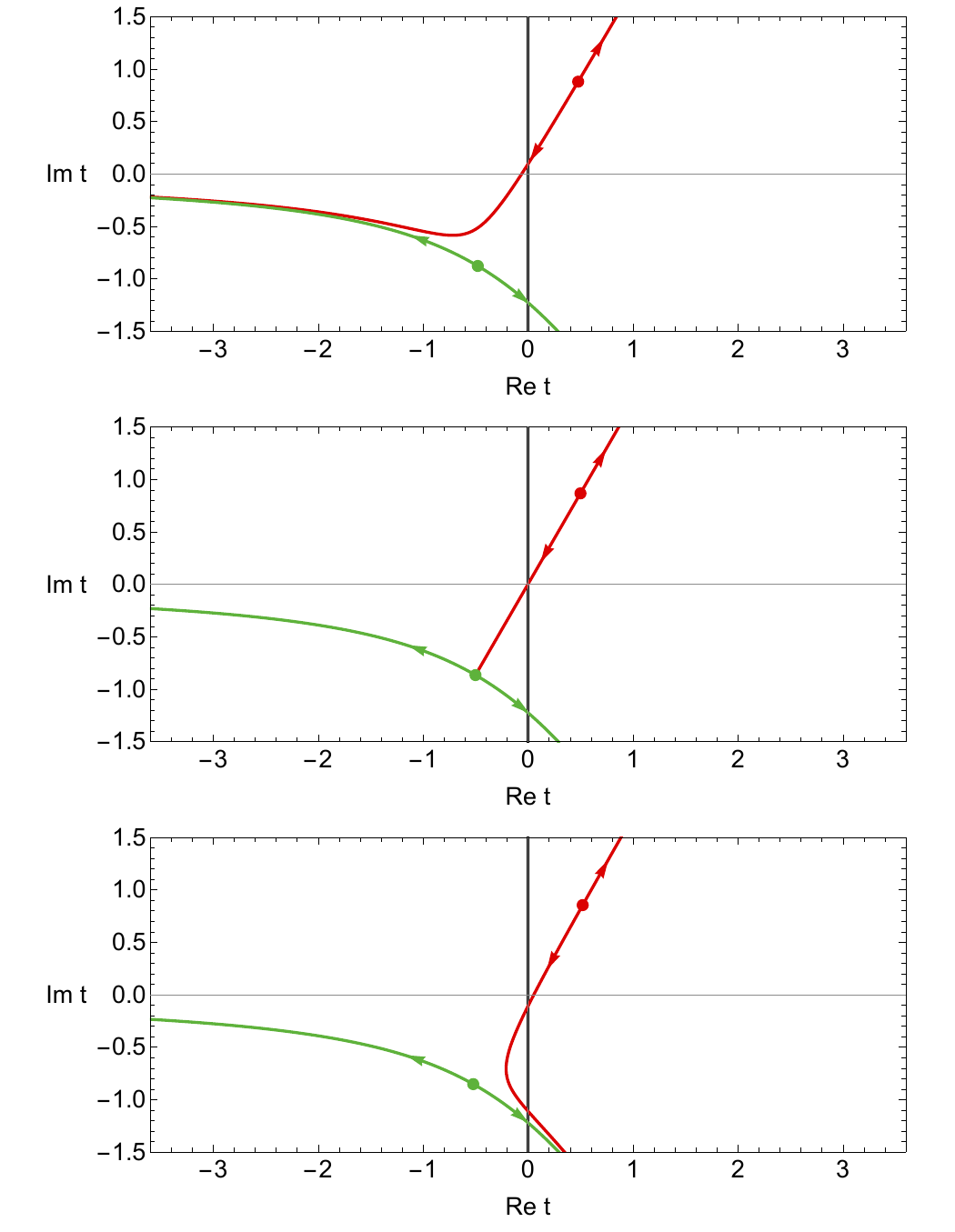}}}
    \caption{Saddles (dots) and the associated ascent thimbles (lines with arrows showing the ascent directions) are shown for the Airy integral \eqref{eq:airy} for values of $z$ near a Stokes' transition, with $z=e^{i(2\pi/3+0.05)}$, $z=e^{2\pi i/3}$, and $z=e^{i(2\pi/3-0.05)}$ in the top, middle, and bottom panels, respectively. Here the red and green saddles are $t=+\sqrt{z}$ and $t=-\sqrt{z}$, and the green saddle catalyzes a transition for the red saddle.  In the region shown, the integration contour ${\cal C}$ coincides with the imaginary axis (heavy vertical black line).
    The (net) red intersection number $n_{red}$ is thus non-zero in the top panel, but vanishes in the bottom panel (where the two local intersections have opposite signs and cancel). The intersection number is ill-defined at the precise moment of transition (middle panel).}
\label{fig:airy}    \end{figure}

A standard example of such a Stokes' transition is shown in figure \ref{fig:airy} below.  This example is associated with the Airy integral 
\begin{equation}
Ai(z) = \int_{\cal C} dt \, e^{-\frac{t^3}{3}}e^{zt},
\label{eq:airy}
\end{equation}
where
for convergence the contour $\cal C$ is defined by choosing some $\phi$ between $\pi/2$ and $5\pi/6$ and taking $\cal C$ to run from infinity with phase $e^{-i\phi}$ (in the lower left quadrant) to infinity with phase $e^{+i\phi}$ (in the upper left quadrant).  For simplicity, however, we take $\cal C$ to coincide with the imaginary axis in the finite region shown in the figure.   

As illustrated by this example, an ascent thimble ${\cal K}_\sigma$ can change discontinuously when it hits another saddle. In figure \ref{fig:airy} this occurs on the part of the red ascent thimble that flows downward and to the left, which changes from following the upper part of the green ascent contour to instead following the lower part of the green ascent contour.  In this context we will say that the green saddle {\it catalyzes} a Stokes' transition for the red saddle.

In the example shown in figure \ref{fig:airy}, the red ascent thimble intersection number changes from +1 (with the proper choice of orientation) in the upper panel to zero in the lower panel (where there are two intersections of opposite sign that cancel to give a net zero).   Note that this change is associated with the fact that the lower part of the green ascent thimble intersects ${\cal C}$ but the upper part does not, giving the full green ascent thimble (the union of these two parts) a net non-zero intersection number with ${\cal C}$.  This is a general feature that will be useful in the analysis of section \ref{subsec:one-dimensional-BPS-integral} (and which holds in all dimensions):  When a saddle $\sigma'$ (having  intersection number $n_{\sigma'}$ with ${\cal C}$) catalyzes a Stokes' transition for the saddle $\sigma$, the intersection number with ${\cal C}$ of the upward thimble ${\cal K}_\sigma$ changes by $\pm n_{\sigma'}$.  In particular, the Stokes' transition leads to a change of intersection numbers only when the ascent contour from the catalyzing saddle  $\sigma'$ (green, in the example) has non-zero intersection number $n_{\sigma'}$ with the contour of integration.  In other words, this occurs only at parameters for which the catalyzing saddle  $\sigma'$ contributes to the desired integral.

Discontinuities in $n_\sigma$ can also clearly occur at parameters where the upward thimble ${\cal K}_\sigma$ flows to the boundary $\partial {\mathcal C}$ of the integration contour.  In this case we say that the transition is catalyzed by the boundary $\partial {\mathcal C}$.  But since either sort of transition is defined by the upward-flow thimble  ${\cal K}_\sigma$, the catalyzing object (a saddle $\sigma'$ or the boundary $\partial {\cal C}$) must be associated with a magnitude of the integrand that is larger than the magnitude at $\sigma$ (and thus with ${\rm Re} \, S_E$ smaller than at $\sigma$). 
In particular, if a given saddle $\sigma$ gives the dominant contribution at any given value of the parameters, then at such parameter values the corresponding $n_\sigma$ is locally-constant and cannot experience jumps.

\section{Fermion determinants and the BPS ansatz}
\label{sec:BPSansatz}

The main goal of this work is to explore how the ansatz \eqref{eq:ansatz}, or a suitable modification thereof, can be used to study the AdS$_5$ supersymmetric index (and supersymmetric indices more generally).  As noted in the introduction, this ansatz is expected to capture  the formal leading-order semiclassical behavior of the desired Lorentzian path integral, but it does not incorporate quantum corrections.   

When used to compute partition functions at real potentials that are not parametrically large, and for $\ell_{Planck} \ll \ell_{AdS}$, one expects quantum corrections to be parametrically small. 
But at certain complex values of the potentials quantum corrections can in fact significantly modify the results of supposedly-leading-order saddle-point calculations.  Famous examples of such effects involve fermion zero modes, which can cause the 1-loop fermion determinant multiplying a semiclassical contribution to vanish (and with corresponding vanishing coefficients at all higher loop orders).  This removes certain saddle-point contributions entirely and, as a result, it can clearly change even the dominant term in any semiclassical expansion. 

Recall that one expects related effects to be important in our context when we choose complex potentials that satisfy \eqref{eq:SUSYPot}, and which thus promote our partition function $Z$ to a supersymmetric index ${\mathcal I}$.   This can be seen in that case from the fact that supersymmetry should render the index
${\mathcal I}$ independent of $\beta$ at fixed $\tau,\sigma, \vec \Delta$.  In particular, as discussed in e.g. \cite{Iliesiu:2021are}, satisfying \eqref{eq:SUSYPot} implies that, in the semiclassical approximation, the linearized theory about any non-BPS bulk saddle will have a  fermion zero mode that indeed sets all contributions from this saddle to zero.  That leaves only contributions from BPS saddles  which, as desired, are independent of $\beta$ at fixed $\tau,\sigma, \vec \Delta$.

It is also clear that quantum corrections from fermions can also have significant effects even without invoking the saddle-point approximation.  This can again be seen from the fact that, for potentials satisfying the supersymmetry condition \eqref{eq:SUSYPot}, the partition function must become independent of $\beta$ at fixed $\tau,\sigma, \vec \Delta$.  In particular, if we neglect quantum corrections, then in direct analogy with the representation \eqref{eq:index_def} for the CFT partition function, the integrand of our bulk ansatz \eqref{eq:ansatz} depends on $\beta$ through a factor $e^{-\beta \tilde E}$ where $\tilde E=E-J_1-J_2-(Q_1+Q_2+Q_3)/2$ is the energy relative to the BPS bound.  At small $\beta$ it would thus receive contributions from a broad range of $\tilde E$, while large $\beta$ it would thus localize on the BPS locus $\tilde E=0$.  

Quantum corrections, and fermionic integrals in particular, must thus conspire to remove the associated apparent dependence on $\beta$.  Nevertheless, localization on the surface $\tilde E=0$ as $\beta \rightarrow \infty$ is an extremely natural property that we expect to survive even large such corrections.  In particular, a further argument for this localization is given by noting that, if we consider potentials satisfying the supersymmetry condition \eqref{eq:SUSYPot},
there is a well-defined generator of an asymptotic supersymmetry transformation. This generator has a non-trivial action on the gravitino, so for general charges it is associated with a fermionic zero mode when studied perturbatively about generic configurations over which we integrate in the ansatz \eqref{eq:ansatz}.  

However, the zero-mode becomes trivial as the charges approach the surface $\tilde E=0$ on which the BPS condition holds.  To see this, recall from \cite{Marolf:2022ybi,Chen:2025leq,Held:2026huj,Kolanowski:2026gii} that for $\tilde E\neq0$, our ansatz \eqref{eq:ansatz} integrates over constrained saddles 
that solve the field equations away from a codimension-2 singularity at what in Lorentz-signature would be called the horizon bifurcation surface.  However,  real Lorentz-signature black holes necessarily become extremal as $\tilde E \rightarrow 0$. And while the actual geometries associated with the integrand of \eqref{eq:ansatz} have Euclidean time-period $\beta$, they are in fact just the 
$T\rightarrow-i\beta$ analytic continuation of periodically-identified real Lorentz-signature black holes.
This means that for $\tilde E=0$ there is in fact no bifurcation surface (see again footnote \ref{foot:inner} for comments on why such black holes are nevertheless included), and thus no violations of the equations of motion.  In other words, the $\tilde E=0$ configurations on our integration contour are smooth BPS solutions. They thus have a globally-defined Killing spinor, so that the above fermion zero-mode becomes trivial.

This provides the additional argument for localization.  Recall that the ansatz \eqref{eq:ansatz} is to apply only after integrating over all degrees of freedom other than $A,\vec J,\vec Q$.
In deriving this ansatz, we must in some sense integrate over the above fermionic zero mode. More precisely, the above structure of perturbative zero modes is precisely what one finds in standard models \cite{Blau:1992pm,Witten:1982im} of supersymmetric localization so, by analogy, one expects the index to localize on the $\tilde E=0$ surface in the limit $\beta \rightarrow \infty$ in just the way discussed above.  It would then also give the same value  at each finite $\beta$ (though in such cases it would do so without being manifestly localized).

Since we see no further argument for fermionic zero modes, it may appear natural to assume that quantum corrections are indeed subleading in the semiclassical expansion  when $\beta$ is  sufficiently large and positive.  In that case we should simply take $\beta \rightarrow \infty$ in \eqref{eq:ansatz}  or, equivalently at leading semiclassical order, insert the Dirac delta-function $\delta(\tilde E)$ into the ansatz \eqref{eq:ansatz} to write:
\begin{eqnarray}
\label{eq:indexansatz}
{\mathcal{I}}(\tau, \sigma, \vec \Delta) &=& 
\sum_{{{n_{J_1},n_{J_2},n_{Q_1},n_{Q_2},n_{Q_3} \in {\mathbb Z}} \atop {n_{J_1}+ n_{J_2}+n_{Q_1}+ n_{Q_2} +n_{Q_3} \in 2{\mathbb Z}}}} 
 \mathcal{I}_{n_{J_1},n_{J_2},n_{Q_1},n_{Q_2},n_{Q_3}}
\ \ \ {\rm with} \nonumber \\
 \mathcal{I}_{n_{J_1},n_{J_2},n_{Q_1},n_{Q_2},n_{Q_3}}
 &\approx&
\int d{\mathcal A} \, dJ_1  \, dJ_2  \, dQ_1 \, dQ_2 \, dQ_3 \, \delta(\tilde E) \, 
 \Bigl[e^{{\mathcal A}/4G_5} 
 e^{2\pi i (\tau J_1 + \sigma J_2 +\frac{1}{2}\sum_{i=1}^3 \Delta_i Q_i)}
 \nonumber \\ 
&\times&
e^{2\pi i \left(\sum_{i=1}^2 n_{J_i} J_i +\sum_{j=1}^3 n_{Q_j} \frac{Q_j}{2}  
\right)} 
\Bigr],
\end{eqnarray}

In interpreting \eqref{eq:indexansatz}, it is important to
recall that, due to the non-linear nature of the BPS condition $\tilde E=0$ when expressed in terms of the variables ${\mathcal A}, \vec J, \vec Q$, the real BPS surface defined by real Lorentz-signature black holes with $\tilde E=0$ will generally have codimension greater than one within the allowed space of real  ${\mathcal A}, \vec J, \vec Q$.  The symbol $\delta(\tilde E)$ should thus be interpreted as the appropriate distribution to localize the above integral to the surface $\tilde E=0$; i.e., it is not just a strict one-dimensional Dirac delta-function.

The observant reader will note that the above argument glosses over the fact that analyses of Euclidean Schwarzian modes about full saddles suggest that there {\it will} in fact be further large corrections in this regime. However, since a Schwarzian mode analysis \cite{Boruch:2022tno} of BPS black holes leads to $e^{S_{BH}}$ exactly-degenerate ground states, with $S_{BH}$ still given approximately by the semiclassical BPS entropy, and then to a gap, such corrections  should not affect the argument that localization occurs and should at most induce a well-defined separation between BPS and non-BPS contributions even at finite $\beta$ (where, in our context, the latter must in fact cancel among themselves).

Let us now finish our discussion of fermionic quantum corrections by providing a sketch of how a proper derivation of \eqref{eq:indexansatz} might proceed.   Since the ansatz \eqref{eq:ansatz} results in an integral over the bosonic parameters ${\cal A}, \vec J, \vec Q$, one may expect that the ansatz can be extended to integrate over a corresponding set of fermionic partners so that the extended ansatz has a manifest supersymmetry in terms of which $\tilde E$ is $Q$-exact.   The result \eqref{eq:indexansatz} would then follow naturally from the associated supersymmetric localization. We hope to study this argument in detail in the future.  

\subsection{Consistency of the localized ansatz in the saddle-point approximation}
\label{subsec:consistency}

For the above reasons, we will simply study the ansatz \eqref{eq:indexansatz} in section \ref{subsec:one-dimensional-BPS-integral} below.  However, before doing so, let us note that there is an important self-consistency condition that should be checked when performing that analysis.

To understand the consistency condition, recall again that our ansatz assumes quantum corrections, and in particular the effects of fermion zero modes, to be perturbatively small in the limit $\beta\rightarrow \infty$.  However, as noted above, even for potentials that satisfy \eqref{eq:SUSYPot}, for any finite $\beta$ the perturbative theory about any non-BPS saddle or boundary contribution should come equipped with a fermionic zero mode that kills all contributions from that saddle (or boundary contribution), thus potentially changing even the leading behavior at small $G_5$.  

Let us therefore further consider the BPS-only ansatz \eqref{eq:indexansatz}.  The factor of $\delta(\tilde E)$ can of course be used to write the ansatz as an integral over the real section (as defined by charges $(E, \vec J, \vec Q)$ associated with real Lorentz-signature black holes) of the BPS surface $\tilde E=0$.  Since a black hole defined by a point on this surface must also be extremal, we denote the surface by ${\mathcal {E}^{BPS}}_{\mathbb R}$.  Any saddles of \eqref{eq:indexansatz} must then lie on the associated complexified surface ${\mathcal {E}^{BPS}}_{\mathbb C}$ defined by complex black hole solutions that are again both extremal (in the sense of having degenerate horizons) and BPS.  A Picard-Lefschetz analysis of \eqref{eq:indexansatz} will then define associated flows within ${\mathcal {E}^{BPS}}_{\mathbb C}$.

Let us now compare these saddles and flows with those that result from the Picard-Lefschetz analysis of the non-BPS ansatz \eqref{eq:ansatz} at large positive values of $\beta$.  In this discussion we assume that (as is true in the AdS$_5$ case of interest below) the function $E({\mathcal A}, \vec J, \vec Q)$ is sufficiently simple that as $\beta \rightarrow \infty$ these structures either converge to points and curves on $\mathcal{E}^{BPS}_{\mathbb C}$ or that they diverge to infinity (and, in particular, that they do not simply oscillate wildly at large $\beta$).  In that case, one can use perturbation theory in the temperature $\beta^{-1}$ around a saddle $\sigma_\infty$ of \eqref{eq:indexansatz} to construct corresponding families of saddles $\sigma_\beta$ for the non-BPS ansatz \eqref{eq:ansatz}.  
Indeed, at least in regions where the action is smooth, one can similarly use perturbation theory around the flows from $\sigma_\infty$ in $\mathcal{E}^{BPS}_{\mathbb C}$ to construct flows from 
$\sigma_\beta$ associated with the non-BPS ansatz \eqref{eq:ansatz} (see appendix \ref{subsec:largebpert})\footnote{The systems we study below will have saddles of the BPS-only action at points where derivatives of the full action diverge.  Nevertheless, the form of the action is such that the relevant flows can again be constructed perturbatively in $\beta^{-1}$ by making an appropriately-singular change of parameterization on the space of allowed black hole horizons.  }.  

In particular, if an upward-flow from $\sigma_\infty$ intersects the integration contour ${\mathcal {E}^{BPS}}_{\mathbb R}$ of \eqref{eq:indexansatz}, then for large $\beta$ there must be\footnote{Here we use the fact that, since the defining contour ${\mathcal C}_{ansatz}$ of \eqref{eq:ansatz} includes both inner and outer horizons, the surface ${\mathcal {E}^{BPS}}_{\mathbb R}$ lies in the interior of ${\mathcal C}_{ansatz}$ except at the locus $\partial{\mathcal {E}^{BPS}}_{\mathbb R}$ at which the horizon area ${\mathcal A}$ vanishes.  However, it is natural to conjecture that finite-$\beta$ corrections always shift the flow in a direction that increases ${\cal A}$, and thus into the interior of the region of ${\mathcal C}_{ansatz}$ associated only with {\it outer} horizons. We also ignore cases where the flow defined by \eqref{eq:indexansatz} intersects ${\mathcal {E}^{BPS}}_{\mathbb R}$ on its boundary $\partial{\mathcal {E}^{BPS}}_{\mathbb R}$, as in such cases the saddle should be considered to lie on a Stokes' ray where the associated intersection number $n_{\sigma_\infty}$ is ill-defined. } a corresponding upward-flow from $\sigma_\beta$ that intersects the defining contour ${\mathcal C}_{ansatz}$ of the original non-BPS ansatz \eqref{eq:ansatz}. Furthermore, since any upward-flow from $\sigma_\beta$ must maintain a constant phase for the integrand of \eqref{eq:ansatz}, at large $\beta$ the imaginary part of $\tilde E$ must remain close to zero over any finite region of the space of charges $\vec J, \vec Q$ and, as a result,  any upward-flow from $\sigma_\beta$ that does not approach the $\tilde E=0$ surface at large $\beta$ will have ${\rm Re} \, \tilde E <0$.  It thus cannot approach ${\mathcal C}_{ansatz}$, on which the magnitude of the integrand is bounded even at $\beta=\infty$; see comments in the first new paragraph below \eqref{eq:ansatz3}.  

This means that at large $\beta$ the intersection number $n_{\sigma_\beta}$  associated with the non-BPS integral \eqref{eq:ansatz} must agree with the intersection number $n_{\sigma_\infty}$ associated with the BPS integral \eqref{eq:indexansatz}.  In much the same way, any boundary contribution to the semiclassical expansion of \eqref{eq:indexansatz} will be associated with the large-$\beta$ limit of either boundary contributions to \eqref{eq:ansatz} or saddle-point contributions to \eqref{eq:ansatz}.

There are now two possibilities.  The first is simply that the family of saddles $\sigma_\beta$ that approaches $\sigma_\infty$ (or the corresponding family of boundary contributions) remains exactly BPS at finite values of $\beta$.  In this case there are no manifest fermion zero modes and it is consistent to believe that quantum corrections are small at large $\beta$.

However, it is also possible that the saddles $\sigma_\beta$ have $\tilde E\neq 0$ for finite $\beta$, but with $\tilde E \rightarrow 0$ as $\beta \rightarrow \infty$.  We will see examples of such saddles in sections \ref{subsec:NonBPSSUSY} and \ref{AdS4}.  At finite $\beta$ there is then a manifest fermion zero mode in perturbation theory around each finite-$\beta$ saddle that should remove all contributions from such saddles.  But the BPS ansatz \eqref{eq:indexansatz} explicitly ignores any such quantum corrections and so would then fail to be a good approximation. 

This observation  defines the consistency check foreshadowed above.  For any saddle $\sigma_\infty$ that contributes non-trivially to the BPS integral \eqref{eq:indexansatz}, we need to check that it is the limit of saddles $\sigma_\beta$ of \eqref{eq:ansatz} that have $\tilde E=0$ even at finite $\beta$.  There is also a corresponding consistency condition on boundary contributions to the semiclassical expansion of \eqref{eq:indexansatz}, and which might involve either boundary contributions at finite $\beta$ or families of finite-$\beta$ saddles $\sigma_\beta$ for which $\tilde E$ vanishes at large $\beta$ while other charges diverge. Note, however, that there is no corresponding constraint on saddles $\sigma_\infty$ whose upward-flow cycles have trivial intersection with $\mathcal{E}^{BPS}_{\mathbb C}$ (since a Picard-Lefschetz analysis already shows that they fail to contribute at either finite or infinite $\beta$).  Note also that we may study this consistency condition separately for each term ${\mathcal I}_{n_{J_1},n_{J_2},n_{Q_1},n_{Q_2},n_{Q_3}}$. 

We may now also ask whether there are additional consistency conditions involving families of saddles $\sigma_\beta$ for the non-BPS ansatz \eqref{eq:ansatz} at finite $\beta$ for which $\tilde E$ instead fails to vanish as $\beta \rightarrow \infty$.  But such saddles have no analogs in the analysis of our BPS ansatz \eqref{eq:indexansatz}, and they also
clearly have fermionic zero modes that will prohibit them from contributing to the desired index.  Thus they cannot raise issues for the consistency of our BPS ansatz\footnote{One could also ask the slightly different question of whether they raise issues regarding the conjecture that the non-BPS ansatz \eqref{eq:ansatz} becomes valid in the limit of large $\beta$.  Consistency of \eqref{eq:ansatz} with the manifest fermion zero-modes would then require that a Picard-Lefschetz analysis of \eqref{eq:ansatz} finds no contribution from such saddles $\sigma_\beta$ at large $\beta$.  This is clear if the magnitude of the integrand at $\sigma_\beta$ diverges as  $\beta \rightarrow \infty$, as then the upward-flow cycles cannot approach the contour ${\mathcal C}_{ansatz}$ (on which the magnitude of the integrand is bounded).  It is also clear if this magnitude vanishes as  $\beta \rightarrow \infty$ as any saddle-point contributions would then also vanish in that limit.  But there might indeed be an additional condition to check if the magnitude of the integrand were to approach a finite value without $\tilde E$ approaching zero.  Luckily, this does not occur for the model studied below.}.

Unfortunately, we will see in section \ref{subsec:one-dimensional-BPS-integral} that there are generally regions of the space of potentials in which interesting systems fail the above consistency test.  This indicates that fermion zero modes have additional large effects for such potentials, and that further modification of \eqref{eq:indexansatz} is required.  While we do not understand such modifications in detail, we give a proposal below for their effect on the semiclassical expansion of the index.  This proposal is again subject to an important self-consistency test, though it that is strictly weaker than the one discusses above and, in particular, the new test turns out to be satisfied by the systems studied below.

\subsection{Truncation of the sum over saddles and a weaker consistency condition}
\label{subsec:truncate}

In section \ref{sec:BPSansatz} we will find values of the potentials that  satisfy \eqref{eq:SUSYPot} and for which the BPS-only ansatz \eqref{eq:indexansatz} receives contributions from saddles that are non-BPS at finite $\beta$.  We will refer to such saddles as being only asymptotically BPS (as opposed to the `truly' BPS saddles that remain BPS at finite $\beta$).  Such contributions violate the self-consistency condition of section \ref{subsec:consistency} for quantum corrections to \eqref{eq:indexansatz} be small, so further modification of \eqref{eq:indexansatz} is required in these regimes.

Luckily, at least for the systems we study below, such violations occur only in relatively small regions of the space of potentials.  Furthermore, in the semiclassical limit of our particular systems,  it is natural to suppose that the required additional large quantum effects act only to truncate the sum over saddles  to the truly-BPS saddles, removing the saddles that are only asymptotically-BPS. 
In some cases one may also propose a similar truncation of the boundary contribution. In particular, both such effects seem likely to be a natural result of the fact that only-asymptotically-BPS saddles do not have Killing spinors at finite $\beta$ (where one understands the relevant boundary conditions at the horizon), so that they may well receive quantum corrections which make them in fact fail to saturate the BPS bound even in the limit $\beta \rightarrow \infty$.

It is important to note, however, that the proposal to truncate the sum over saddles in this way is again subject to an important self-consistency check, though one that is weaker than the condition discussed in section \ref{subsec:consistency} above.  The point here is that one should in principle be able to perform the fermion integrals at an early stage and to then use the results to modify the integrand of the BPS-only ansatz \eqref{eq:indexansatz}.  We imagine that doing so will yield a good approximation to the desired index.  We are thus proposing that the resulting saddle-point expansion will agree with that given by simply studying the original BPS-only ansatz \eqref{eq:indexansatz} and then truncating the sum over saddles to those that are truly-BPS.  Consistency of this proposal then requires that the truncated sum over saddles is in fact the saddle-point approximation to {\it some} integral over the real BPS contour ${\mathcal E}^{BPS}_{\mathcal R}$.

However, the truncated sum will generally be associated with intersection numbers $n_\sigma$ that undergo jumps.  These jumps arise from Stokes' transitions of the full BPS-ansatz \eqref{eq:indexansatz}.  But recall from section \ref{sec:PLO} that a jump in $n_\sigma$ for some saddle $\sigma$ must be catalyzed by a saddle $\sigma'$ that actually contributes to the desired integral (or by the boundary $\partial {\mathcal C}$ of the integration contour).  As a result, if in the semiclassical expansion of \eqref{eq:indexansatz} a truly-BPS saddle $\sigma$ experiences a jump catalyzed by a saddle $\sigma'$ that is only asymptotically BPS, removing the $\sigma'$ term from the sum may leave a result that is no longer the saddle-point approximation to any integral.

The consistency condition for the truncated-sum proposal is thus that, for each truly-BPS saddle $\sigma$ for the BPS-only ansatz \eqref{eq:indexansatz},  every jump in $n_\sigma$ is catalyzed either by another truly-BPS saddle or by the boundary $\partial {\mathcal C}$.    This weaker consistency condition will in fact be satisfied for the systems we study below.  

Interestingly, however, the results of appendix \ref{sec:thimbles} show that even this weaker condition would fail for our systems if it were applied to the non-BPS ansatz \eqref{eq:ansatz} described in the introduction.
At finite $\beta$, non-BPS saddles for \eqref{eq:ansatz} do in fact catalyze jumps in $n_\sigma$ for BPS saddles $\sigma$.  Thus the restriction of the contour of integration for \eqref{eq:indexansatz}
to the real BPS locus ${\mathcal E}^{BPS}_{\mathbb R}$ remains an important ingredient of our proposal, even though we must still remove by hand contributions from saddles that are only asymptotically BPS.

\section{Brief Review of AdS$_5$ Black Holes}
\label{sec:BHreview}

The last ingredient we will need to study the above proposal for the semiclassical expansion of the index ${\mathcal I}$ is an understanding of the black holes that we wish to include.   Our brief review here largely  follows that of  \cite{Chong:2005hr} and \cite{Aharony:2021zkr}, working in the five-dimensional gauged supergravity obtained as a consistent truncation of
type IIB supergravity on $S^5$. Using this truncation requires imposing $\tau=\sigma$ and $\Delta_1=\Delta_2=\Delta_3$.  We do so for simplicity, but this restriction may also be motivated by the analysis of \cite{Aharony:2021zkr} suggesting that no black hole saddles (or quotients thereof) can contribute to the (unrefined) index when this condition fails to be satisfied\footnote{This was due to an instability of all BPS black hole saddles to emit perturbative D3 branes, in the sense that perturbatively adding a positive-real amount of M5-brane charge lowers the real part of the action.  While it would be useful to rederive this criterion from a Lorentzian starting point, we save such an analysis for future work. \label{foot:braneinstability}}.

We thus consider maximal supergravity with  gauge group $SO(6)$.
The three Cartan generators of $SO(6)$ correspond to three independent $U(1)$ gauge fields,
and the associated conserved charges are the three R-charges of the dual $\mathcal{N}=4$ SYM.
Restricting to the $U(1)^3$ truncation yields the STU model \cite{Cvetic:1999xp}. 

In this paper we further impose
the \emph{equal-charge} condition, i.e.\ we set the three $U(1)$ charges equal, which consistently
reduces the theory to minimal $\mathcal{N}=2$ gauged supergravity with a single gauge field $A$.
The bosonic action is then\footnote{Following \cite{Aharony:2021zkr}, we have rescaled the gauge fields of \cite{Chong:2005hr} by a factor of $\frac{\sqrt{3}}{2}$.}
\begin{align}
\frac{1}{16\pi G_5}\int \Big[
(R+12)\,\star 1
-\frac{2}{3}\,F\wedge \star F
+\frac{8}{27}\,F\wedge F\wedge A
\Big],
\label{eq:5d-action}
\end{align}
where $F=dA$ and in \eqref{eq:5d-action} (and throughout this work) we have set the AdS radius to one. 

Charged rotating stationary (real) Lorentz-signature AdS$_5$ black hole solutions of this theory, with general
angular momenta, were found in \cite{Chong:2005hr}.
Here we focus on the equal-angular-momentum special case, for which the metric and gauge field can be written
\begin{align}
ds^2
={}&
-\frac{\left(1+r^2\right)}{\Xi_a} d t^2-\frac{2 q}{\Xi_a \rho^2} \nu \varpi+\frac{f_t}{\Xi_a^2 \rho^4} \varpi^2+\frac{\rho^2}{\Delta_r} d r^2
\nonumber\\
&+\frac{\rho^2}{\Xi_a}\left(d \theta^2+\sin ^2(\theta) d \phi^2+\cos ^2(\theta) d \psi^2\right)
\nonumber
\\[1ex]
A={}&
\frac{3 q}{2 \rho^2 \Xi_a} \varpi-\alpha d t,
\label{eq:metric}
\end{align}
and where we have introduced the one-forms and functions
\begin{equation}
\begin{aligned}
\nu & =a\left(\sin ^2(\theta) d \phi+\cos ^2(\theta) d \psi\right), & \varpi & =d t-\nu, \\
\Delta_r & =\frac{\rho^4\left(1+r^2\right)+q^2+2 a^2 q}{r^2}-2 m, & \rho^2 & =r^2+a^2, \\
f_t & =2\left(m+a^2 q\right) \rho^2-q^2, & \Xi_a & =1-a^2.
\end{aligned}
\end{equation}
The solutions are parameterized by $q\in {\mathbb R}$,  $a^2<1$, and by appropriate positive $m$. At fixed $q,a$ the  allowed range of $m$ is of the form $m \ge m_E(a,q)$, with $m_E(a,q)$  being the mass at which the horizon becomes extremal.

The angular coordinates $\phi$ and $\psi$ on the three-sphere have period $2\pi$.
Taking $r$ to be large and defining 
$z=\sqrt{\Xi_a}/r$ shows that the metric asymptotically approaches $d s^2 = z^{-2}\bigl(-dt^2 + d z^2 + d\Omega_3^2\bigr) + \mathcal{O}(z^0)$.  The metric is thus asymptotically AdS$_5$.  In particular,  in this conformal frame the boundary metric is just that of  ${\mathbb R} \times S^3$, where the $S^3$ factor has been written in coordinates for which
$d\Omega_3^2 = d\theta^2 + \sin^2\theta\, d\phi^2 + \cos^2\theta\, d\psi^2$.


From the metric~\eqref{eq:metric} we may read off the conserved charges.
The solution is parametrized by three constants $(a,q,m)$. It is often convenient to trade $m$ for a horizon radius 
$r_0$ that satisfies  $\Delta_r(r_0)=0$, in terms of which (and for any such root $r_0$) we have
\begin{equation}
m=\frac{(r_0^2+a^2)^2(1+r_0^2)+q^2+2a^2 q}{2r_0^2}\,.
\label{eq:m-rplus}
\end{equation}
Recalling that the domain of integration in the ansatz \eqref{eq:ansatz} is determined by the space of real stationary  Lorentz-signature black hole horizons, for any $q$ and any $|a|<1$ we should consider all positive real values of $r_0$.  This domain is the union of the set of triples $(r_0,q,a)$ associated with outer horizons, the set of triples associated with inner horizons, and those defined by degenerate (extremal) horizons; see again \cite{Kolanowski:2026gii}  for comments on the inner horizon case. 

 The angular momentum and electric charge are 
\begin{equation}
J=\frac{\pi\bigl(2am+qa(1+a^2)\bigr)}{4 G_5 (1-a^2)^3}\,,
\qquad
Q=\frac{\pi q}{2 G_5(1-a^2)^2}\,,
\label{eq:JQ}
\end{equation}
and the ADM energy and horizon area are given by
\begin{equation}
E=\frac{\pi\bigl((3+a^2)m+4a^2 q\bigr)}{4 G_5(1-a^2)^3}\,,
\qquad
\mathcal{A}=\frac{2\pi^2\bigl((r_0^2+a^2)^2+a^2 q\bigr)}{(1-a^2)^2\,r_0}\,.
\label{eq:EA}
\end{equation}
We may therefore regard $(r_0,a,q)$ as independent variables and view $(A,J,Q,E)$ as functions of $(r_0,a,q)$ through
\eqref{eq:m-rplus}--\eqref{eq:EA}.

Below, we will use the above set of black holes to study the ans\"atze  \eqref{eq:ansatz} and \eqref{eq:indexansatz}.  To do so, we restrict to potentials that satisfy 
$\Omega_1=\Omega_2=: \Omega$ and
$\Phi_1=\Phi_2=\Phi_3=:\Phi$, so that saddles with equal charges are natural.  However, since equality of the potentials need not necessarily require that saddles have equal charges, we must also assume that saddles with unequal charges do not give important contributions to \eqref{eq:ansatz}.  We will similarly assume that all important contributions come from terms in the sum over shifts  that satisfy both $n_{J_1}=n_{J_2}$ and $n_{Q_1}=n_{Q_2}=n_{Q_3}$.  When this is the case, we may approximate
the ansatz \eqref{eq:ansatz} by 
\begin{equation}
Z  \approx \sum_{{n, n' \in {\mathbb Z}}, \,  {n' \rm{even}}}\int \mathrm{d}\mathcal{A} \, \mathrm{d} J\, \mathrm{d}Q ~ \text{exp}\left[\frac{\mathcal{A}}{4G_5}-\beta (E-2\Omega_n J-\frac{3}{2}\Phi_{n'} Q)\right],
    \label{eq:ZgravAJQ}
\end{equation}
with $\Omega_n = \Omega +\frac{2n \pi }{\beta}i$ and 
$\Phi_{n'} = \Phi +\frac{2 n' \pi }{\beta}i$, 
Here $\mathcal{A}$ denotes the horizon area, while $(E,J,Q)$ are the ADM energy, angular momentum,
and electric charge, respectively. We may of course alternatively parameterize our black holes by $r_0,a,q$ to write
\begin{equation}
Z \approx \sum_{n,n' \in {\mathbb Z}{\,  {n' \rm{even}}}}
\int_{{r_0 >0, \, q \in {\mathbb R }}\atop {1> a> -1}} \! dr_{0}\, da\, dq \;
\det\!\left[\frac{\partial(\mathcal{A},J,Q)}{\partial(r_0,a,q)}\right]
\exp\!\left[\frac{\mathcal{A}}{4G_5}-\beta\bigl(E-2\Omega_n J-\frac{3}{2}\Phi_{n'} Q\bigr)\right] ,
\label{eq:Zgrav}
\end{equation}
where the integral runs over all $r_0\in {\mathbb R}^+$, $a \in (-1,1)$, and $q\in {\mathbb R}$.  The BPS-only ansatz \eqref{eq:indexansatz} is of course given by similar expressions with insertions of $\delta(\tilde E)$. Here, despite the fact that we have discarded other one-loop effects, we have chosen to explicitly write the Jacobian $\frac{\partial(\mathcal{A},J,Q)}{\partial(r_{+},a,q)}$ as any zeros of this Jacobian can still have important effects (as will be seen in appendix \ref{sec:thimbles}).

For fixed $n,n'$,  the usual Legendre transform analysis shows that saddles of \eqref{eq:ZgravAJQ} are defined by (perhaps complex)  black hole solutions for which the period of {\it Euclidean}-signature time is $\beta$ and for which the asymptotic values of the $U(1)$ gauge field component $A_t$ and the metric components $g_{t\phi}=g_{t\psi}$ are determined by $\beta, \Phi, \Omega$.   This then gives the standard relations\footnote{Note that our definition of the chemical potential $\Phi_{n'}$ is, for historical reasons, non-standard. In our conventions, $\Phi_{n'}$ is equal to $\frac{2}{3}$ of the potential difference between the horizon and infinity.} 
\begin{equation}
    \begin{aligned}
        &\beta=\frac{2\pi}{\kappa}=\frac{2\pi r_0 \left[ (r_0^2 + a^2)^2+ a^2q \right]}{r_0^4 \left[ 1 +  2(r_0^2 + a^2) \right] - (a^2 + q)^2}\\
        &\Omega_n = \frac{a (r_0^2 + a^2)(1 + r_0^2) + aq}{(r_0^2 + a^2)^2+ a^2q},\qquad\Phi_{n'}=\frac{q r_0^2}{(r_0^2 + a^2)^2 + a^2 q }
        \label{eq:saddle eq}
    \end{aligned}
\end{equation}
as found in \cite{Chong:2005hr,Aharony:2021zkr}.
In particular, for real potentials and positive $\beta$ the saddle-point values of $r_0,a,q$ are always those of an outer horizon.   
Saddles of the BPS-only ansatz \eqref{eq:indexansatz} satisfy \eqref{eq:saddle eq} with $\beta=\infty$, though they must also satisfy the BPS condition discussed below.


\subsection{Supersymmetry}
\label{sec:SUSY}

Because we wish to compute the gravitational path integral dual to the supersymmetric index of the boundary CFT, bulk saddles that preserve the appropriate supersymmetry will play a special role (even when the associated metric is complex). We will, in particular, follow the recent literature in using
the term ``supersymmetric black hole'' to refer to a general \emph{complex} black-hole saddle that obeys the BPS relation 
\begin{equation}
\label{eq:BPS}
E - 2J - \frac{3}{2}Q= 0,
\end{equation}
which is the analytic continuation to (potentially) complex charges and energies of the usual BPS relation for our theory as defined at real $E,J,Q$; see e.g. \cite{Chong:2005hr}. 
Here we have followed \cite{Aharony:2021zkr} in choosing the BPS relation associated with preservation of a particular supersymmetry, and that the choice of that supersymmetry breaks the original invariance under changing the sign of the angular momentum (though it has nothing to do with the right-hand side of \eqref{eq:SUSYPot}).

In terms of the variables $(q,a,r_0)$, the BPS condition takes the form
\begin{equation}
\label{eq:BPS2}
q=\frac{m}{1+2a}\,.
\end{equation}
Substituting this relation into the horizon equation $\Delta_r(r_0)=0$, one obtains two possible solutions for $q$ in terms of $a,r_0$,
\begin{equation}
q=-a^2+(1+2a)r_0^2 \;\mp\; i r_0\bigl(r_0^2-r_*^2\bigr),
\label{eq:q-complex}
\end{equation}
where we have defined $r_*^2 \equiv 2a+a^2$. Note that {\it real} BPS black holes, which in particular define the integration contour ${\mathcal E}^{BPS}_{\mathbb R}$ of our BPS-only ansatz \eqref{eq:indexansatz}, must also satisfy
\begin{equation}
\label{eq:BPSE}
r_0 = r_* = \sqrt{2a+a^2}
\end{equation}
and must thus have $0 \le a< 1$.  Note also that using both \eqref{eq:q-complex} and \eqref{eq:BPSE} in \eqref{eq:BH_saddle} gives $\beta =\infty$, showing that such solutions are extremal as expected.

The astute reader may note that from \eqref{eq:q-complex} it appears that there may be another branch of real BPS solutions with $r_0=0$ and $q=-a^2$.  However, limits approaching that point are highly direction-dependent.  Writing the charges in terms of $r_0, Y = q+a^2$ and $a$, one finds 
\begin{equation}
    Q = \frac{\pi(Y-a^2)}{2G_5(1-a^2)^2},
    \end{equation}
    so $Q$ has a finite real limit as $Y \rightarrow 0$ if and only if $a$ converges to a real value with $a^2<1$.  One then finds 
    \begin{equation}
        E-2J -\frac{3}{2}Q=
\frac{\pi(3-a)}{8G_5(1-a)^2(1+a)^3}
\left[
Y^2-2(1+2a)Y r_0
+a^2(a+2)^2
+(1+2a^2)r_0^2
+r_0^4 \right],
    \end{equation}
    whence it follows that satisfying the BPS condition in a limiting sense along a path through the real contour requires $a\rightarrow 0$ and $Y/r_0 \rightarrow 0$.  This yields $E=J=Q={\cal A}=0$, which describes the thermal AdS endpoint and which is already included in the branch discussed above.

Returning to the more general complex case, let us momentarily choose the upper sign in~\eqref{eq:q-complex}, saving exploration of the other sign for later. The charge parameter can then be written
in the factorized form
\begin{equation}
q = -(a-i r_0)^2(1-i r_0)\,.
\label{eq:q-factorized}
\end{equation}
Introducing
\begin{equation}
\tau = \frac{\beta\left(\Omega-1\right)}{2\pi i}, \ \Delta = \frac{\beta\left(\Phi-1\right)}{2\pi i},
\end{equation}
and imposing  both \eqref{eq:saddle eq}  and \eqref{eq:BPS}, we find the supersymmetric black hole (Euclidean) action 

\begin{equation}
    S_{\text{SUGRA}} = - \frac{\mathcal A}{4G_5} +\beta (E-2\Omega_n J -\frac{3}{2}\Phi_{n'} Q)
=
-\frac{i\pi^{2}}{2G_5}
\frac{a\,(a-i r_0)^{3}}
{(1-a)^{2}(r_{*}^{2}-3 i a r_0)}
=
\frac{i\pi^{2}}{2G_5}
\frac{\Delta^{3}}{\tau^{2}}.
\label{eq:SUSYS}
\end{equation}

However, imposing a relation between $q$, $a$, and $r_0$ and using \eqref{eq:saddle eq} generally also imposes a relation between the resulting potentials $\beta, \Omega, \Phi$.  In particular, using~\eqref{eq:q-factorized} imposes
\begin{equation}
\beta\bigl(1+2\Omega_n-3\Phi_{n'}\bigr)=2\pi i\,.
\label{eq:susy-constraint}
\end{equation}
Thus $\tau$ and $\Delta$ should not be viewed as independent variables in \eqref{eq:SUSYS}.

It is important to emphasize that, while the manipulations above lead to a unique action $S_{SUGRA}$ for given $\tau,\Delta$ satisfying \eqref{eq:susy-constraint}, we have not yet actually solved for the relevant values of $(a,r_0)$.  In doing so one finds that there are in fact {\it two} BPS saddles $(a,r_0)$, for each choice of sign in \eqref{eq:q-complex}.

Let us now consider using the lower $(+)$ sign in 
\eqref{eq:q-complex}.  Doing so yields the opposite sign on the right-hand-side of \eqref{eq:susy-constraint}, so that the general constraint for BPS solutions is in fact
\begin{equation}
\beta\bigl(1+2\Omega_n-3\Phi_{n'}\bigr)=\pm 2\pi i\,.
\label{eq:susy-constraint2}
\end{equation}
Note also that if \eqref{eq:susy-constraint2} is satisfied for some $n=n_0,n'=n_0{}'$ with the upper sign on the right-hand-side, then it is satisfied with the lower sign on the right-hand side for $n=n_0-1,n'=n_0{}'$ ; i.e., the sign can be changed by allowed shifts of the potentials.

The $\pm$ branches of \eqref{eq:q-complex} can be related by noting that, with ${\mathcal A}, J, Q$ defined by \eqref{eq:JQ} and \eqref{eq:EA},
 the effective (Euclidean) action  
\begin{equation}
S_{\beta, \Omega, \Phi}(r_0,a,q):=  \beta(E-2\Omega_n J -\frac{3}{2}\Phi_{n'} Q)- \frac{\mathcal A}{4G_5},
\end{equation} 
is a real function in the sense that we have
\bal
S_{\beta, \Omega, \Phi}(r_0,a,q) \equiv \overline{S_{\bar\beta, \bar\Omega, \bar\Phi}(\bar r_0,\bar a,\bar q)} \, .
\eal
Substituting the values of $\Phi$ associated with the two branches, we may define
\bal
\label{eq:sym1}
S_{\beta, \Omega}^{(\pm)}(r_0,a,q) = 
S_{\beta, \Omega, \Phi^{(\pm)}}(r_0,a,q),~~~\Phi^{(\pm)}=\frac{1}{3} \(1+2\Omega \mp \frac{2\pi i }{\beta}\) \, ,
\eal
to find that the on-shell actions of the two branches are related by complex conjugation:
\bal
\label{eq:sym2}
S_{\beta, \Omega}^{(-)}(r_0,a,q) = \overline{S_{\bar\beta,\bar\Omega}^{(+)}(\bar r_0,\bar a, \bar q)} \, .
\eal
Alternatively, we may similarly define $S_{\beta, \tau}^{(\pm)}(r_0,a,q)$ to obtain
\bal
S_{\beta, \tau}^{(-)}(r_0,a,q) = \overline{S_{\bar\beta,-\bar{\tau}}^{(+)}(\bar r_0,\bar a, \bar q)} \, .
\label{eq:pmsym}
\eal

In particular, the above symmetry implies that for any supersymmetric choice of boundary conditions there are in fact {\it four} BPS black hole saddles, with two such saddles associated with each of the $\pm$ branches.
It also follows that the saddle-point contributions and the Lefschetz thimble analysis for the second branch can be obtained directly from the corresponding results for the first branch.

\subsection{Non-BPS saddles for supersymmetric potentials}
\label{subsec:NonBPSSUSY}

Importantly, the supersymmetric-potential condition~\eqref{eq:susy-constraint2} by itself does  not guarantee that a smooth saddle satisfying \eqref{eq:saddle eq} actually satisfies the BPS condition \eqref{eq:BPS}.  Instead, even after choosing a sign in~\eqref{eq:susy-constraint}, the conditions~\eqref{eq:saddle eq} still admit
multiple branches of solutions. While the branch characterized by \eqref{eq:q-factorized} (associated with the positive sign in \eqref{eq:susy-constraint2}) satisfies the BPS relation \eqref{eq:BPS}, and similarly for the branch associated with the other choice of sign in \eqref{eq:q-complex} which then leads to the negative sign in \eqref{eq:susy-constraint2}, for either sign in \eqref{eq:susy-constraint2} there is also a non-BPS branch.  

In particular, choosing $2\tau-3\Delta= 1$  allows non-BPS saddles with
\begin{equation}
q = -a^2 - r_0^2 - 2a r_0^2 - 2i r_0^3\,.
\label{eq:nonsusy-branch}
\end{equation}
As noted in section \ref{subsec:consistency}, we will need to analyze such solutions to check either the consistency of the assumption that quantum corrections can be neglected in writing the BPS-only ansatz \eqref{eq:indexansatz}, or the proposal that the sum over saddles can be consistently truncated to just the truly-BPS saddles.

At finite $\beta$, for each choice of potentials there are generally two BPS saddles and two non-BPS saddles. We saw above that the BPS saddles have the same action \ref{eq:SUSYS}, and that this action is independent of $\beta$.

The explicit expressions for the non-BPS saddles are complicated, though one can expand them in series at large $\beta$.  Doing so yields
\bal
a=-\frac{1}{5} +\mathcal O(\beta^{-1}),\qquad r_0=\frac{3}{5}i + \mathcal O(\beta^{-1})\,,
\eal
and
\bal
a=1 -\frac{2\pi i \tau}{3\beta}+\mathcal O(\beta^{-2}),\qquad r_0=\frac{\pi}{3\beta} + \mathcal O(\beta^{-2})\,.
\eal
In both cases $q$ is then given by \eqref{eq:nonsusy-branch}.

The respective Euclidean actions are 
\bal
S_{\text{non-BPS}} = \frac{i\pi^2}{54G_5} (8\tau-3) + \mathcal O (\beta^{-1})\,,
\eal
and
\bal
\label{eq:nBPS2}
S_{\text{non-BPS}} = -\frac{27}{128\pi G_5}\frac{\beta^3}{\tau^2} + \mathcal O (\beta^2)\rightarrow \infty\,.
\eal
In the limit ${\rm Re} \, \beta \rightarrow +\infty$, the action of the 2nd saddle diverges with a sign determined by the phase of $\tau$.  When the real part of the divergence is positive, any contributions of the saddle become negligible in the limit.    When the real part of the divergence is negative, the fact that the integrand of our ansatz is bounded on the contour of integration prevents an ascent contour from the saddle from reaching the integration contour at large ${\rm Re} \, \beta$, and thus also prohibits the saddle from contributing.  However, there is a possibility that this saddle could contribute to our ansatz at large $\beta$ when $\tau^2$ is purely imaginary\footnote{Though it should not contribute to the true index since the action has non-trivial dependence on $\beta$.}.

On the other hand, there is no immediate obstacle to the first non-BPS saddle contributing to our integral, even at large $\beta$.  Interestingly, the limiting values of $a,r_0,q$  for that saddle define charges $E,J,Q$ that saturate the BPS bound;  i.e., the saddle becomes BPS in the limit $\beta \rightarrow \infty$.  Indeed, we will see in section \ref{subsec:one-dimensional-BPS-integral} that this limiting value is also a saddle for our BPS-only ansatz \eqref{eq:indexansatz}.

As a final comment, we note that the full finite-$\beta$ expressions for the two saddles involve a square root, and that the two non-BPS saddles correspond to the two branches of the square root.  As a result, these saddles coincide when the argument of the square root vanishes.  In fact, the locus where it vanishes defines branch points, and circling either branch point exchanges the two non-BPS saddles.  There is thus no invariant distinction between the above two branches of non-BPS saddles.

\section{A one-dimensional BPS integral}
\label{subsec:one-dimensional-BPS-integral}

After the above preliminaries, we are finally ready to study the semiclassical approximation to the BPS-only ansatz \eqref{eq:indexansatz} for potentials that satisfy \eqref{eq:SUSYPot}.   As in section \ref{sec:BHreview}, for simplicity we restrict to the case
 $\Omega_1=\Omega_2=: \Omega$ and
$\Phi_1=\Phi_2=\Phi_3=:\Phi$ and assume for such cases that the important contributions come from black holes with $J_1=J_2= :J$ and with $Q_1=Q_2=Q_3=:Q$ and with shifts that satisfy both $n=n_{J_1}=n_{J_2}$ and $n'=n_{Q_1}=n_{Q_2}=n_{Q_3}$. The corresponding restricted ansatz takes the form
\begin{eqnarray}
\label{eq:indexansatz2}
{\mathcal{I}}(\tau, \Delta) &=& 
\sum_{n, n'  \in {\mathbb Z}, \, n' \, \rm{even}} 
 \mathcal{I}_{n, n'}
\ \ \ {\rm with} \nonumber \\
 \mathcal{I}_{n, n'}
 &\approx&
\int d{\mathcal A} \, dJ   \, dQ\, \delta(E-2J-\frac{3}{2}Q) \, 
 e^{{\mathcal A}/4G_5} 
 e^{2\pi i (2\tau_n J +\frac{3}{2}\Delta_{n'} Q)},
\end{eqnarray}
where we have defined
\begin{equation}
\tau_n = \tau +n  \ \ \ {\rm and} \ \ \ 
\Delta_{n'} = \Delta + n' 
\end{equation}

Recall that supersymmetry requires us to impose
\begin{equation}
\label{eq:SUSYpot52}
2\tau-3\Delta =2m+1
\end{equation}
for some integer $m$.  We thus find
\begin{equation}
\label{eq:SUSYpot53}
2\tau_n-3\Delta_{n'} =2(m+ n - \frac{3}{2}n')+1,
\end{equation}
where, since $n'$ is even, the right-hand-side ranges over all odd numbers.
So, although at finite $\beta$ only the cases with right-hand-side equal to $\pm 1$ allow BPS black hole saddles, there will nevertheless be an infinite set of sectors in which this condition holds.  One then expects the contribution from all other sectors to be given only by thermal AdS.  

It is thus convenient to first consider the case $2\tau_n-3\Delta_{n'}=+1$ in section \ref{subsec:+1ase} below. In doing so, we will find that the consistency condition of section \ref{subsec:consistency} is violated even in this sector when $\tau$ lies in a certain region near $\tau=0$.  However, the weaker consistency condition of section \ref{subsec:truncate} is satisfied. Corresponding results for $m=-1$ then follow from relations \eqref{eq:sym1} and \eqref{eq:sym2} above. 

Since there are no truly-BPS black holes for other values of $n,n'$,  this then provides all of the information we need to discuss black hole saddles in the full sum over $n,n'$ in section \ref{sec:sum}.  However, as discussed in \cite{Aharony:2021zkr}, there are also additional bulk saddles associated with quotients and orbifolds of black hole spacetimes.  As discussed in section \ref{subsec:orbifolds}, at least in our current context it is straightforward to extend our methods to analyze the relevance of these saddles as well.

\subsection{The case $2\tau_n-3\Delta_{n'}=1.$}
\label{subsec:+1ase}
This section will focus on shifts $n, n'$ for which we have
\begin{equation}
\label{eq:anotherSUSYpoteq}
2\tau_n-3\Delta_{n'} =1.
\end{equation}
For such $n, n'$ we may write 
\begin{eqnarray}
\mathcal{I}_{n, n'} \approx
\int d{\mathcal A} \, d\tilde J   \, d\tilde Q\, \delta(E-2J-\frac{3}{2}Q) \, 
 e^{-\frac{1}{G_5}\tilde S_{BPS}},
\end{eqnarray}
where
\begin{equation}
\tilde S_{\rm BPS}
=
-\frac{\mathcal A}{4}
-2\pi i\left(2\tau_n \tilde J+\frac{3}{2}\Delta_{n'} \tilde Q\right).
\label{eq:BPS_action_general}
\end{equation}
Here $G_5^{-1}\tilde S_{BPS}$ is the Euclidean action and the tildes on $\tilde S_{\rm BPS}, \tilde J, \tilde Q$ indicate that we have stripped off factors of $G_5^{-1};$ i.e., $\tilde J = G_5 J$ and $\tilde Q = G_5 Q$. 

As discussed in section \ref{sec:SUSY}, the space of real BPS  Lorentz-signature black holes is given by the 
triples $(r_0,a,q)$ that satisfy
\begin{equation}
r_0^2=a^2+2a,
\qquad
q=-a^2+(1+2a)r_0^2=2a(1+a)^2,
\qquad
0\leq a<1 .
\label{eq:real_BPS_locus}
\end{equation}
The BPS locus  
is thus the one-dimensional manifold ${\mathcal E}^{BPS}_{\mathbb R}$ on which all black holes are extremal. The endpoint $a=0$ has $r_0=q=0$.  While in our discussion this represents the limit of small black holes, its action coincides with that of thermal AdS.  As a result, at leading-order in the semiclassical expansion we can think of this point as thermal AdS in lieu of adding a separate explicit term to our ansatz representing contributions from bulk geometries with the topology of thermal AdS.

Restricting to the BPS locus, the reduced integral takes the schematic form 
\begin{equation}
{\mathcal I}_{n,n'}
\approx
\int_{C_a} da\,
{\cal J}_a(a)\,
\exp\left[-\frac{1}{G_5}\tilde S_{\rm BPS}(a;\tau_n)\right],
\qquad
C_a=\{a\in\mathbb{R}:0\leq a<1\},
\label{eq:BPS_reduced_a_integral}
\end{equation}
where ${\cal J}_a(a)$ denotes the induced measure on the BPS locus, including
the Jacobians from the change of variables to $r_0,a,q$ and from the restriction to the
BPS sector. Its explicit form will not be needed for the leading semiclassical
analysis. Substituting \eqref{eq:real_BPS_locus} into
\eqref{eq:BPS_action_general} gives
\begin{equation}
\tilde S_{\rm BPS}(a;\tau)
=
\frac{\pi^2 a}{(1-a)^3}
\left[
(a-1)\sqrt{a(a+2)}
-i\left(a-1+2(1+a)^2\tau\right)
\right],
\label{eq:BPS_action_a}
\end{equation}
where the square root is chosen to be positive on the contour of integration ${C}_a={\mathcal E}^{BPS}_{\mathbb R}$. Note that the rescaled action $\tilde S_{\rm BPS}$ and so also ${\mathcal I}_{n,n'}(\tau)$ are $\beta$-independent. Moreover, the integral converges as long as ${\rm Im} \, \tau >0$.  This is the same condition as for the convergence of the index  in the dual CFT.

To remove branch points associated with the square root, it is useful to map the interval $0\leq a<1$ to a finite interval in a new
coordinate $t$ by
\begin{equation}
a=-1+\frac{1}{2}\left(t+\frac{1}{t}\right),
\qquad
C_t=\{t\in\mathbb{R}:1\leq t<2+\sqrt{3}\}.
\label{eq:a_to_t_map}
\end{equation}
The reduced integral becomes\footnote{We could have decided instead to take $2-\sqrt{3}<t\le1$. In that case the relevant $\tilde S_{BPS}(t;\tau)$ would have a different functional form. Nevertheless the following Picard-Lefschetz analysis would be exactly the same. In particular, we would find that at most one truly-BPS black-hole saddle can contribute and it would be the same one as in the main text.}
\begin{equation}
{\mathcal I}_{n, n'}\approx
\int_{C_t} dt\,
{\cal J}_t(t)\,
\exp\left[-\frac{1}{G_5}\tilde S_{\rm BPS}(t;\tau)\right],
\label{eq:BPS_reduced_t_integral}
\end{equation}
with
\begin{equation}
\tilde S_{\rm BPS}(t;\tau)
=
-\pi^2
\frac{(t-1)^2(t-i)^2
\left[
t^2-4t+1
-2i(t+i)^2\tau
\right]}
{(t^2-4t+1)^3},
\label{eq:BPS_action_t}
\end{equation}
where here and below we simplify the notation by taking the second argument of $\tilde S_{\rm BPS}$ to be simply $\tau$ (without the subscript $n$). 

The two $\tau$-dependent critical points are
\begin{equation}
t_{\pm}
=
\frac{
2-\tau
\pm
\sqrt{3-6\tau-9\tau^2}
}
{1+(1-3i)\tau}.
\label{eq:t_pm}
\end{equation}
These $t_\pm$ are the $\beta \rightarrow \infty$ limits of the two finite-$\beta$ BPS saddles described in section \ref{sec:SUSY}, here both using the upper sign in \eqref{eq:q-complex} since we used \eqref{eq:anotherSUSYpoteq}.  
Since we require ${\rm Im} \, \tau >0$, while the square root has branch points only at real values of $\tau$, we may treat the square-root as a single-valued function, taking it to be positive when its argument is real and positive. 

The on-shell action turns out to be the same for both the $t_+$ and $t_-$ saddles and yields 
\begin{equation}
\tilde S_{\rm BPS}(t_{\pm};\tau)
=
\frac{i\pi^2(2\tau-1)^3}{54\tau^2},
\label{eq:BPS_action_on_shell}
\end{equation}
in agreement with the $\beta$-independent result \eqref{eq:SUSYS}.  As a result, if the upward-flow cycle from the $t_\pm$ saddle has non-trivial intersection with the defining contour $C_t$,  the corresponding leading-order semiclassical contribution to ${\mathcal I}_{n,n'}$ is
\begin{equation}
\exp\left[-\frac{1}{G_5}\tilde S_{\rm BPS}(t_{\pm};\tau)\right]
=
\exp\left[
-\frac{i\pi^2}{54G_5}
\frac{(2\tau-1)^3}{\tau^2}
\right].
\label{eq:BPS_black_hole_contribution}
\end{equation}
Using the standard AdS$_5$/CFT$_4$ normalization
$N^2=\pi/(2G_5)$, \eqref{eq:BPS_black_hole_contribution} takes the form
\begin{equation}
\exp\left[
-N^2\frac{i\pi(2\tau-1)^3}{27\tau^2}
\right],
\label{eq:BPS_black_hole_contribution_N}
\end{equation}
which agrees with the so-called unshifted black-hole saddle of \cite{Kinney:2005ej,Cabo-Bizet:2018ehj,Choi:2018vbz,Benini:2018ywd,Aharony:2021zkr}. 

There are also $\tau$-independent stationary points at
$t=1$,
$t=\pm i$, and 
$t=\frac{4+3i}{5}$.
The point $t=1$ is the thermal AdS endpoint and has
$\tilde S_{\rm BPS}(1;\tau)=0$. As discussed in section \ref{sec:PLO}, the fact that it is an endpoint makes its saddle-point nature irrelevant and it contributes for all values of $\tau$. The points $t = \pm i$ map to $a=-1$, though $\tilde S_{\textrm{BPS}}(i;\tau) \neq \tilde S_{\textrm{BPS}}(-i;\tau)$ since these two points lie on separate sheets of the Riemann surface associated with a square root.\footnote{A careful reader may remember that $a=-1$ was a pole of the original action on our contour. As can be seen, the residue of that pole vanishes after  imposing the BPS and extremality conditions; i.e., after restricting to ${\mathcal E}^{BPS}_{\mathbb C}$. } The point $t=\frac{4+3i}{5}$ has $a=-1/5$.  This last point turns out to be more interesting in that it is a limit of non-BPS finite-$\beta$ saddles for the non-BPS ansatz \eqref{eq:ansatz}, and is thus relevant to the consistency checks described in sections \ref{subsec:consistency} and \ref{subsec:truncate}, and which have yet to be performed.

To analyze contributions from saddles with $t\neq 1$, let us recall that we expect no other saddles to contribute at large ${\rm Im}\,\tau$.  In addition, let us also recall from section \ref{sec:PLO} (see figure \ref{fig:airy}) that, at a given value of $\tau$, only saddles that contribute at that $\tau$ can catalyze Stokes' phenomena that actually change the intersection numbers associated with any saddle.  This means that, as we decrease ${\rm Im}\, \tau$, contributions of saddles with $t\neq 1$ should appear only after a Stokes' transition catalyzed by the thermal AdS saddle.\footnote{Note that Stokes' phenomena are never catalyzed by the endpoint at $t=2+\sqrt{3}$ (where ${\rm Re}\, \tilde S_{BPS} = +\infty$ when the endpoint is approached along the defining contour) since the catalyzing saddle or endpoint must have smaller action than the saddle whose intersection number experiences the discontinuity.}

We therefore begin by noting that thermal AdS cannot catalyze Stokes' transitions of the saddles at $t=\pm i$ or $t=\frac{4+3i}{5}$ at any value of $\tau$ with ${\rm Im} \, \tau >0$. 
Such Stokes' phenomena would occur only when the upward flow from $t=\pm i, \frac{4+3i}{5}$  reaches the thermal $\textrm{AdS}$ saddle or a contributing $t_\pm$ saddle.
But $S_\textrm{BPS}$ is strictly decreasing  along such a flow, while
thermal AdS has $\tilde S_\textrm{BPS} = 0$.  We have  ${\rm Re} \, \tilde S_\textrm{BPS}(i, \tau) ={\rm Re} \, \tilde S_\textrm{BPS}(-i, \tau)= 0$ as well, and that for ${\rm Im} \, \tau >0$ we have ${\rm Re} \, \tilde S_\textrm{BPS}(\frac{4+3i}{5}, \tau) <0$.  So such transitions cannot occur.

\begin{figure}[h]
    \centering
    {{\includegraphics[width=0.6\linewidth]{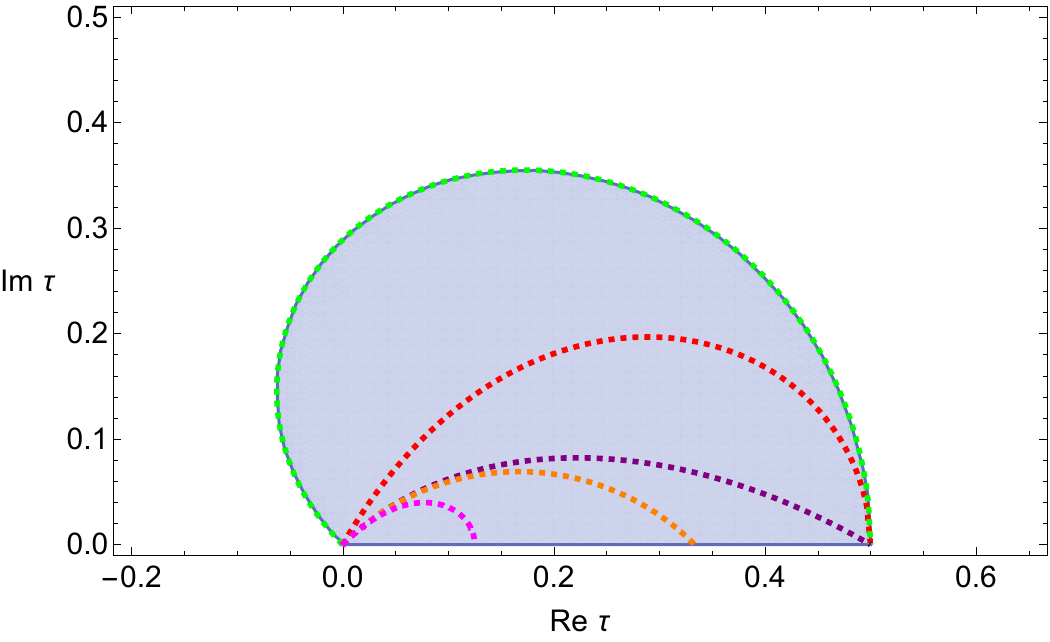}}}
    \caption{As we show below, the $t_+$ saddle contributes to the restricted BPS ansatz \eqref{eq:indexansatz2} in the shaded region of the $\tau$-plane. The dashed green curve indicates the locus with $\operatorname{Im} \tilde S_{\rm BPS}=0$ and $\operatorname{Re} \tilde S_{\rm BPS}>0$,
    at which the thermal AdS saddle could catalyze Stokes' phenomena involving $t_\pm$. 
    The figure also shows the loci where the supersymmetric saddle
    lies on the real BPS contour and corresponds to a~real extremal black hole (dashed red), and where the $t=i$, $t=\frac{4+3i}{5}$ and $t=-i$ saddles satisfy \eqref{eq:neccond} (dashed purple, orange, and magenta respectively) which is necessary for a Stokes' transition catalyzed by the $t_+$ saddle.}\label{fig:contributing_tau}
\end{figure}

On the other hand, the $t_\pm$ saddles {\it might} experience Stokes' transitions catalyzed by thermal AdS. Since $\tilde S_{BPS}$ vanishes for thermal  $\textrm{AdS}$, such phenomena can occur only at values of $\tau$ where ${\rm Im}\, \tilde S_{BPS}(t_+) ={\rm Im}\, \tilde S_{BPS}(t_-)$ also vanishes.  The corresponding curve is shown as a dashed green line in figure \ref{fig:contributing_tau}, whose shading we will shortly explain.

\begin{figure}[h]
    \centering
    {{\includegraphics[width=0.6\linewidth]{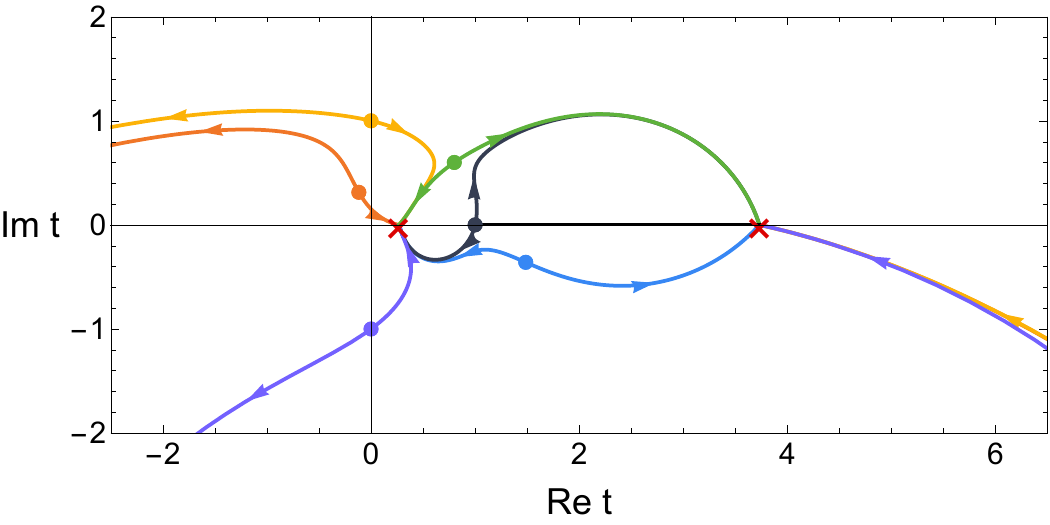}}}
    {{\includegraphics[width=0.6\linewidth]{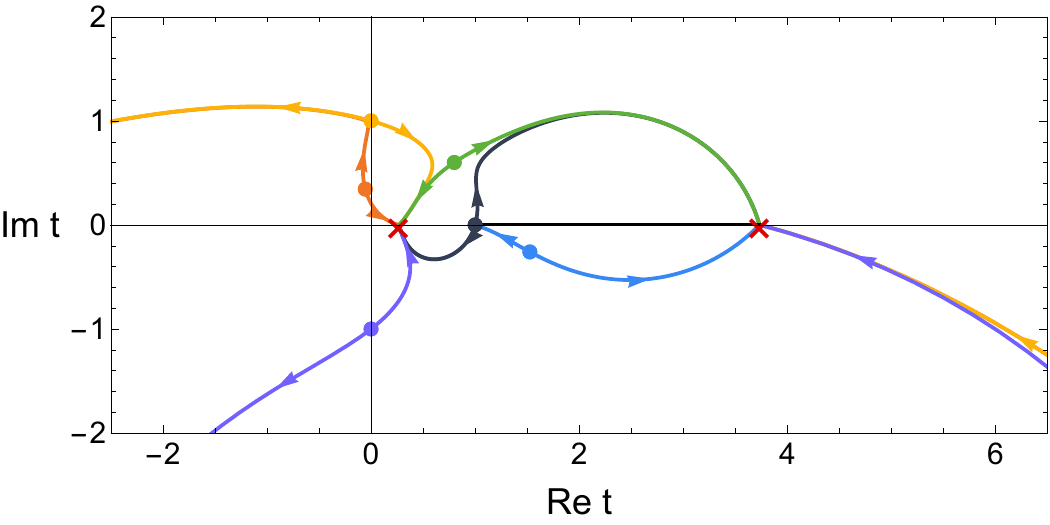} }}
    {{\includegraphics[width=0.6\linewidth]{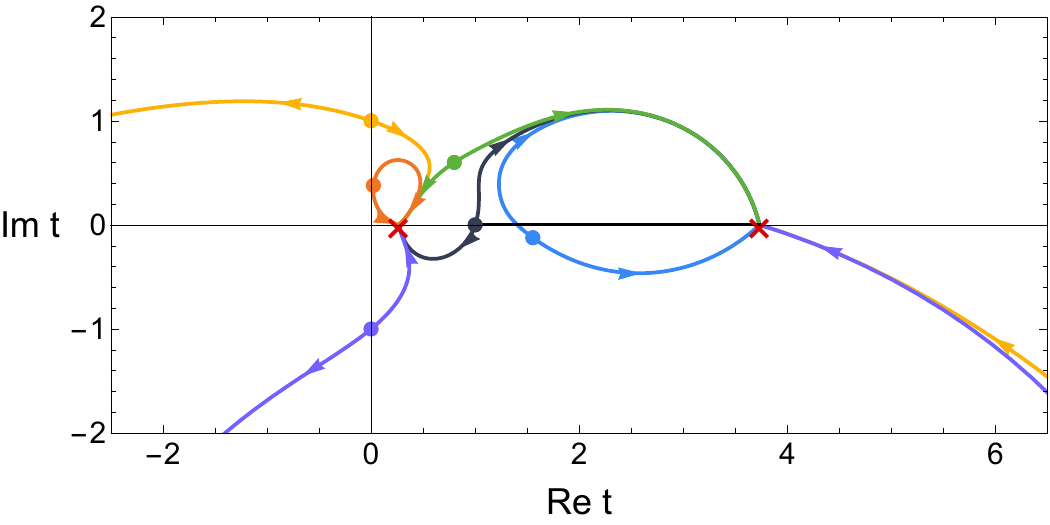} }} 
    \caption{ Steepest-ascent contours (colored lines with arrows) are shown for the saddles (dots)  $t=1$ (black), $t_+$ (blue), $t_-$ (orange), $t=(4+3i)/5$ (green), $t=i$ (yellow), and $t=-i$ (purple) at values of $\tau$ outside, on, and inside the dashed green curve of figure \ref{fig:contributing_tau}. The yellow and purple lines exiting the figure on the left each form a large loop (not shown) and then reenter the figure on the right as shown.  The red $\times$ symbols mark the poles of the Euclidean action.  The defining contour $C_t$ of our integral is the heavy black segment of the real axis running from the black dot (thermal AdS saddle) to the rightmost $\times$.  Arrows indicate the direction in which $\operatorname{Re}(-S_{\rm E})$, the real part of the exponent of the integrand, increases. \textbf{Top:} $\tau=0.3+0.4i$. The saddle $t_+$ does not contribute to the integral. \textbf{Middle:} $\tau=0.3+0.326i$. The saddle $t_+$ undergoes a Stokes phenomenon catalyzed by the thermal AdS saddle $t=1$. \textbf{Bottom:} $\tau=0.3+0.25i$. The saddle $t_+$ contributes to the integral. }
    \label{fig:1dStokes}
\end{figure}

Let us now directly check the contributions of the various saddles on either side of this dashed green line.  Numerical results for the saddles and their flows at such $\tau$ are displayed in  \ref{fig:1dStokes}. As shown in the upper panel, outside the dashed green curve  we see that the only ascent contour  to intersect $C_t$ comes from the $t=1$ thermal AdS saddle (black).  Given the above analysis of possible transitions, seeing this at this one value of $\tau$ guarantees that the same must be true in the entire region outside the dashed green curve. 

Once we move inside the dashed green curve we must also consider possible Stokes' transitions catalyzed by the $t_+$ saddle.    As above, a necessary condition for this to occur at some $\tau$ for a saddle at $t_*$ is 
\begin{equation}
\label{eq:neccond}
{\rm Im} \tilde S_{BPS}(t_+) =     {\rm Im} \tilde S_{BPS}(t_*) \ \ \ {\rm with} \ \ \ {\rm Re} \tilde S_{BPS}(t_+) <    {\rm Re} \tilde S_{BPS}(t_*). 
\end{equation} 
Since the $t_\pm$ saddles have the same action but never coincide, this condition is never satisfied for  $t_*=t_-$. 
However, it remains to study the loci where \eqref{eq:neccond} is satisfied for $t_*=\pm i,\frac{4+3i}{5}$, which are also shown in figure \ref{fig:contributing_tau}. 

\begin{figure}[ht]
    \centering
    {{\includegraphics[width=0.65\linewidth]{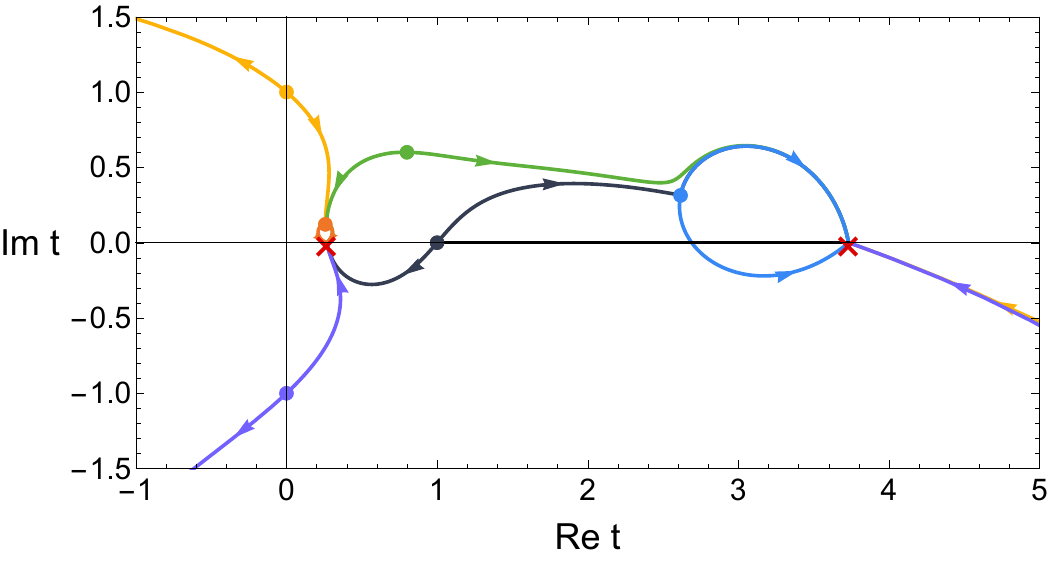}}}
    {{\includegraphics[width=0.65\linewidth]{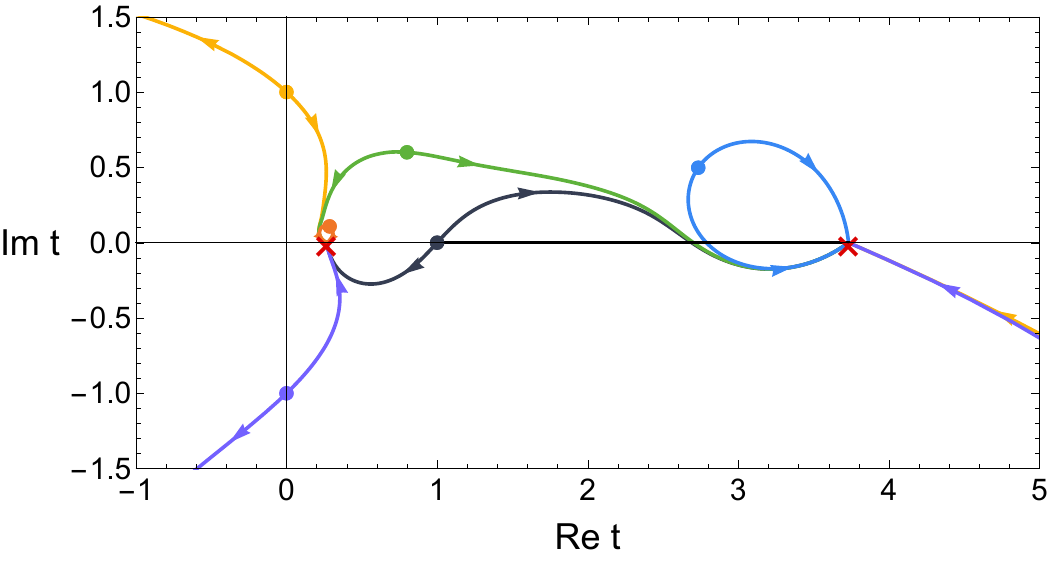} }}
    \caption{Steepest-ascent contours, with colors and labels as in figure \ref{fig:1dStokes}. \textbf{Top:} $\tau=0.1+0.062i$, on the locus where the thermal AdS and $t=+i$ saddles satisfy the necessary condition \eqref{eq:neccond} for a Stokes transition catalyzed by $t_+$. The thimble ${\cal K}_{+i}$ remains well separated from $t_+$, so no such transition occurs. \textbf{Bottom:} $\tau=0.1+0.035i$, on the corresponding locus for the $t=-i$ saddle. Again, ${\cal K}_{-i}$ remains well separated from $t_+$, so no Stokes transition occurs.}
    \label{fig:1dStokes_2}
\end{figure}

\begin{figure}[H]
    \centering
    {{\includegraphics[width=0.65\linewidth]{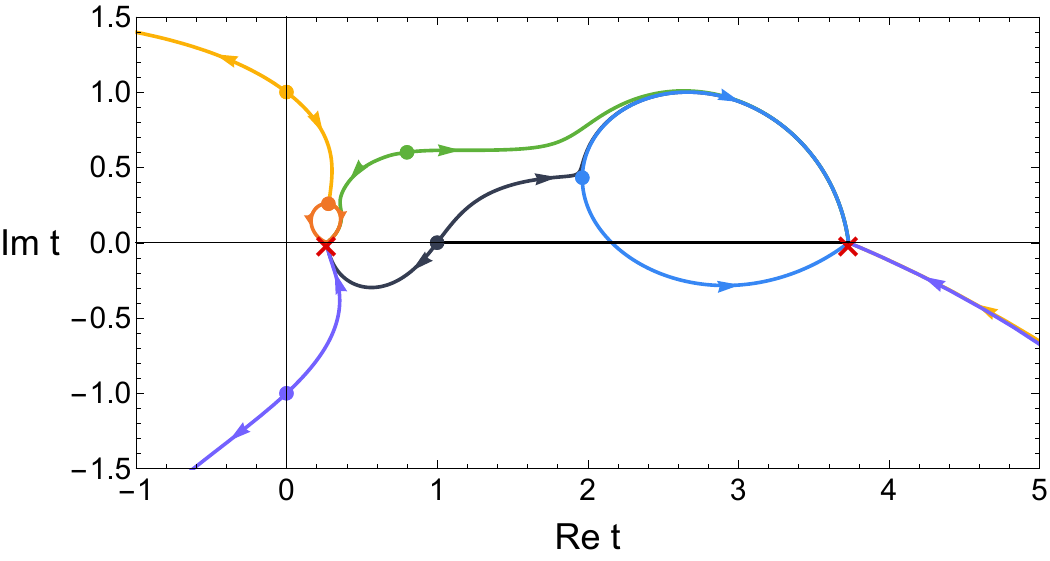}}}
    {{\includegraphics[width=0.65\linewidth]{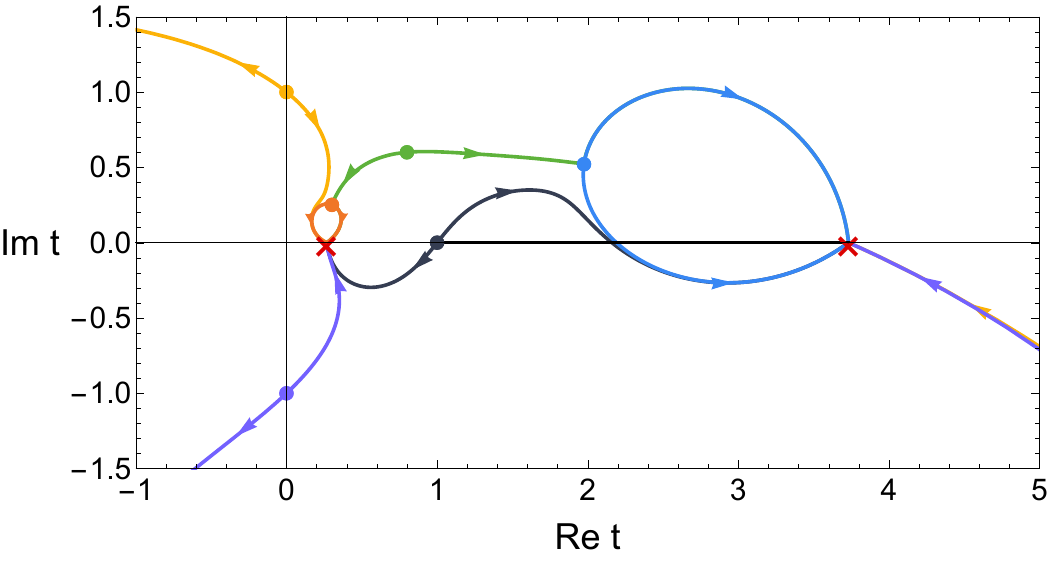} }}
    {{\includegraphics[width=0.65\linewidth]{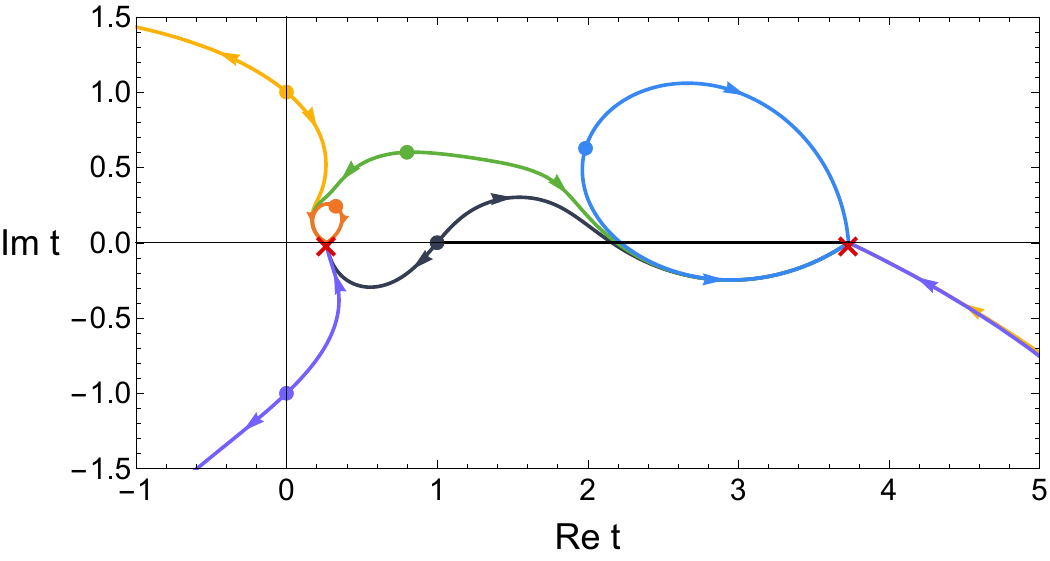} }} 
    \caption{The Stokes' transition for the $t=\frac{4+3i}{5}$ saddle catalyzed by the $t_+$ saddle. Saddles and steepest-ascent contours are shown with colors and labels as in figure \ref{fig:1dStokes}. \textbf{Top:} $\tau=0.20+0.082i$, above the dashed orange line of figure \ref{fig:contributing_tau}. \textbf{Middle:} $\tau=0.20+0.0667i$, on the dashed orange line of figure \ref{fig:contributing_tau}.  \textbf{Bottom:} $\tau=0.20+0.05i$, below the dashed orange line of figure \ref{fig:contributing_tau}.}
    \label{fig:1dStokes_3}
\end{figure}

That the thimbles ${\mathcal K}_{\pm i}$ associated with the $t=\pm i$ saddles experience no Stokes' transitions at these loci can then be seen from figure \ref{fig:1dStokes_2}.\footnote{The top panel does show a Stokes' transition for ${\mathcal K}_{\rm{thermal\, AdS}}$.  Since $\tilde S_{BPS}(+i)$ and the action of thermal AdS both vanish,   the locus at which $t_+$ might catalyze transitions of the $t=+ i$ saddle also allows transitions for thermal AdS. However, as noted above, the fact that thermal AdS is an endpoint makes such transitions irrelevant; thermal AdS contributes to our integral for all values of $\tau$.}  On the other hand, as shown in figure \ref{fig:1dStokes_3} the $t_+$ saddle does in fact catalyze a Stokes' transition that gives the ascent thimble for  $t=\frac{4+3i}{5}$ a non-zero intersection number inside the dashed orange curve in figure \ref{fig:contributing_tau}

It turns out that we may then complete our analysis by showing that the $t=\frac{4+3i}{5}$ saddle never catalyzes a change of intersection number.  Since only relevant saddles can catalyze such transitions, we need only analyze the part of the upper half-plane below the dashed orange curve in figure \ref{fig:contributing_tau}.  But the imaginary parts of $\tilde S_{BPS}\!\left(\frac{4+3i}{5}\right)$ and $\tilde S_{BPS}(+i)$ agree only at $\rm{Re} \, \tau = \frac{3}{8}$, while the imaginary parts of $\tilde S_{BPS}\!\left(\frac{4+3i}{5} \right)$ and $\tilde S_{BPS}(-i)$ agree only at $\rm{Re} \, \tau = -3,$ both of which lie outside the above region (as can be seen from figure \ref{fig:contributing_tau}). 
So the $t = \frac{4+3i}{5}$ saddle cannot catalyze changes in the intersection numbers of the $t_\pm$ saddles.    Furthermore, some simple numerics\footnote{Or some more complicated and rather tedious algebra.} then shows that ${\rm Re} \, \tilde S_{BPS}(t_+) = {\rm Re} \, \tilde S_{BPS}(t_-)< {\rm Re} \, \tilde S_{BPS}(\frac{4+3i}{5})$ everywhere in the relevant region.  Thus the $t=\frac{4+3i}{5}$ saddle cannot induce Stoke's transitions of the $t=\pm i$ saddles.  This then completes the list of possible Stokes' transitions and concludes the desired analysis. 

\begin{figure}[H]
    \centering
    {{\includegraphics[width=0.8\linewidth]{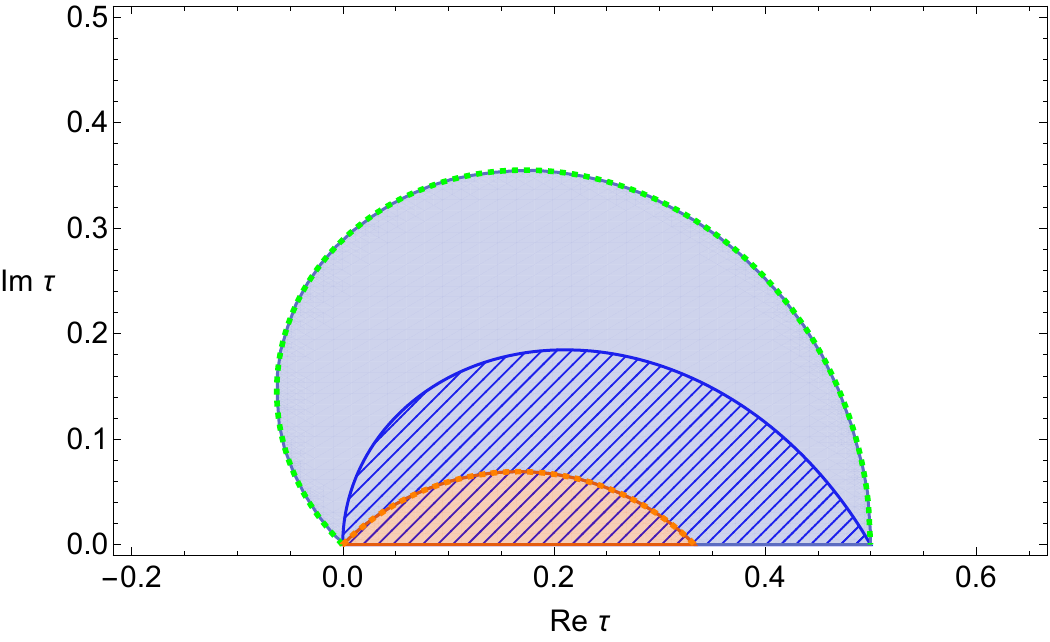}}}
    \caption{Phase diagram for $\mathcal{I}_{n,n'}$ in the complex $\tau$-plane. The thermal AdS endpoint contributes everywhere. In the lavender region inside the green dashed Stokes curve, the $t_+$ saddle contributes in addition to thermal AdS; the corresponding Stokes transition is catalyzed by thermal AdS. In the red subregion inside the orange dotted Stokes curve, the $t=(4+3i)/5$ saddle also contributes; the corresponding Stokes transition is catalyzed by $t_+$. The blue hatching marks the region where $t_+$ contributes and dominates over thermal AdS.}\label{fig:AdS5_phase_diagram}
\end{figure}

The complete phase diagram for this ${\mathcal I}_{n,n'}$ is then described by figure \ref{fig:AdS5_phase_diagram}.  As discussed in section \ref{subsec:NonBPSSUSY}, the saddle at $t=\frac{4+3i}{5}$ is only asymptotically BPS.   Since it contributes, for our system even the BPS-only ansatz of section \ref{sec:BPSansatz} continues to suffer from large quantum corrections.  However, since it catalyzes no changes of intersection number, and since all other contributing saddles are truly-BPS in the terminology of section \ref{subsec:truncate}, our system {\it does} satisfy the weaker condition of section \ref{subsec:truncate} for removing contributions of the 
$t=\frac{4+3i}{5}$ saddle by hand.  We thus predict that the semiclassical expansion of the actual index in this sector is given only by the thermal AdS saddle (which contributes everywhere) and the $t_+$ saddle (which contributes in the original shaded region of figure \ref{fig:contributing_tau}).

In particular, we have shown that only the thermal AdS and $t_+$ saddles can contribute to this ${\mathcal I}_{n,n'}$.  Since both are the $\beta \rightarrow \infty$ limits of BPS finite-$\beta$ saddles, our BPS-only ansatz passes the consistency check described in section \ref{subsec:consistency}.  We thus expect that quantum corrections are small, and that leading-order saddle-point results will be a good approximation to the above ${\mathcal I}_{n,n'}$.

As a small aside, let us mention that the possibility that the inclusion of more degrees of freedom in our ansatz might by itself remove the contributions of only-asymptotically-BPS saddles, so that we no longer needed to do so by hand.  In particular, following \cite{Aharony:2021zkr}, we have computed an action of a D3-brane placed on the horizon (such a brane does not break supersymmetry). We found that the Euclidean action of this brane is purely imaginary\footnote{This should be contrasted with a class of saddles considered in \cite{Aharony:2021zkr} for which the Euclidean action of a~brane was negative, signalizing an instability.}. This suggests that this saddle is only marginally stable to D3-brane nucleation. It is possible that taking into account back-reaction would render it unstable. Alternatively, there could be another brane-related emission channel, perhaps akin to that of \cite{Choi:2024xnv}, which renders the saddle unstable. 

\subsection{The sum over shifted sectors}
\label{sec:sum}

The full ansatz \eqref{eq:indexansatz2} involves a sum over $n,n'$ with $n'$ constrained to be even.  Let us recall from \eqref{eq:SUSYpot53} that for any supersymmetric choice of potentials, we have 
\begin{equation}
    \label{eq:SUSY4}
    2\tau_n - 3\Delta_{n'} =  2(m+n-\frac{3}{2}n')+1,
\end{equation} 
As a result, we can reorganize the sum over $n,n'$ as a sum over those pairs $n,n'$ with fixed $k:= m+ n - \frac{3}{2}n'$, together with a sum over $k$. 

The contribution of sectors with $2\tau_n - 3\Delta_{n'}=1$ was analyzed in detail in section \ref{subsec:+1ase}.  There it was found that the thermal AdS saddle always contributes, so there will be an infinite set of thermal AdS contributions from the sum over $n,n'$ that preserve this condition.  The convergence of this sum clearly depends on one-loop contributions that we have not computed.  But we will make the natural assumption that this sum converges and simply gives an order-one coefficient for the thermal AdS saddle. 

Let us now note that the shifts $\Delta n, \Delta n'$ that preserve the right-hand-side of \eqref{eq:SUSY4} have $\Delta n = \frac{3}{2} \Delta n'$.  Since $\Delta n'$ is required to be even, such shifts have $\Delta n \in 3 {\mathbb Z}$.  On the other hand, after applying the truncation recipe of section \ref{subsec:truncate}, section \ref{subsec:+1ase}  found that for $2\tau_n -3\Delta_{n'} = +1$ there was precisely one contributing black hole saddle for values of $\tau_n$ that lie in the shaded region of figure  \ref{fig:contributing_tau} (and no contributions for $\tau_n$ in the unshaded region).  
Since the diameter of this region is smaller than $1$ (and thus much smaller than $3$), given any $\tau_{n_0}$ in the shaded region, the other $\tau_{n}$ obtained by such shifts must lie outside.  This means that in the entire sum over all sectors with $2\tau_n - 3 \Delta_{n'}=1$, at most one black hole saddle can contribute. 

Since saddles with $2\tau_n - 3 \Delta_{n'}=-1$ are related to those with $2\tau_n - 3 \Delta_{n'}=1$ by relations \eqref{eq:sym1} and \eqref{eq:sym2}, there is again at most one truly-BPS black-hole saddle in this class of sectors and, in particular, it again contributes for $- \bar \tau$ in the shaded region of \ref{fig:contributing_tau}.   Moreover, if $X$ is the shaded region of figure \ref{fig:contributing_tau} (where the truly-BPS saddle contributes for $2\tau_n - 3 \Delta_{n'}=1$), and if $X^*$ is its image under $\tau \rightarrow -\bar \tau$, we see from the figure that the diameter of $X \cup X^*$ is still less than $1$.  As a result, if a black hole saddle contributed in any sector with $2\tau_n - 3 \Delta_{n'}=1$, then no black hole saddle can contribute in any sector with
$2\tau_n - 3 \Delta_{n'}=-1$ (and vice versa). 

On the other hand, as noted in section \ref{sec:SUSY}, there are no BPS black hole saddles\footnote{Except for the saddle that we have called thermal AdS, which is better counted as an endpoint contribution, and which always contributes in that sense.} at finite $\beta$ in sectors with $2m+1\neq \pm 1$.  There are thus no truly-BPS black hole saddles in such sectors, so that we may consistently truncate the semiclassical expansion to just the endpoint contribution associated with thermal AdS.

It is then straightforward to sum over the values of $m$.  We again assume that one-loop factors make the sum of the thermal AdS contributions converge and to just give some order-one coefficient.  When i) no $\tau_n$ with $2\tau_n - 3\Delta_{n'} =+1$ lies in the shaded region of figure \ref{fig:contributing_tau} for any integer $n$ and ii) 
 no $- \bar \tau_n$ with $2\tau_n - 3\Delta_{n'} =-1$ lies in the shaded region of figure \ref{fig:contributing_tau}, 
this gives the full semiclassical result for the BPS-only ansatz \eqref{eq:indexansatz2}.  In particular, this is manifestly the case for large ${\rm Im}\, \tau$, resolving the puzzle raised in the introduction.
When some value of $\tau_n$ with $2\tau_n - 3\Delta_{n'} =+1$
does lie in the shaded region, or when some value of $-\bar \tau_n$ with $2\tau_n - 3\Delta_{n'} =-1$ lies in this region (and there are no original values of $\tau,\Delta$ that allow both of these to be true), we find an additional contribution from precisely one truly-BPS saddle. 

\subsection{Consistency Checks for $2\tau_n - 3 \Delta_{n'} \neq \pm 1$}\label{consistency_AdS5}

We saw in section \ref{subsec:+1ase} that sectors with 
$2\tau_n - 3 \Delta_{n'} = 1$ satisfy the consistency conditions for the ansatz \eqref{eq:indexansatz2} that we described in section \ref{subsec:consistency}.  Corresponding results for $2\tau_n - 3 \Delta_{n'} = -1$ then follow from the symmetry relations of section \ref{sec:SUSY}.  

While we already know that for other sectors we will simply remove any contributions that do not come from thermal AdS,
we now take a moment to verify that such removal is in fact needed.  This provides further evidence that even the BPS-only ansatz \eqref{eq:indexansatz} generally suffers from large quantum corrections.

\begin{figure}[H]
    \centering
    {{\includegraphics[width=0.49\linewidth]{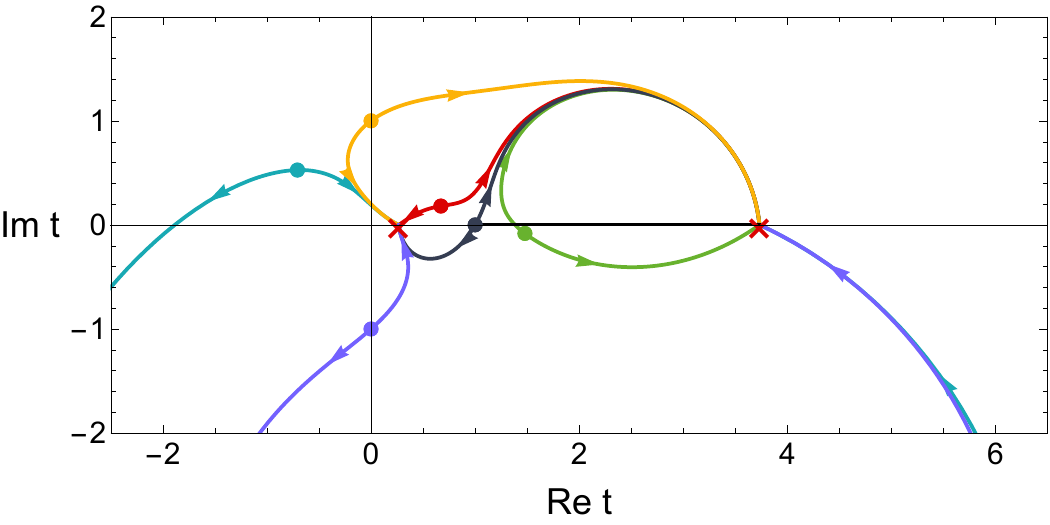}}}
    {{\includegraphics[width=0.49\linewidth]{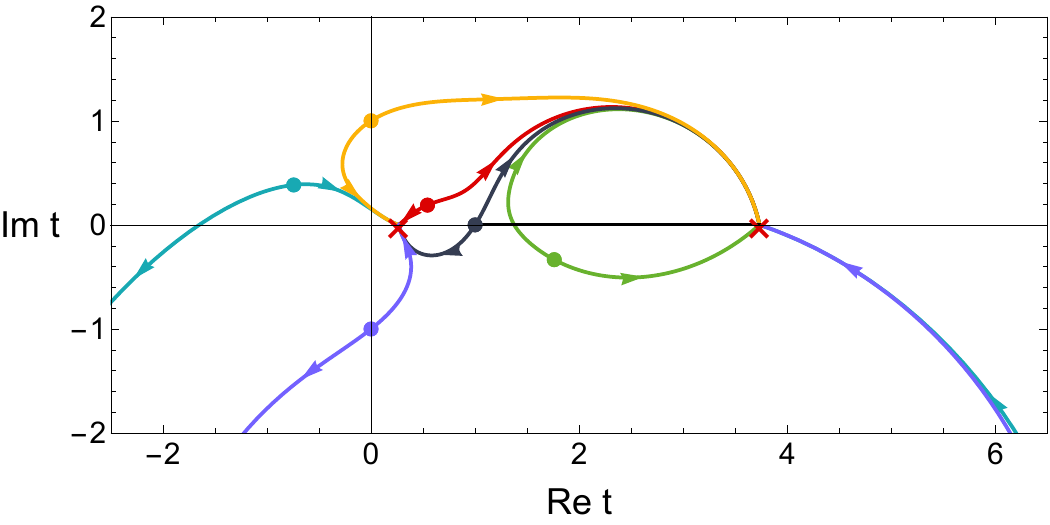}}}\\
    {{\includegraphics[width=0.49\linewidth]{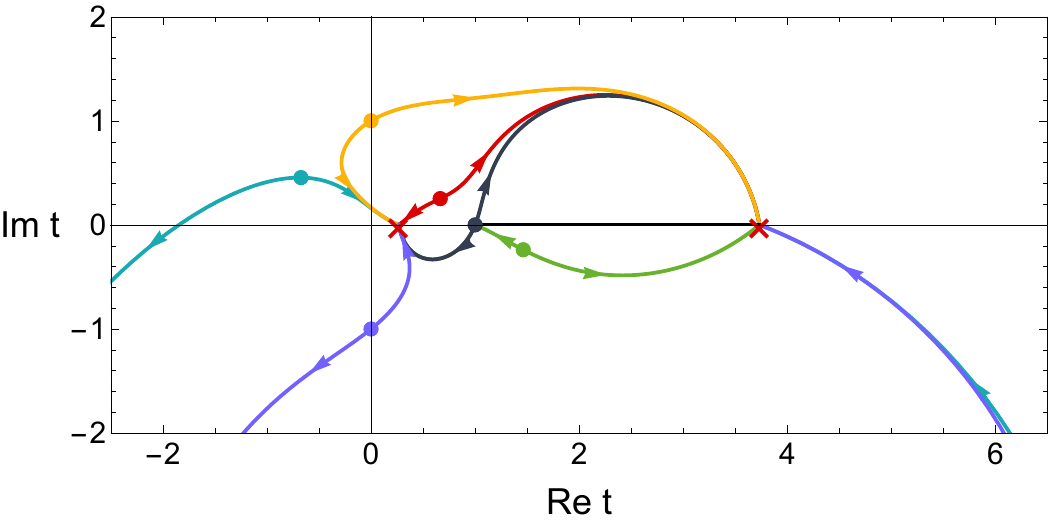} }}
    {{\includegraphics[width=0.49\linewidth]{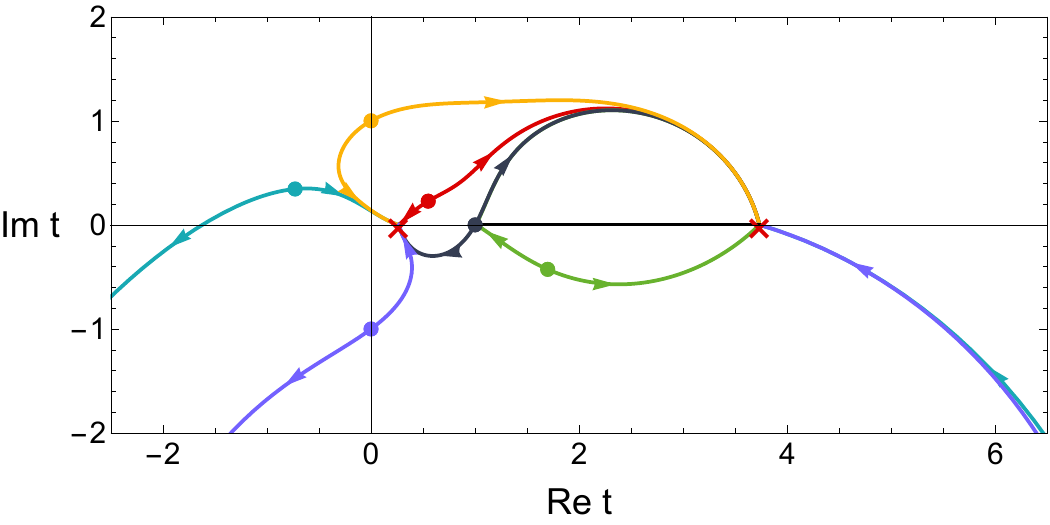} }}\\
    {{\includegraphics[width=0.49\linewidth]{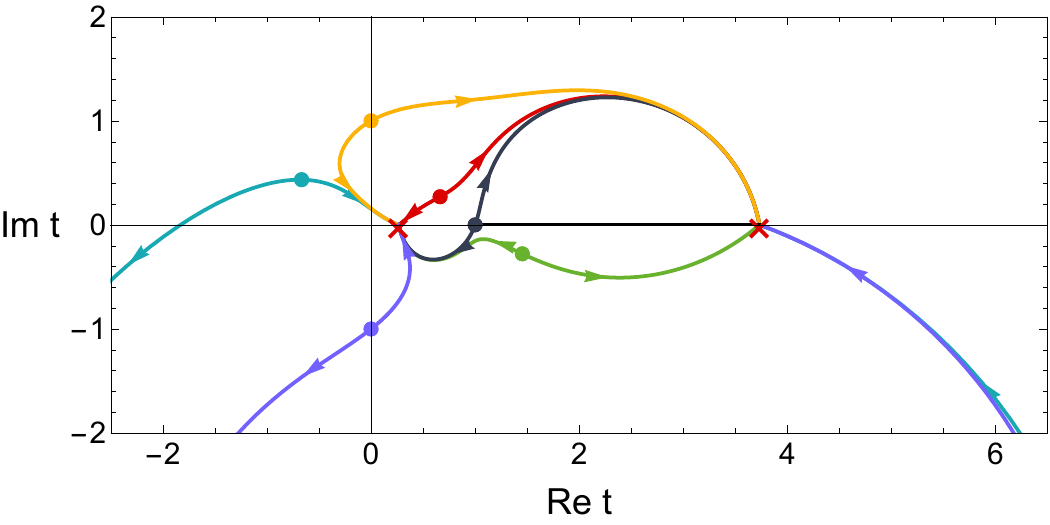} }}
    {{\includegraphics[width=0.49\linewidth]{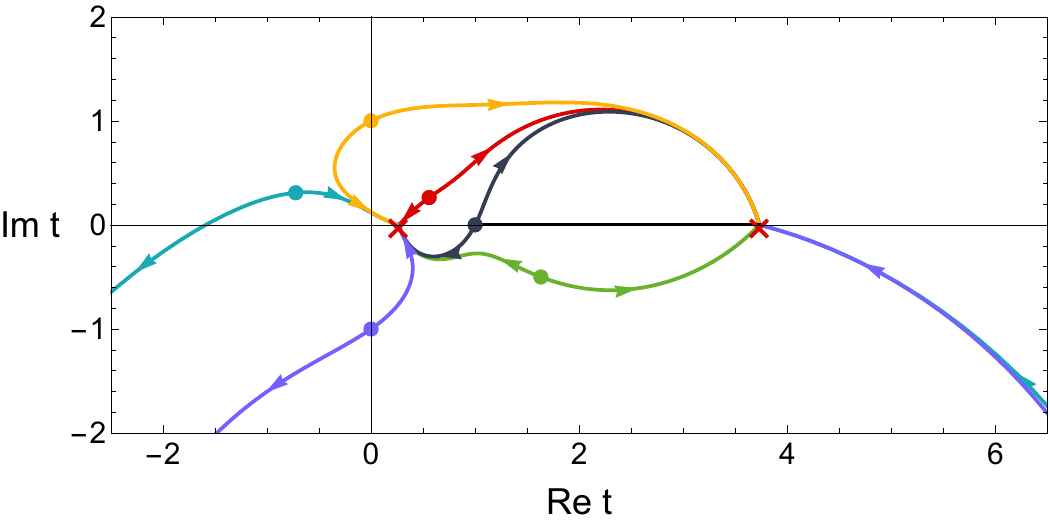} }}
    \caption{Examples of Stokes phenomena for selected values of $\tau$ with $2\tau_n-3\Delta_{n'}\neq 1$. The colored curves with arrows are steepest-ascent contours through the saddles (dots) at $t=1$ (black), $t=i$ (yellow), $t=-i$ (purple), and the three additional saddles $t_k$ (red, green, and cyan). The red $\times$ symbols mark the poles of the Euclidean action. The defining integration contour $C_t$ is the thick black segment along the real axis, extending from the black dot (the thermal AdS saddle) to the rightmost pole. The arrows point in the direction of increasing $\operatorname{Re}(-S_{\rm E})$, the real part of the exponent of the integrand. \textbf{Left column, from top to bottom:} $2\tau_n-3\Delta_{n'}=3$, with $\tau=1+0.3i$, $1+0.533i$, and $1+0.6i$. \textbf{Right column, from top to bottom:} $2\tau_n-3\Delta_{n'}=5$, with $\tau=1+0.7i$, $1+0.9i$, and $1+1.1i$.}
    \label{fig:1dStokesshiftedsectors}
\end{figure}

Let us now examine the saddles that arise from sectors with $2\tau_n-3\Delta_{n'} \neq \pm 1$. One again finds saddles at $t=1,\pm i$, with $t=1$ being the thermal AdS endpoint.  For the values of $\tau, \Delta$ that we have checked, the saddles at $t=\pm i$ do not contribute; see figure \ref{fig:1dStokesshiftedsectors}. Furthermore, as in section \ref{subsec:+1ase} above, all 3 of these saddles have ${\rm Re} S_E=0,$ so none of these saddles can catalyze a Stokes' transition for any other from this set.    We therefore expect that the $t=\pm i$ saddles do not contribute for any $\tau_n, \Delta_{n'}.$
  
However, there are also three saddles 
$t_k$ (for $k=1,2,3)$ that depend on $\tau_n,\Delta_{n'}$ and which arise from solving a cubic equation.  While this is awkward in general, it is straightforward to solve the cubic  perturbatively in $1/\tau$.  This allows one to compute the Euclidean actions of $t_k$ at large ${\rm Im} \, \tau > |{\rm Re} \, \tau |$, and to show that all 3 saddles have ${\rm Re} \, S_E < 0$ in that regime.  We may thus again use the above observation that the magnitude of the BPS-only integrand is bounded above by its thermal AdS value $1$, whence it follows that these saddles cannot contribute at large ${\rm Im} \tau$.  

Nevertheless, there remains the possibility of a Stokes transition at some finite $\tau$ catalyzed by the thermal AdS saddle $t=1$.  A straightforward numerical check for $\tau$ near the corresponding Stokes' loci then shows that such transitions do in fact  occur; see figure \ref{fig:1dStokesshiftedsectors}.  This verifies the necessity of truncating the semiclassical expansion in these sectors by discarding all contributions other than that of the thermal AdS endpoint.  However, unlike in the $2\tau -3\Delta = \pm1$ case, we found these configurations to be stable to the emission of D3-branes at the horizon.

\subsection{Orbifolds and other quotients}
\label{subsec:orbifolds}

According to \cite{Aharony:2021zkr}, black holes are not the only saddles contributing to the index; one needs to include also orbifolds of black hole spacetimes. In the formalism used in this paper, we integrate over configurations with conical deficits (and various generalizations) so it might seem  unsurprising that more general orbifolds are included as well. However, there is a certain caveat. Constructions leading to our path integral, in the saddle point approximation, would necessarily pick up only contributions from smooth configurations since saddle-point approximation imposes equations of motion and these are equivalent to smoothness at the fixed points of the orbifold action. It is thus far from clear what prescription for the gravitational path integral that would produce a saddle that could be interpreted as a non-smooth orbifold\footnote{At least without including additional UV stringy degrees of freedom.}. Fortunately, \cite{Aharony:2021zkr} found that all of the quotients that they find to contribute in the saddle-point approximation to the unrefined index are in fact smooth (after an uplift to 10d). We therefore postpone consideration of more general non-smooth cases for future work.

We will now very briefly describe the class of orbifolds discussed in \cite{Aharony:2021zkr} and how they can be reproduced in our formalism. We begin by discussing them in  Lorentzian signature, taking the time coordinate $t$ to be periodic with period $T$.  This fits with the approach to gravitational partition functions described in \cite{Marolf:2022ybi}, which uses an integral transform to trade the Lorentzian period $T$ for the inverse temperature $\beta$.  The Euclidean orbifolds of \cite{Aharony:2021zkr} then emerge as saddle points of the associated integrals in direct parallel with the treatment of black holes in \cite{Marolf:2022ybi}. 

We will need to work in 10 dimensions. The uplift of our 5-dimensional solutions\footnote{For simplicity, we will only prescribe metric degrees of freedom, for more details see \cite{Cvetic:1999xp}.} to $\textrm{AdS}_5 \times S^5$ reads \cite{Cvetic:1999xp}
\begin{subequations}
\begin{equation}
    {\rm d}s^2_{10} = {\rm d}s^2_5 + \sum_a \left({\rm d}\mu_a^2 + \mu_a^2 \left({\rm d}\phi_a + \frac{2}{3}A \right)^2 \right),
\end{equation}
where we embed $S^5$ in $\mathbb{C}^3$ as $z_a = \mu_a e^{ i \phi_a}$ ($a \in \lbrace 1,2,3 \rbrace$) subject to the constraint
\begin{equation}
    \sum_a \mu_a^2 = 1.
\end{equation}
It is convenient to replace the real coordinates $\phi_a$ by twisted\footnote{Note that, unlike in \cite{Aharony:2021zkr}, all coordinates remain real. We follow a different normalization for $\Phi$ than \cite{Aharony:2021zkr} which means there is no additional factor of $\frac{2}{3}$ in our case. Moreover, it appears that their definition contained a typo in used sign.} ones:
\begin{equation}
    \hat{\phi}_a = \phi_a - \Phi t. 
\end{equation}
\end{subequations}
In these coordinates it is easy to see that the asymptotic metric is $\textrm{AdS}_5 \times S^5$ and that we have the following identifications:
\begin{subequations}
\begin{equation}
    (\phi, \psi) \sim (\phi + 2\pi, \psi) \sim (\phi, \psi + 2\pi)
\end{equation}
    \begin{equation}
        \hat\phi_a \sim \hat\phi_a + 2\pi,
    \end{equation}
    \begin{equation}
        (t, \phi, \psi, \hat{\phi}_a) \sim \left(t + \tilde{T}, \phi  +\tilde{\Omega} \tilde{T}, \psi + \tilde{\Omega} \tilde{T}, \hat{\phi}_a - \tilde{\Phi} \tilde{T} \right),
    \end{equation}
\end{subequations}
where, for reasons that will soon be apparent, we have chosen to decorate the usual potentials of these solutions with tildes (so that they become $\tilde{T},\tilde{\Omega},\tilde{\Phi}$).

On this spacetime, let us act with
\begin{equation}
    g_{m,r,s}: (t, \phi, \psi, \hat{\phi}_a) \mapsto \left(t + \frac{\tilde{T}}{m}, \phi + \tilde{\Omega} \frac{\tilde T}{m} - \frac{2\pi r}{m}, \psi + \tilde{\Omega} \frac{\tilde{T}}{m} - \frac{2\pi r}{m}, \hat{\phi}_a -  \tilde{\Phi} \frac{\tilde{T}}{m} + \frac{2\pi s}{m} \right),
\end{equation}
where $m \in \mathbb{N}_{>0}, r \in \lbrace 0, ...m-1 \rbrace, s \in \lbrace 0,..., 2m-1 \rbrace$. Clearly, $g^m_{m,r,s}$ is an identity and so it defines (non-equivalent for various $r,s$) actions of $\mathbb{Z}_m$.  
So long as $r$ and $s$ are not simultaneously zero, the map $g_{m,r,s}$ has no fixed points and the quotient is a smooth manifold (up to a codimension-two conical singularity that was already present before taking the quotient).

The above spacetimes are again asymptotically $\textrm{AdS}_5 \times S^5$, but now with different potentials:
\begin{subequations}
\begin{equation}
    T = \frac{\tilde{T}}{m}, \quad \Omega = \tilde{\Omega} - \frac{2\pi r}{\tilde T}, \quad \Phi = \tilde\Phi - \frac{2\pi s}{\tilde T}.
\end{equation}
We can now perform the integral transform of \cite{Marolf:2022ybi} from Lorentzian period $T$ to the Euclidean period $\beta$.  This essentially amounts to mapping $T \mapsto -i \beta$. 

In Euclidean signature, it is a familiar story that in black hole spacetimes the conical singularity disappears when the parameters are properly tuned. It thus follows immediately from the above quotient construction that this is also the true here in the case where $g_{m,r,s}$ has no fixed points (we will soon return to the case $r=s=0$).  

The required tuning turns out to be one of the saddle point conditions.  The full set of saddle point conditions show that the Euclidean spacetime just described provides a saddle for the integral that computes the partition function with parameters
\begin{equation}
    \beta =: \frac{\tilde\beta}{m}, \quad \Omega = \tilde{\Omega} - \frac{2\pi i r}{\tilde \beta}, \quad \Phi = \tilde\Phi - \frac{2\pi i s}{\tilde \beta},
\end{equation}
or equivalently
\begin{equation}
\label{eq:dictionary}
    \tau = \frac{\tilde\tau}{m}-\frac{r}{m}, \quad \Delta = \frac{\tilde\Delta}{m}- \frac{s}{m}.
\end{equation}
\end{subequations}
In other words we find that, from a black hole saddle with parameters $(\tilde\beta,\tilde\tau,\tilde\Delta)$, we can produce an orbifold saddle with parameters $(\beta, \tau, \Delta)$. 

Note, however, that the latter has a different topology (since it has  $\pi_1 = {\mathbb Z}_m$ instead of being simply connected). 
Since these (constrained) saddles lie in different topological sectors than our black holes, the two sets of saddles will not interact in any way. In particular, even in the full path-integral, neither type of saddle can catalyze Stokes phenomena for the other. More generally, saddles with $\pi_1 = {\mathbb Z}_m$ can catalyze Stokes' transitions for saddles with $\pi_1 = {\mathbb Z}_{m'}$ only when $m=m'.$ The same is true for saddles with distinct values of $r,s$, as, at least from the perspective of the 10-dimensional geometry, both $r$ and $s$ describes which boundary cycle is contractible in the bulk.

As has already been discussed at length, we are most interested in saddles that are supersymmetric. Of course, for a quotient to be supersymmetric, we need the original spacetime to be supersymmetric, and so
\begin{subequations}
    \begin{equation}
        2\tilde\tau -3\tilde\Delta = \pm 1.
    \end{equation}
But this condition is not sufficient, as we must also ensure that the covering-space Killing spinor survives the quotient\footnote{In particular, as opposed to more standard contexts in which spacetimes can saturate the BPS bound only when they have a well-defined Killing spinor, the claim of \cite{Aharony:2021zkr} is that in the context of the above orbifolds the BPS bound can be saturated even when the Killing spinor fails to be globally defined.}. In other words, we must show that the Killing spinor is preserved by $g_{m,r,s}$. This turns out to be the case precisely when \cite{Aharony:2021zkr}
    \begin{equation}
        2 \tau - 3\Delta \in 2\mathbb{Z}+1.
    \end{equation}
    In particular, for any $r \in \lbrace 0,...,m-1 \rbrace$, there is exactly one $s \in \lbrace 0, ...,2m-1 \rbrace$ with this property, provided $m$ is not divisible by three. In particular, when $r=0$, we have
    \begin{equation}
        2 \mathbb{Z}+1\ni 2 \frac{\tilde\tau}{m}- 3 \frac{\tilde{\Delta}-s}{m} = \pm \frac{1}{m} + \frac{3s}{m}
    \end{equation}
    and so $s \neq 0$. Thus, all the supersymmetric quotients are smooth.
    Thus, there are $m$ supersymmetric $\mathbb{Z}_m$-orbifolds  for $m$ not divisible by $3$; for other $m$ there are less\footnote{For $m\in 3 {\mathbb Z}$ with some values of $r$ there is no SUSY orbifold which preserves the natural symmetry between the relevant circles on the $S^5$.  On the other hand,  one {\it can} find a SUSY orbifold by starting with a black hole with $\tilde\Delta_1, \tilde{\Delta}_2, \tilde{\Delta}_3$ that are not all equal,   but which are chosen in such a way that, for a certain $\mathbb{Z}_m$ action, the resulting $\Delta_i$ are equal. However, based on the analysis of \cite{Aharony:2021zkr}, such configurations would be at least marginally unstable to $\mathrm{D}3$-brane emission.}). 
\end{subequations}



Let us now discuss how these configurations contribute to the partition function. As already mentioned, the partition function should be seen as a sum over different topological sectors. Within a sector with 
fixed $m,r,s$, we may find constrained saddles (with one codimension-two defect) parametrized by the area $A$, angular momentum $J$ and charge $Q$ in the manner described above.   
As in the black hole sector, we leave the integrals over $A,Q,J$ for last, defining ans\"atze analogous to \eqref{eq:ansatz3} and \eqref{eq:indexansatz2}. The action $S_{E,m}$ of the constrained saddle reads
\begin{equation}
    S_{E,m,r,s}\left(A, J, Q; \beta, \tau, \Delta\right) = \frac{1}{m} S_E\left(mA, J, Q; \tilde{\beta}, \tilde{\tau}, \tilde{\Delta}\right),
\end{equation}
where the relations between $\beta, \tau, \Delta$ and $\tilde{\beta}, \tilde{\tau}, \tilde{\Delta}$ were stated above. That this is the correct mapping of the arguments can be seen from the fact that the above quotient preserves the periodicities of the coordinates on the asymptotic $S^3$ so that the energy, charge,  and angular momenta of the quotient are given by precisely the same ADM or Gauss' law expressions as in the original black hole, while the quotient acts non-trivially on the horizon bifurcation surface and thus reduces its area by a factor of $m$.

In particular, we see that the critical points (as functions of $\beta, \tau,\Delta)$ are in one-to-one correspondence with those of the black hole sector\footnote{However,  due to contributions from twisted sectors, this relation does not generally extend to quantum corrections.} (on which $m=1)$, with the mapping again being given by the dictionary \eqref{eq:dictionary}. In particular, for BPS orbifolds we have
\begin{equation}
    S_{E,m,r,s}(\tau)= \frac{1}{m} \frac{i \pi^2 (2m\tau+2r-1)^3}{54 (m\tau+r)^2}
\end{equation}
Since the same relation also applies away from the saddle points, 
they also define identical Picard-Lefschetz thimbles up to the additional simple change of variables $\tilde{A}=mA$. Thus, a given saddle for the orbifold defined by the quotient $g_{m,r,s}$ contributes to the either the non-BPS \eqref{eq:ansatz} or BPS-only \eqref{eq:indexansatz} partition function if and only if the corresponding black hole saddle contributes with suitably changed fugacities.

However, as discussed in section \ref{subsec:truncate}, we must then consider which of the saddles are truly-BPS as opposed to being only-asymptotically-BPS, after which we wish to truncate our sum over saddles accordingly.  As described above,  supersymmetry of the orbifold requires, but does not  follow directly from, supersymmetry of the black hole cover.  We thus discard most of the $r,s$ sectors for each $m$ in their entirety.  However, for each $m$ not divisible by 3 there are $m$ sectors with truly-BPS saddles.  Since the structure of Stokes' transitions in such sectors is determined by that of the black hole sector, we may then consistently discard all but the truly-BPS orbifold contribution from each such sector.  In particular, the weaker consistency condition of section \ref{subsec:truncate} remains satisfied.

In particular, let us recall that black hole saddles contribute only when ${\rm Im}\, \tau$ is sufficiently small. But here ${\rm Im} \, \tau = \frac{1}{m} {\rm Im} \, \tilde\tau< {\rm Im} \, \tilde\tau$, so the region of ${\rm Im} \, \tau$ for which orbifolds contribute is even smaller.   As a result, at sufficiently large ${\rm Im} \, \tau$ it remains true that the only available saddle is thermal AdS.  Furthermore,  at fixed $\tau$ we again find at most a finite number of contributing saddles.

\section{The $\textrm{AdS}_4$ Superconformal Index}\label{AdS4}

Let us now change the number of AdS dimensions from five to four. Recently, an analysis similar to \cite{Aharony:2021zkr} appeared for supersymmetric black holes in $\textrm{AdS}_4$ \cite{Suh:2026ikt}; see also \cite{BenettiGenolini:2023rkq} for earlier foundational work. In this case, there are two $U(1)$ fields and only one axis of rotation. BPS black holes saddles were shown to arise only when the associated fugacities satisfy 
\begin{equation}
    \tau - \Delta_1 - \Delta_2 = \pm 1,
    \label{eq:AdS4SUSYPot}
\end{equation}
though again one obtains a supersymmetric index whenever the right-hand side is an odd integer $2m+1$.

The condition \eqref{eq:AdS4SUSYPot} is invariant under the joint shifts\footnote{We follow a slightly different conventions for shifts than \cite{Suh:2026ikt}. In particular, it seems that the conclusions of their Sec. 3 implicitly depend on the convention we followed here} 
\begin{equation}
    \tau \mapsto \tau + 2 n_\Omega, \quad \Delta_1 \mapsto \Delta_1 +n_\Omega + n_e, \quad \Delta_2 \mapsto \Delta_2 + n_\Omega - n_e, 
\end{equation}
where $n_\Omega, n_e \in \mathbb{Z}$.
On the other hand, the unrefined ABJM index from the dual theory does not seem to have saddle-point contributions corresponding to a sum over the shifts associated with $n_e$ \cite{BenettiGenolini:2023rkq}.This tension was resolved by uplifting the solutions to $11$ dimensions and noticing that most of them are in fact unstable to the emission of $M5$-branes (see again footnote \ref{foot:braneinstability}). At the end of the day, it was argued that only configurations with $\Delta_1 = \Delta_2$ contribute to the unrefined index.  Or, to be more precise, they did full refined version but showed that in the unrefined case, that happens.

This still leaves one family of shifts to be performed. The on-shell Euclidean action for $\Delta_1 = \Delta_2 = \frac{1}{2} \left(\tau \mp 1\right)$ reads
\begin{subequations}
\begin{equation}
\label{eq:AdS4BHaction}
    S_E = \mp \frac{\pi}{G_4} \frac{(\tau \mp 1)^2}{4\tau}.
\end{equation}
Thus, at large shifts, it becomes
\begin{equation}
    S_E = \mp \frac{\pi}{4G_4} \left(
    2n + \tau \mp 2 + O\left(\frac{1}{n}\right)
    \right).
\end{equation}
\end{subequations}

This may look very similar to the situation in $\textrm{AdS}_5$, but it is in fact worse. Since we are to sum over all integers $n$, the real part of the on-shell BPS action is not bounded from below with either choice of sign.  We would thus find a divergent index\footnote{This is worse than what one would have found  in $\textrm{AdS}_5$ if one supposed that all shifted saddles contribute.  There real part of the action remains bounded and only the imaginary part grows without bound.  It might then be possible that one-loop effects render the full sum convergent, though this would not have resolved the tension associated with the $\tau \to \infty$ limit emphasized in our discussion.} if all of these saddles were to contribute at any value of $\tau.$

Moreover, although thermal AdS should again dominate in the limit ${\rm Im} \,  \tau \to \infty$, we see that even the unshifted $n=0$ saddle will be exponentially large (and thus larger than the thermal AdS saddle) for $|{\rm Re}\, \tau|>2$. Both of these issues should be resolved by a careful Picard-Lefschetz analysis of the sort performed in section \ref{subsec:one-dimensional-BPS-integral} for the AdS$_5$ case. 

We thus consider the AdS$_4$ analogue of the BPS-only ansatz \eqref{eq:indexansatz2}, again imposing the equal-charge constraint.  As before, this ansatz is associated with a one-dimensional real contour obtained by imposing the BPS conditions directly on real Lorentz-signature solutions. 

To understand the this contour, we will need to discuss the thermodynamics of four dimensional black holes in more detail. Following \cite{Caldarelli:1999xj}, let us recall that the equal-charge system is described in the bulk by a minimal supergravity theory with bosonic action
\begin{equation}
    \frac{1}{16\pi G_4} \int \textrm{d}^4 x \sqrt{-g} \left( R + 6 - F^2 \right).
\end{equation}
There is a well-known class of real Lorentz-signature Kerr-Newman $\textrm{AdS}$ black holes that can be used to define \eqref{eq:indexansatz}.   The associated horizons (again including inner horizons, outer horizons, and degenerate horizons) are parametrized by real values of $r_0, a, q$ with $r_0>0$ and $|a|<1$. The associated extensive parameters can be written as
\begin{subequations}
    \begin{equation}
        A = \frac{4\pi(r_0^2 +a^2)}{1-a^2}, \quad E = \frac{m}{G_4(1-a^2)^2},
    \end{equation}
    \begin{equation}
        J = \frac{a m}{G_4(1-a^2)^2}, \quad Q = \frac{q}{G_4(1-a^2)}
    \end{equation}
    where $m$ is
    \begin{equation}
        m = \frac{(r_0^2 + a^2)(1+r_0^2) + q^2}{2 r_0}.
    \end{equation}
\end{subequations}
    The Euclidean action reads
    \begin{equation}
        -S_E = \frac{A}{4 G_4} - \beta(E - \Omega J - \Phi Q)
    \end{equation}
    and fugacities are defined as
    \begin{equation}
        \tau = \beta \frac{\Omega-1}{2\pi i}, \quad \Delta = \beta \frac{\Phi -1}{2\pi i}.
    \end{equation}
    
The BPS condition is
\begin{subequations}
    \begin{equation}
        E = J + Q
    \end{equation}
    which translates to
    \begin{equation}
        m = (1+a)q.
    \end{equation}
    Substituting this back into the equation that connects $r_0$ to $m$, we find
    \begin{equation}
        (q-(1+a)r_0)^2 + (a-r_0^2)^2 = 0.
    \end{equation}
Since we integrate over real configurations, both terms must be zero on their own. Thus we have
\begin{equation} \label{eq:bps_act_ads4}
    q = (1+r_0^2)r_0, \quad a = r_0^2.
\end{equation}
Since the real solution exists only when $a<1$, we see that we should also restrict to $r_0<1$. The action on the contour then reads\footnote{From now on, we set $G_4=1$}
\begin{equation}
\label{eq:AdS4IE}
    S_E = \frac{\pi  r_0 (r_0 \mp i) \left(r_0^2-i r_0 \tau \pm \tau -1\right)}{\left(r_0^2-1\right)^2},
\end{equation}
\end{subequations}
which is manifestly independent of $\beta$ at fixed $\tau$.  Here the choice of sign is determined by the RHS of \eqref{eq:AdS4SUSYPot}.

As for AdS$_5$, we find that two choices of sign are related by the map 
\begin{equation}
\label{eq:Isym}
    (r_0,\tau)\rightarrow (\bar r_0, -\bar \tau), \ \ \ {\rm and} \ \ \ S_E \mapsto \bar{S}_E
\end{equation} where the bar denotes complex conjugation.

One can easily check that this action has saddles at
\begin{subequations}
    \begin{equation}
        r_0 = \frac{-\sqrt{1-2 \tau ^2}\pm i \tau }{\tau \pm 1} \quad \textrm{and} \quad r_0 = \frac{\sqrt{1-2 \tau ^2}\pm i \tau }{\tau \pm 1},
        \label{eq:tdepr+4}
    \end{equation}
    while there are also saddles at
    \begin{equation}
        r_0 = -i (\sqrt{2} \pm 1) \quad \textrm{and} \quad r_0 = i (\sqrt{2}\mp 1).
        \label{eq:onlyas4}
    \end{equation}
\end{subequations}
In particular, in this case the thermal AdS endpoint $r_0=0$ is not a saddle, though it nevertheless always contributes as an endpoint.

The action of the first two saddles is given by \eqref{eq:AdS4BHaction}. The latter two saddles are $\tau$-independent and correspond to $\beta \to \infty$ limit of non-BPS saddles; i.e., they are only asymptotically BPS. Any contributions from these latter saddles will thus be removed by the truncation proposal of section \ref{subsec:truncate}.

\begin{figure}[H]
    \centering
    {{\includegraphics[width=0.65\linewidth]{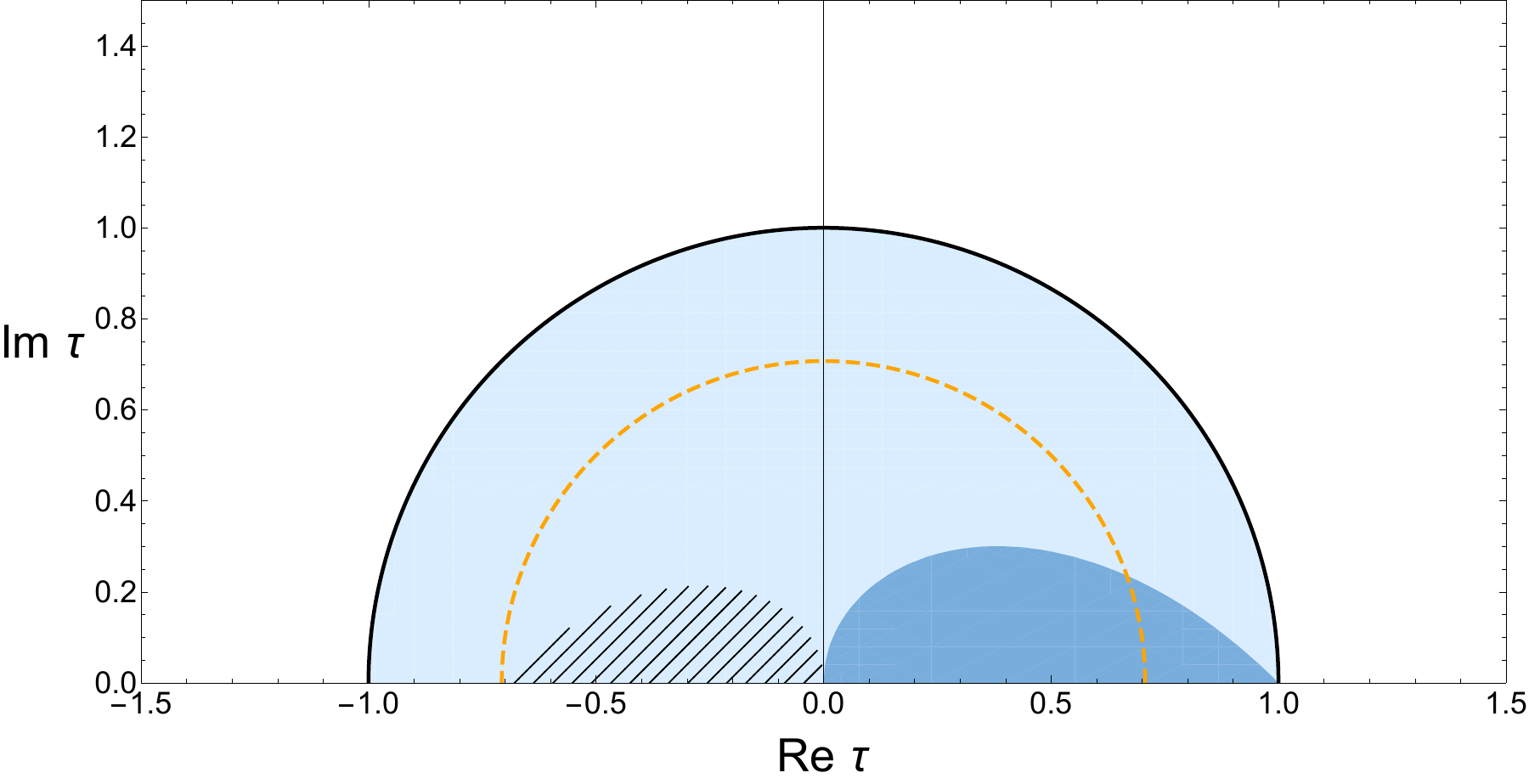}}}
    \caption{The phase diagram for the AdS$_4$ equal-charge version of our BPS-only ansatz \eqref{eq:indexansatz} in sectors with $\tau-2\Delta = 1$. In the bulk semiclassical limit,  the only contribution in the white region comes from the thermal AdS endpoint.  Light blue shading indicates the region where the truly-BPS saddle at $r_0 = \frac{\sqrt{1-2 \tau ^2}+ i \tau }{\tau + 1}$ also contributes but is subdominant to thermal AdS.  Darker blue shows the region where these two saddles again contribute but where   the $r_0 = \frac{\sqrt{1-2 \tau ^2}+ i \tau }{\tau + 1}$ black hole dominates.  While we expect its effects to be removed in a more complete treatment of the bulk path integral, the only-asymptotically-BPS saddle at $r_0 = -i (\sqrt{2}+1)$  also contributes inside the yellow dashed semi-circle, Its action is positive throughout this region, so it is always subdominant to thermal AdS, but it dominates over the above truly-BPS black hole in the cross-hatched region. }
\label{fig:AdS4_phase_diagram}
\end{figure}

\begin{figure}[h!]
    \centering
    {{\includegraphics[width=0.47\linewidth]{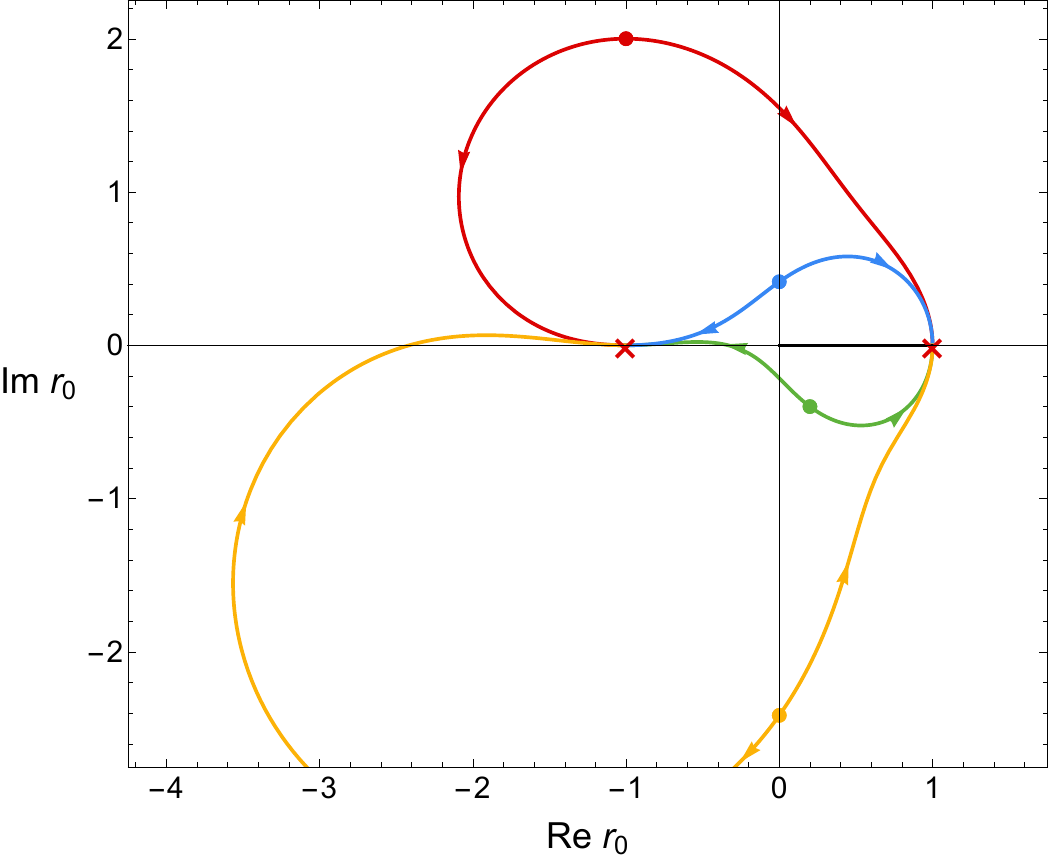}}}
    ~~~{{\includegraphics[width=0.47\linewidth]{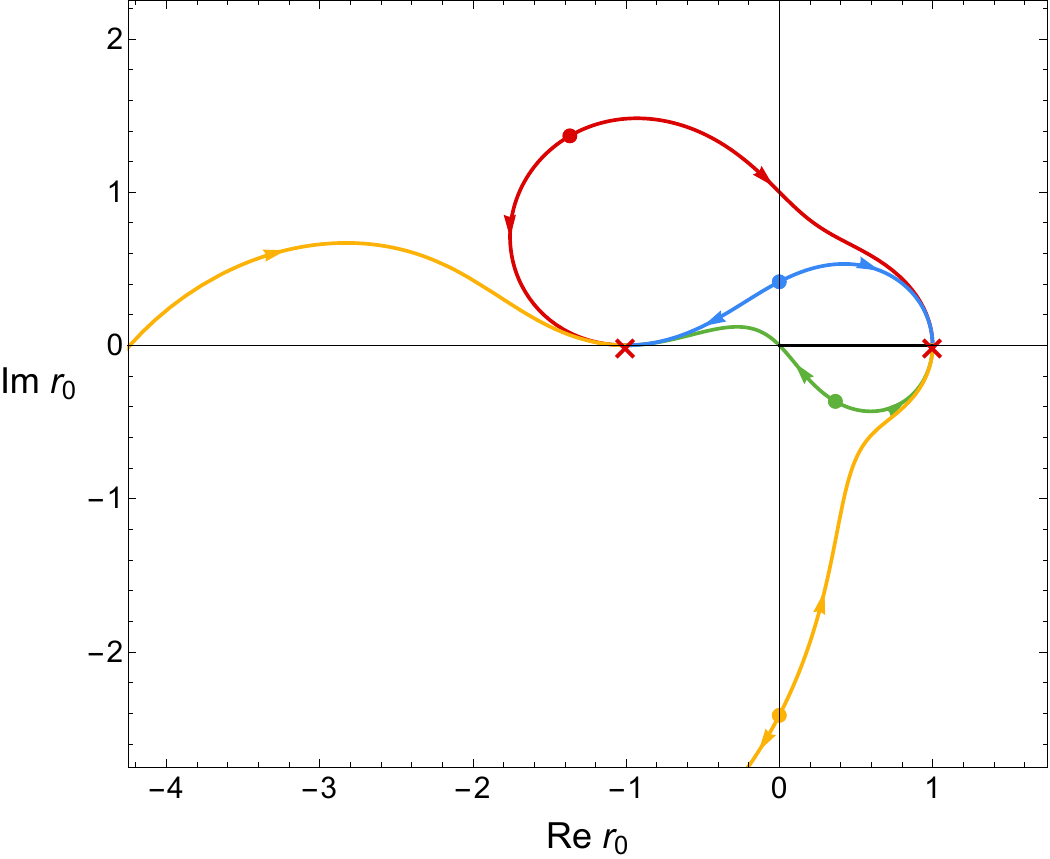}}}
    {{\includegraphics[width=0.47\linewidth]{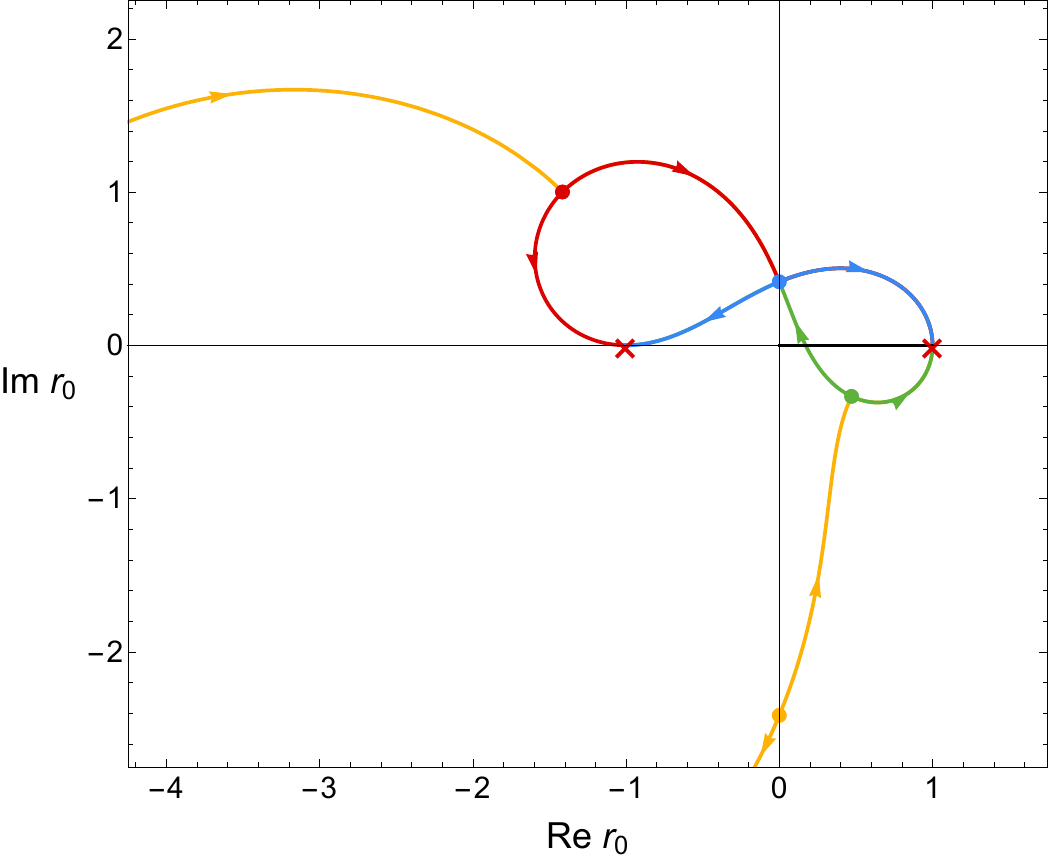}}}
    ~~~
    \caption{Saddles and ascent contours for our AdS$_4$ system at $\tau=2i$ (top left), $i$ (top right), and $i/\sqrt{2}$ (bottom). The red and green dots mark the two truly BPS saddles with $r_0= \frac{i\tau-\sqrt{1-2\tau^2}}{\tau + 1}$ and $r_0= \frac{i\tau+\sqrt{1-2\tau^2}}{\tau + 1}$, while the blue and yellow dots mark the asymptotically BPS saddles at $r_0=i(\sqrt{2}-1)$ and $r_0=-i(1+\sqrt{2})$.  The top left panel shows that no saddles contribute above the two semicircles shown in figure \ref{fig:AdS4_phase_diagram}.  The top right panel shows that, on the upper semicircle of figure \ref{fig:AdS4_phase_diagram}, the thermal AdS endpoint ($r_0=0$) catalyzes a transition for the truly-BPS  saddle (shown in green) at $r_0= \frac{i\tau+\sqrt{1-2\tau^2}}{\tau + 1}$.  This just means that the green flow passes through $r_0=0$.
    The bottom panel shows that, at the lower semicircle in figure \ref{fig:AdS4_phase_diagram}, the green saddle in turn catalyzes a transition of the saddle (shown in yellow) at $r_0 = -i(1+\sqrt{2})$, which is only asymptotically BPS.     }
\label{fig:AdS4Stokes}
\end{figure}

The only remaining question is which of these various saddles contribute at each $\tau$. Due to the symmetry \eqref{eq:Isym}, it suffices to analyze only the upper sign in \eqref{eq:AdS4IE} (corresponding to $\tau-2\Delta =1$), whence we can read off results for the other sign by applying \eqref{eq:Isym}. Note that for this choice of sign the saddles in \eqref{eq:onlyas4} have actions
\begin{equation}
S_E(-i(\sqrt{2} + 1))=\frac{\pi}{4}\left(\tau + 2 + 2 \sqrt{2} \right), \ \ \ S_E(i(\sqrt{2} - 1))=\frac{\pi}{4}\left(\tau + 2 - 2 \sqrt{2} \right).
\end{equation}
The Picard-Lefschetz analysis proceeds much as in section \ref{subsec:one-dimensional-BPS-integral}.  We begin by analyzing Stokes transitions that might be catalyzed by thermal AdS.  This cannot occur for the $\tau$-independent saddles since they both have ${\rm Im} \, S_E >0$ everywhere in the upper half-plane.  However, in the upper half-plane the other two saddles (which have equal action) satisfy the condition
\begin{equation}
    {\rm Im} \, S_E =0 \ \ \ {\rm with} \ \ \
{\rm Re} \, S_E >0
\end{equation}
precisely on the semicircle $|\tau|=1$ (with ${\rm Im}\, \tau >0$).  Numerically studying the ascent curves then shows that no saddles contribute for $|\tau|>1$ (where the only contribution comes from the thermal AdS endpoint), but that the second saddle in \eqref{eq:tdepr+4}  (and only this saddle) contributes after one moves inside this semicircle.  One then finds that this saddle can, and does, catalyze a Stokes' transition that makes the saddle at $r_0 = -i(\sqrt{2}+1)$ contribute for $|\tau|< 1/\sqrt{2}$ (though it catalyzes no other transitions in the relevant region $|\tau|<1$).  But this completes the analysis since for ${\rm Im} \, \tau>0$ the imaginary part of the action for another saddle $r_*$ can never agree with the action of the saddle at 
$r_0 = -i(1+\sqrt{2})$ unless the real part of the action for $r_*$ is smaller  than that for  the saddle at 
$r_0 = -i(1+\sqrt{2})$.   As a result, 
the saddle at $r_0 = -i(1+\sqrt{2})$ cannot catalyze any jumps in intersection numbers. 
The full phase diagram is presented in figure \ref{fig:AdS4_phase_diagram} below for sectors with $\tau -2\Delta = 1$ (with $\Delta_1=\Delta_2=\Delta$).  The relevant Stokes' transitions are then shown in figure \ref{fig:AdS4Stokes}. 

As in the AdS$_5$ case, we do find a contribution from a saddle that is only asymptotically BPS (the saddle at $r_0 = -i(1+\sqrt{2})$), shown in yellow in figure \ref{fig:AdS4Stokes}.  However, since it catalyzes no changes in intersection number, this system again satisfies the weaker consistency condition of section \ref{subsec:truncate}, so that we may simply truncate the sum over saddles so as to remove its contributions by hand. Let us mention that, similarly, to the 5-dimensional case, asymptotically-BPS black holes are only marginally stable to M5-brane emission. Thus, we may expect that a more detailed analysis could remove them in a more systematic manner.

As in the AdS$_5$ case, we should now sum over shifted sectors and bulk topologies associated with orbifolds.  The results are similar.   Allowed shifts that preserve $\tau-2\Delta$ shift $\tau$ by at least $2$.  So, since the truly-BPS black hole saddle contributes only within the semicircle $|\tau| \le 1$ (and since we take ${\rm Im}\, \tau >0$), the full sum over all sectors with $\tau-2\Delta =+1$  gives at most one black-hole contribution at any value of $\tau$.  

Other shifts change the value of $\tau-2\Delta$. Sectors with $\tau-2\Delta = -1$ are related by a symmetry to the those studied above that maps $\tau \rightarrow -\bar \tau$.  Since the shaded region of figure  \ref{fig:AdS4_phase_diagram} is invariant under this operation, such sectors again contribute for $|\tau| <1$.  As a result, in contrast to the AdS$_5$ case, for small enough ${\rm Im} \, \tau$ we can find two truly-BPS black hole contributions in our sum over shifted sectors, one of which arises in a sector with $\tau-2\Delta = 1$ and 
one of which arises in a sector with $\tau-2\Delta = -1$.

As in the AdS$_5$ case, sectors with
$\tau-2\Delta \neq \pm 1$ have no truly-BPS black hole saddles.  However, additional truly-BPS saddles can be constructed from quotients of black hole spacetimes for general  
$\tau-2\Delta = 2m+1$ \cite{BenettiGenolini:2023rkq}.  Also as in the AdS$_5$ case, the equal-charge SUSY quotients of \cite{BenettiGenolini:2023rkq} can be described by  three non-negative integers $k,r,s$ (with $r,s$ bounded by linear functions of $k$) and, in direct analogy with the discussion in section \ref{subsec:orbifolds}, the regime where they contribute can be deduced by scaling the black hole phase diagram (figure \ref{fig:AdS4_phase_diagram}) by a factor of $k$ (and including an appropriate $r,s$-dependent shifts in the real directions). For a given $k >1$, there are $k$ inequivalent orbifolds that preserve supersymmetry. The sum over $k$ can thus yield only a finite number of supersymmetric orbifold contributions at any given ${\rm Im}\, \tau >0$.

\section{Discussion}
\label{sec:disc}

Our work above focused on using inspiration from the Lorentzian formulation of the gravitational path integral to better understand the set of complex black hole saddles that contribute to the AdS$_4$ and AdS$_5$ superconformal indices.  In particular, up to quantum corrections, results from \cite{Marolf:2022ybi,Chen:2025leq} suggest that the index will be described by a finite-dimensional integral of the form \eqref{eq:ansatz}.  Arguments related to fermion zero modes then suggest that the same should be true of the more restrictive BPS-only ansatz \eqref{eq:indexansatz}, which integrates only over real Lorentz-signature BPS black holes.  Interestingly, however, we find that even the BPS-only ansatz appears to suffer from large quantum corrections.  Nevertheless, for both the AdS$_5$ and AdS$_4$ systems studied here, such corrections enter in a relatively tame way that allows us to consistently truncate the resulting sum-over-saddles to saddles that are truly-BPS in the sense that they are the large-$\beta$ limit of saddles that remain BPS at finite-$\beta$. As discussed in section \ref{subsec:truncate}, we conjecture this truncated sum to give a good semiclassical expansion of the desired index (up to the limitations of our model discussed further below).

We implemented the above program for AdS$_5$ and AdS$_4$ systems corresponding to ${\mathcal N}=4$ SYM$_4$ and the ABJM theory, though with certain caveats. In particular,  we studied only saddles that preserve homogeneity on the internal $S^5$ or $S^7$. 
Furthermore, 
in both contexts we considered the so-called unrefined index in which certain fugacities were assumed to be equal.  We then further simplified the problem by considering only contributions from bulk spacetimes in which the corresponding charges and angular momenta were also set to be equal.  While we did so in order to make the analysis more tractable, at least in the AdS$_5$ case, other black hole saddles are expected to exhibit instabilities associated with D3-branes \cite{Aharony:2021zkr} (for the AdS$_5$ case) or 
M5-branes \cite{Suh:2026ikt} (for the AdS$_4$ case)
and thus to have no contribution to the index.   
Nevertheless, an important goal of future work would be to generalize our treatment to the full set of black hole saddles and to analyze effects associated with D3/M5-branes in order to see such instabilities from the current point of view, and in fact to have a full derivation from a Lorentzian path integral.

With the above restriction, the index depends only on a single fugacity $\tau$.  At each $\tau$ our results in both $\textrm{AdS}_4$ and $\textrm{AdS}_5$ suggest this restricted index to be given by a very limited sum over truly-BPS bulk saddles and the thermal AdS endpoint.  For some $\tau$ it received contributions only from the thermal AdS endpoint.  At other values of $\tau$,  there is a contribution from a single black hole saddle in the AdS$_5$ case, and from either one or two black hole saddles in the AdS$_4$ case.  In particular, while we studied the sum over `shifted' black hole sectors (more precisely, over the original unshifted saddle as well as all possible nontrivial shifts), for a given $\tau$ we found for the AdS$_5$ case that at most one saddle in this sum will  contribute, and for AdS$_4$ we found at most two.

The main technical step in our analysis was the exclusion of contributions from an infinite set of saddles by performing a careful Picard-Lefschetz study of saddle-point contributions to the restricted BPS ans\"atze \eqref{eq:indexansatz2} and its AdS$_4$ analogue.  However, this analysis {\it did} find contributions from saddles that become non-BPS saddles of the more general ansatz \eqref{eq:ansatz} at finite temperature $\beta^{-1}$. We refer to such saddles as being only asymptotically BPS, in contrast to the truly-BPS saddles of the BPS-only ansatz which are large $\beta$ limits of BPS finite-$\beta$ saddles for \eqref{eq:ansatz}.
On general grounds fermion zero modes should exclude these only-asymptotically-BPS saddles  from contributing to the index\footnote{Though it is always interesting to ask if this argument might have some subtlety that offers opportunities for exceptions.}. 
This then indicates that quantum corrections to our BPS-only ansatz \eqref{eq:indexansatz} can remain large for both our AdS$_5$ and AdS$_4$ systems at the level at which they were studied in the main text.

Luckily, our Picard-Lefschetz analysis found that the only-asymptotically-BPS saddles had no effect on the contributions of truly-BPS saddles.  By this we mean that the only-asymptotically-BPS saddles never catalyzed a Stokes' transition that changed the contribution from any truly-BPS saddle (though this can certainly happen at finite $\beta$ in the non-BPS ansatz \eqref{eq:ansatz}; see appendix \ref{sec:thimbles}).  As a result, for both the AdS$_4$ and AdS$_5$ systems, it is consistent to simply ignore the contributions of only-asymptotically-AdS saddles at all $\tau$.  Occam's razor then suggests  that doing so gives the correct index up to small quantum corrections.  As noted in section \ref{subsec:truncate}, this truncation may be a natural result of the fact that only-asymptotically-BPS saddles do not have Killing spinors at finite $\beta$ (where one understands the relevant boundary conditions at the horizon), so that  they may well receive quantum corrections which make them in fact fail to saturate the BPS bound even in the limit $\beta \rightarrow \infty$.

It would clearly be of great interest to compare this prediction with computations in the dual CFT.  However, the relevant results do not yet appear to be available in a clean form.  In particular, while the Bethe ansatz approach of \cite{Closset:2017bse,Benini:2018mlo,Benini:2018ywd,Aharony:2021zkr} for the AdS$_5$ case is quite powerful, and while it should in principle determine which saddles contribute for which values of $\tau$, it was already clear from \cite{Copetti:2020dil} that subtleties remain to be understood. Furthermore, as noted in the introduction, the results of \cite{Closset:2017bse,Benini:2018mlo,Benini:2018ywd,Aharony:2021zkr} are in explicit tension with the ${\rm Im}\, \tau \rightarrow +\infty$ behavior of the matrix integral representation of the CFT index, at least at finite $N$.  So we cannot yet take such results as a point of comparison.

There are similarly interesting analytic results for saddle-point approximations to the ${\mathcal N}=4$ SYM$_4$ matrix integral itself; see e.g. ~\cite{Cabo-Bizet:2019eaf,Choi:2021rxi}.
However,  since the analogue of our Picard-Lefschetz analysis has not yet been performed for such saddles, it is again difficult to directly compare such results with ours.\footnote{Furthermore, the saddle point approximation of \cite{Cabo-Bizet:2019eaf,Choi:2021rxi} does not take the standard form discussed in section \ref{sec:PLO}, as the dimension of the domain of integration in \cite{Cabo-Bizet:2019eaf,Choi:2021rxi} is not fixed.  Instead, the saddles are expected to describe the limit $N\rightarrow \infty$ in which the number of integration variables itself diverges. As a result, it is not immediately clear precisely how the relevance of such saddles should be determined or, in other words, what analogue of our Picard-Lefschetz analysis should be performed.} On the other hand, our results seem to be consistent with the perturbative analysis of \cite{Copetti:2020dil}.

The situation for $\textrm{AdS}_4$ is similar. There, the dual theory is the ABJM theory of \cite{Aharony:2008ug}, and the problem of computing the index can once more be recast as a certain matrix integral \cite{BenettiGenolini:2023rkq}.  Various saddle contributions to this matrix integral are known, and their `actions' match the actions of BPS black holes in the bulk.  However, we again lack clear criteria to determine which saddles contribute for given fugacities. In both the AdS$_4$ and AdS$_5$ cases, our analysis strongly suggests that only a finite set of these saddles should contribute.

Let us also mention a rather intriguing work \cite{Cabo-Bizet:2020ewf} which computed the {\it microcanonical} index. The main conclusion of that work was that, even though infinitely-many saddles could potentially contribute to this quantity, at charges of order $N^2$ only two remained relevant.  Since the Atiyah-Bott-Berline-Vergne equivariant integration formula used in \cite{Cabo-Bizet:2020ewf} is similar to (albeit different from) the Bethe ansatz, further exploration of such results may  shed light on the aforementioned issues.

Meanwhile, there remains much to do to better understand the bulk side of such computations.   We have already remarked on the importance of including e.g. D3/M5-brane degrees of freedom in our formalism to obtain a more complete understanding of which saddles are truly relevant.  Including such degrees of freedom may also lead to new saddles, perhaps corresponding to grey galaxies \cite{Bajaj:2024utv} or black holes dressed with dual giant gravitons \cite{Choi:2024xnv}.

In particular, a quick first check suggests that the inclusion of branes may have significant effects on the only-asymptotically-BPS saddles that contributed to our BPS-only ansatz \eqref{eq:indexansatz}.  Following \cite{Aharony:2021zkr} and \cite{Suh:2026ikt}, we computed the action associated with a probe $\textrm{D}3$-brane  or $\textrm{M}5$-brane (for the AdS$_5$ and AdS$_4$ cases respectively) embedded in the background of these saddles (or, to be more precise, of their $10$- and $11$-dimensional uplifts). Interestingly, for the AdS$_5$ sectors with $2\tau-3\Delta = \pm 1$ and the AdS$_4$ sectors with $\tau-2\Delta = \pm 1$,  we found that the resulting Euclidean brane action is purely imaginary, whereas \cite{Aharony:2021zkr} and \cite{Suh:2026ikt} identified the positivity of the Euclidean action as a necessary condition for stability of a black hole saddle under nucleation. The only-asymptotically-BPS saddles in these sectors are thus only marginally stable in this sense, and it is possible that additional corrections would in fact render its contribution to the path integral vanishing. Furthermore, whatever the result of that particular computation, and regardless of whether a Lorentzian perspective might alter the stability criterion of \cite{Aharony:2021zkr} and \cite{Suh:2026ikt}, 
this marginal stability it strongly suggests that more detailed analyses involving brane degrees of freedom may generally be crucial to properly compute the desired superconformal indices from the gravitational path integral.


While it is very important to understand the above issues subtleties for particular theories, in the dual theories, there also remains work to be done in setting up the desired formalism for general systems. First, in order 
to fully establish even at leading order in the semiclassical approximation that the ansatz \eqref{eq:ansatz} follows from the Lorentzian path integral, one still needs to generalize the treatment of rotation in \cite{Chen:2025leq} to arbitrary dimensions.  
 Second, and perhaps more significantly, we should understand how to properly include fermionic zero modes rather than simply relying on the consistency checks of section \ref{sec:BPSansatz} and then attempting to remove by hand pollution from only-asymptotically-BPS  saddles.

Let us finish this discussion by pointing out that the AdS$_4$ and AdS$_5$ systems studied above are not the only ones in which complex BPS black holes appear in bulk computations of partition functions with complex potentials. Recently discussed supersymmetric examples appeared in \cite{Larsen:2026sav} and \cite{Nanda:2026mbp}, which are based on the
construction of multi-centered asymptotically-flat solutions of  \cite{Boruch:2023gfn, Boruch:2025sie}.  
Examples without supersymmetry can be found in \cite{Chen:2023mbc, Grabovsky:2024vnb, Goker:2026tct}. 

Further non-supersymmetric examples can also be constructed from the indices studied here by inserting operators so that we instead study objects of the form
\begin{equation}
    \textrm{Tr} \left(
e^{-\beta\lbrace Q, \bar{Q} \rbrace} \zeta \hat{O}
    \right),
\end{equation}
where $\hat{O}$ at most polynomial in $N$, since when $\hat{O}$ breaks supersymmetry it can soak up any fermion zero modes and allow non-BPS saddles to contribute.  With an eye toward such applications, we thus exhibit some explorations of saddle-point contributions to our more general non-BPS ansatz \eqref{eq:ansatz} in appendix \ref{sec:thimbles}.  That appendix also illustrates the possibility that only-asymptotically-BPS saddles may in some cases catalyze finite-$\beta$ Stokes' transitions for truly-BPS saddles. 

Let us now close by recalling that, 
in all of the above situations, it remains crucial to determine the physically-correct contour prescription for the gravitational path integral. The Lorentzian prescription used here shows promise, but should be further  developed and tested as described above.


\section*{Acknowledgements}
We are grateful to Jan Boruch, Yiming Chen, Matt Heydeman,    Luca Iliesiu, Jingru Lu, Ohad Mamroud,  Victor Rodriguez, Joaquin Turiaci, and  Zhenbin Yang for useful discussions.  
The work of MK, DM, and WZ was supported by NSF grant PHY-2408110. ZW is supported by Heising-Simons Foundation grants \#2024-5307.  This work was also supported by
funds from the University of California.

\appendix

\section{Index Asymptotics}
\label{sec:bound}

This section provides a brief discussion of the superconformal index ${\mathcal I}$ of $\mathcal N=4$ $SU(N)$ SYM \cite{Romelsberger:2005eg,Kinney:2005ej} in a regime where the bulk Hawking-Page transition suggests that we should find ${\mathcal I}$ to be of order $N^0$ at large $N$.  The index 
was defined in the above works as a special case of the partition function
\begin{equation}
Z(\beta, \tau, \sigma, \vec \Delta)
=\mathrm{Tr}_{\mathcal H_{\rm phys}}
\!\left(e^{-\beta\{\mathcal Q,\bar{\mathcal Q}\}}\zeta\right) \ \ \ {\rm with} \ \ \ 
\zeta = 
\,e^{2\pi i(\tau J_1+\sigma J_2+\frac{1}{2}\sum_i \Delta_i Q_i)},
\label{eq:index_def2}
\end{equation} 
which has been written in terms of the supercharge ${\mathcal Q},$ angular momenta $J_1,J_2$, and integer-normalized $R$-charges $Q_i$ for $i=1,2,3$ and with a constraint on $\tau,\sigma, \vec \Delta$ that we discuss below.  This partition function counts states in the physical Hilbert space of ${\mathcal N}=4$ $SU(N)$ super Yang-Mills theory, with states 
weighted by the given functions of their conserved charges. We specialize to the case $\sigma=\tau, \Delta_1 = \Delta_2 = \Delta_3 = \Delta$ with
\begin{equation}
2\tau-3\Delta = 1. 
\label{eq:chem_constraint}
\end{equation}

As described in  \cite{Copetti:2020dil, Aharony:2021zkr,Chang:2013fba,CMV}, after introducing the `single-letter index' function
\begin{equation}
f(\tau)
=1-\sum_{n_i=0}^{1}\sum_{n_{\pm}=0}^{\infty}(-1)^{\sum_i n_i}
\exp\!\Bigl[2\pi i\bigl(\tau\,(n_+ + \,n_-) + \Delta \sum_{i=1}^3  n_i\bigr)\Bigr] =  1- \frac{\(1-e^{2\pi i (2\tau - 1)/3}\)^3}{\(1-e^{2\pi i \tau}\)^2},
\label{eq:f_def}
\end{equation}
the index is computed by the matrix integral
\begin{equation}
\mathcal I(\tau)
=\frac{\int [DU]\,
\exp\!\left(\sum_{n=1}^{\infty}\frac{1}{n}\,f(n\tau)\,
\tr U^n\,\tr (U^\dagger)^n\right)}{\exp\!\left(\sum_{n=1}^{\infty}\frac{1}{n}\,f(n\tau)\right)},
\label{eq:matrix_integral}
\end{equation}
where $[DU]$ denotes the normalized Haar measure ($\int[DU]=1$) on $U(N)$ and the denominator cancels the contributions from the overall $U(1)$ in the $U(N)$ matrix integral in the numerator to leave us with the desired result for $SU(N)$. 

In the limit $\mathrm{Im}\,\tau\to+\infty$, the single-letter index $f(\tau)$ is exponentially suppressed.  In particular, expanding \eqref{eq:f_def} yields
\bal
f(\tau)= - 3 (-1)^{\frac{1}{3}} e^{4\pi i \tau/3} -2e^{2\pi i \tau}  +{\mathcal O}( e^{8\pi i \tau/3}).
\eal
It then follows that the $f(n\tau)$ are even more suppressed. Keeping only the leading $n=1$ term  and using the Haar measure Weingarten formula,
\begin{equation}
\int [DU]\; U_{ij}\,U^\dagger_{kl}=\frac{1}{N}\,\delta_{il}\delta_{jk},
\label{eq:weingarten}
\end{equation}
we obtain
\begin{equation}
\int [DU]\;\tr U\,\tr U^\dagger
=\sum_{i,j}\int [DU]\;U_{ii}\,U^\dagger_{jj}
=\sum_{i,j}\frac{1}{N}\delta_{ij}=1,
\label{eq:trace_contract}
\end{equation}
and thus
\begin{equation}
\mathcal I(\tau)=1+\mathcal O(e^{-8\pi\,\mathrm{Im}\tau/3}).
\label{eq:I_asymptotic}
\end{equation}

Hence, in the large $\operatorname{Im}\tau$ limit with $N$ fixed, the index tends to unity in accord with expectation that it be dominated by vacuum contributions on the AdS side of the correspondence, and in particular by contributions from a thermal AdS bulk saddle.   This also fits with the fact that, from the explicit expression for $f(\tau)$, we see that the leading contribution arises from the sector without any BPS excitations, $n_i = n_\pm = 0$.  It also agrees with the results of \cite{Copetti:2020dil} which used essentially the same argument but did not explicitly and analytically take the limit of large $\operatorname{Im}\tau$. However, as noted in the introduction, it is in distinct tension with the Bethe ansatz 
analyses of e.g. \cite{Benini:2018ywd,Aharony:2021zkr}, which suggested that contributions exponentially large in $N$ would remain relevant at large ${\rm Im}\, \tau$.  It would be interesting to further investigate this tension, perhaps by finding additional solutions to the Bethe ansatz equations, or by better understanding the behavior of the existing solutions at large $N$ (and especially the ones that lie on the Stokes ray in the analysis of \cite{Aharony:2021zkr}).

\section{Finite $\beta$ computations}
\label{sec:thimbles}

This appendix studies the finite $\beta$ non-BPS ansatz \eqref{eq:ansatz}.  As discussed in the main text, due to large corrections associated with Fermion zero modes it will {\it not} give a good approximation to the supersymmetric index defined by potentials that satisfy \eqref{eq:SUSYPot}.  We nevertheless include a preliminary study of the saddle-point approximation to this ansatz in order to better appreciate the impact of such fermionic quantum corrections, and also to understand the behavior of related quantities that break supersymmetry either by violating \eqref{eq:SUSYPot} or by including additional insertions of operators that soak up fermion zero modes, see section \ref{sec:disc}.

We begin by  analyzing the convergence properties of \eqref{eq:Zgrav} in section \ref{subsec:conv}.
Section \ref{sec:a=0} then considers a toy model defined by setting $\Omega=0$ (and by dropping the sum over shifts in $\Omega$).   The fact that section \ref{sec:SUSY} found that BPS black hole saddles with $\Omega=0$ are non-rotating (i.e., that they have $a=0$) thus suggests that we may then simplify the analysis by simply setting $a=0$ rather than performing the integral over $a$.  Doing so allows us to study the relevant contours and to determine the phase diagram for this truncated model.   While an analogous treatment of the full integral \eqref{eq:Z-eff} is beyond the scope of this work, we then return to the general rotating case in section \ref{subsec:gen} and find that simple constraints can provide insights into the resulting phase diagram.

\subsection{Convergence of the ansatz}
\label{subsec:conv}

 The Euclidean action has poles at $a = \pm 1$. Near these poles we find
\begin{equation}
\left.S_{\text{eff}}\right|_{a\rightarrow \pm 1}= \frac{\pi (1 + r_0^2)^2 \, \beta \, ( \Omega\mp1)}{16 (a \mp 1)^3} +O \left((a\mp 1)^{-2} \right).
\end{equation}
Convergence of our integral at $a=\pm 1$ thus requires
\begin{equation}
\label{eq:condition_convergence}
    0 < \text{Im }\tau < \frac{\text{Re }\beta}{\pi}.
\end{equation}
At large $r_0$, we find
\begin{equation}
   S_{\text{eff}} =\pi r_0^4\frac{ \left(\left(a^2-4 a+3\right) \beta -8 i \pi  a \tau \right)}{8 \left(1-a^2\right)^3} + O(r_0^3).
   \label{eq:poleaction}
\end{equation}
It is easy to check that if \eqref{eq:condition_convergence} is satisfied, the integral at $r_0$ is convergent.

One may be worried about the consistency of \eqref{eq:condition_convergence} with CFT picture where the most right-hand side inequality is not needed to ensure that the index is finite. To get a better understanding of this condition, let us start by considering the special case of real $\beta$ and $\Omega$. The condition \eqref{eq:condition_convergence} then reduces to $\Omega>-1$ which is a well-known condition for the existence of the thermal partition function on both sides of the duality. Of course, the index is not exactly partition function because it involves $(-1)^F$ which leads to a lot of cancellations between non-BPS states. Thus, the index should remain finite, even with $\Omega<-1$, if we understand it as the sum only over BPS states.  However, when written as the trace \eqref{eq:index_def} over the entire Hilbert space, it is not absolutely convergent. 

The same interpretation should be made for the gravitational path integral. When 
\begin{equation}
    \text{Im }\tau < \frac{\text{Re }\beta}{\pi}
\end{equation}
stops being satisfied, the ansatz \eqref{eq:ansatz} is no longer absolutely convergent. 
In contrast, our BPS-only ansatz \eqref{eq:indexansatz} remains convergent for $    \text{Im }\tau > \frac{\text{Re }\beta}{\pi}$ since (as noted below \eqref{eq:q-complex}) the contour of integration for \eqref{eq:indexansatz} involves only $a\ge 0$. 

\subsection{Thimbles with $a=0$}
\label{sec:a=0}

Despite simplifications from setting equal the two angular momenta and the three charges, the basic structure of the integral \eqref{eq:Zgrav} remains rather cumbersome.  The primary issue is that $E$ is not a single-valued function of general complex ${\mathcal A},J,Q$ so that the integrand lives on complicated multi-sheeted Riemann surface.  While, the original contour, as defined over real configurations, lives only on one-sheet, intensive effort is nevertheless required to carefully track contours (and in particular the upward-flow dual thimbles) relevant to the desired Picard-Lefschetz analysis of general saddles. 

Since the index is independent of $\beta$ at fixed $\tau, \Delta$, by tuning $\beta$ appropriately we can always choose to evaluate it for $\Omega_0=0$.  While the shifted sectors will have $\Omega_n \neq 0$ for $n\neq 0$, for simplicity we assume here that the $n=0$ sector dominates.  Since $\Omega_0$ vanishes, it is natural to also assume that the integral is dominated by configurations with $a=0.$   In this case we may approximate \eqref{eq:Zgrav} by
\begin{equation}
Z  \approx \sum_{m \in {\mathbb Z}}
\int_{r_0 \ge 0, \, q \in {\mathbb R}} \! dr_{0}\, dq \;
\det\!\left[\frac{\partial(\mathcal{A},Q)}{\partial(r_0,q)}\right]
\exp\!\left[\frac{\mathcal{A}}{4G_5}-\beta\bigl(E-\frac{3}{2}\Phi_m Q\bigr)\right].
\label{eq:Zgrav2}
\end{equation}
Here we use the symbol $Z$ even though we have tuned to supersymmetric potentials satisfying \eqref{eq:SUSYPot} since it is clear that our model's neglect of fermionic quantum corrections will result in large deviations from the true index.

Before we will dive into detailed analysis of \eqref{eq:Zgrav2}, let us explain why it is consistent to restrict to the $a=0$ hypersurface within the context of the semiclassical approximation. In a general setting, this restriction might require the violation of one of the equations for the ascent flow, namely
\begin{equation}
    \frac{da}{d\lambda} = -\overline{ \partial_a S}.
\end{equation}
However, one can check that when $\Omega = 0$ we have $\partial_a S|_{a=0} = 0$. In other words, the $a=0$ hypersurface is preserved under the flow. In particular, if any (non-critical) point of the thimble crosses this surface, it follows that its upward and backward evolution under the flow will remain on the $a=0$ surface. Thus, if we impose the $a=0$ constraint and then find a steepest-ascent curve that intersects our contour, we know that this contour corresponds to a steepest-ascent curve in the larger analysis as well. Having said that, we must acknowledge the limitations of our approach. Even if there is no flow from the saddle to the integration contour that maintains $a=0$, it cannot be excluded that such a flow exists in the larger model, but that away from the  critical point it has $a \neq 0$.

In the $a=0$ sector the ADM energy, horizon area, and electric charge simplify to
\bal
E=\frac{3\pi(r_0^4(1+r_0^2)+q^2)}{8G_5r_0^2},~~~\mc{A}=2\pi^2r_0^3,~~~\mc{Q}=\frac{\pi}{2G_5} q \, .
\eal
We may thus write \eqref{eq:Zgrav2} in the explicit form
\begin{equation}
Z \approx \sum_{m\in {\mathbb Z}} \int_{{\mathbb R}^+} dr_0\,\int_{\mathbb R} dq\;
\left(\frac{3}{4}\,\pi^3 r_0^2\right)
\exp\!\left[\frac{1}{G_5}\left(
\frac{\pi^2}{2} r_0^3
-\beta\!\left(
\frac{3\pi\bigl(r_0^4(1+r_0^2)+q^2\bigr)}{8r_0^2}
-\frac{3}{2}\Phi_m\,\frac{\pi}{2} q
\right) \right)
\right] .
\end{equation}

Note that the dependence on $q$ is purely Gaussian.  Furthermore, since $r_0$ is real on the integration contour, the integral over $q$ converges for ${\rm Re} \, \beta > 0$.  As a result, we may perform the $q$-integral explicitly to find
\begin{equation}
{\mathcal I} \approx \sum_{m\in {\mathbb Z}} \int_{{\mathbb R}^+}  dr_0\;
\left(\pi^3r_0^3\sqrt{\frac{3G_5}{2\beta}}\right)
\exp\!\left[\frac{1}{G_5} \left(
\frac{\pi^2}{2} r_0^3
+\frac{3}{8}\,\beta\pi r_0^2
\bigl(\Phi_m^2-1-r_0^2\bigr) \right)
\right] .
\label{eq:Zgrav_rplus}
\end{equation}

Furthermore, 
the dominant contribution to \eqref{eq:Zgrav_rplus} arises from the saddle points of the exponent.
Varying with respect to $r_0$ yields the saddle-point equation
\begin{equation}
r_0\left(2 \pi r_0 + \beta(\Phi_m^2-2r_0^2 -1)
\right)=0 .
\label{eq:rplus_saddle_BPS}
\end{equation}

We now wish to impose the supersymmetric relation \eqref{eq:SUSYPot}.  However, as described in section \ref{sec:SUSY}, our \eqref{eq:rplus_saddle_BPS} will admit BPS saddles only for $m=0$ or $m=-1$ (the two cases satisfying \eqref{eq:susy-constraint2}).  Since contributions from non-BPS saddles are removed by the integrals over fermion zero modes, we simply impose $\Phi_0=\Phi^{(\pm)} := \frac{1}{3}-\frac{2i\pi}{3\beta}$ and $\Phi_{-1}=\Phi^{(-)} := \frac{1}{3}+\frac{2i\pi}{3\beta}\ $.  Note that this can be satisfied for at most one value of $m$ (i.e., for only one of the two possible signs on the right-hand-side).  And since we saw in section \ref{sec:SUSY} that the $\pm$ actions are related by a symmetry, it suffices to choose just one of these signs to study in detail.  We shall write explicit formulae only for $\Phi^{(+)}$ below.

Substituting this value into \eqref{eq:rplus_saddle_BPS} yields the following possible radii of the horizon:
\begin{equation}
r_0^{(0)}=0,
\qquad
r_0^{(1)}=\frac{2(\pi-i\beta)}{3\beta},
\qquad
r_0^{(2)}=\frac{\pi+2i\beta}{3\beta}.
\label{eq:rplus_solutions}
\end{equation}
It is then straightforward to check that $r_0^{(1)}$ lies on the supersymmetric branch selected by \eqref{eq:q-factorized}, while $r_0^{(2)}$ fails to satisfy the BPS relation \eqref{eq:BPS}.
Note also that the saddle $r_0^{(0)}$ always lies at the endpoint of our contour.  Since its action vanishes identically,  it is associated with an $O(N^0)$ endpoint contribution for all values of $\beta, \Phi^{(\pm)}$; see e.g. the discussion in \cite{Kolanowski:2026gii}. 

\begin{figure}[h!]
    \centering
    {{\includegraphics[width=0.48\linewidth]{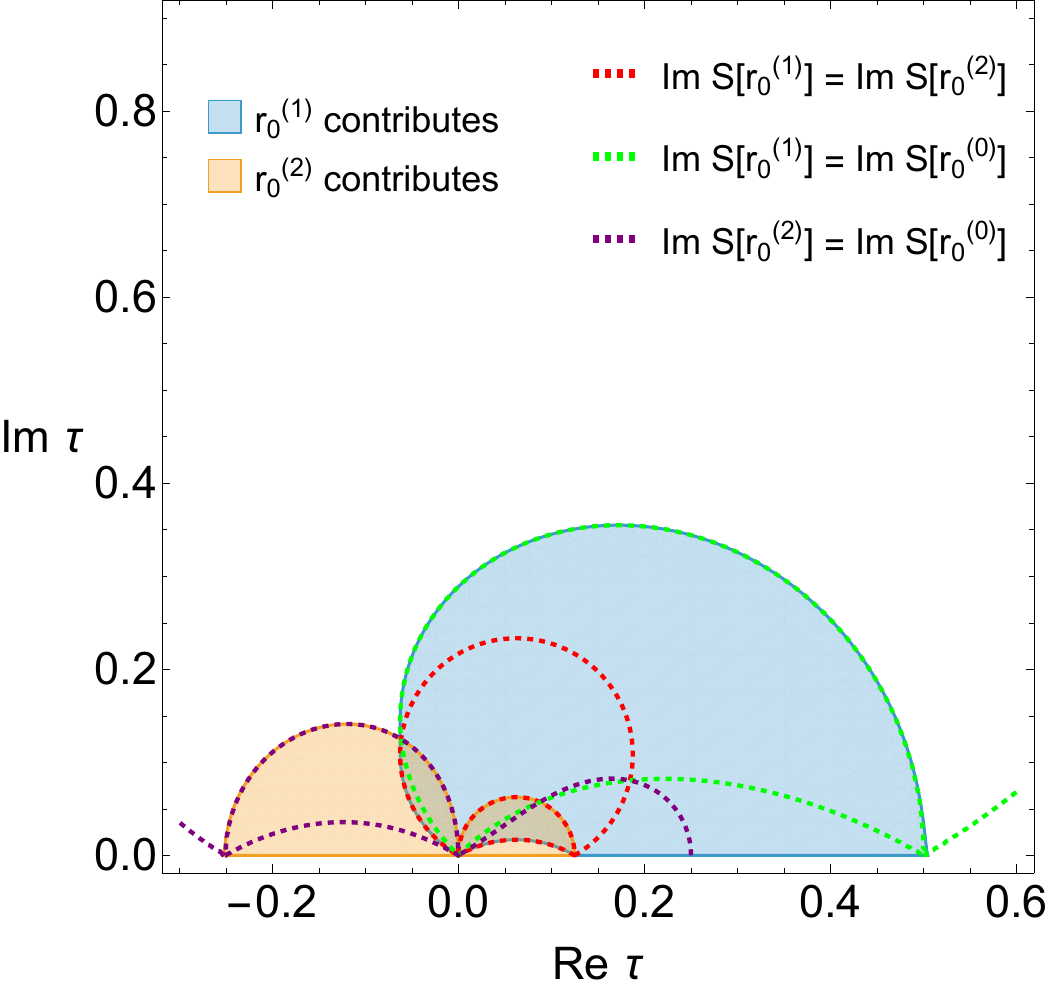}}}
    \quad
    {{\includegraphics[width=0.48\linewidth]{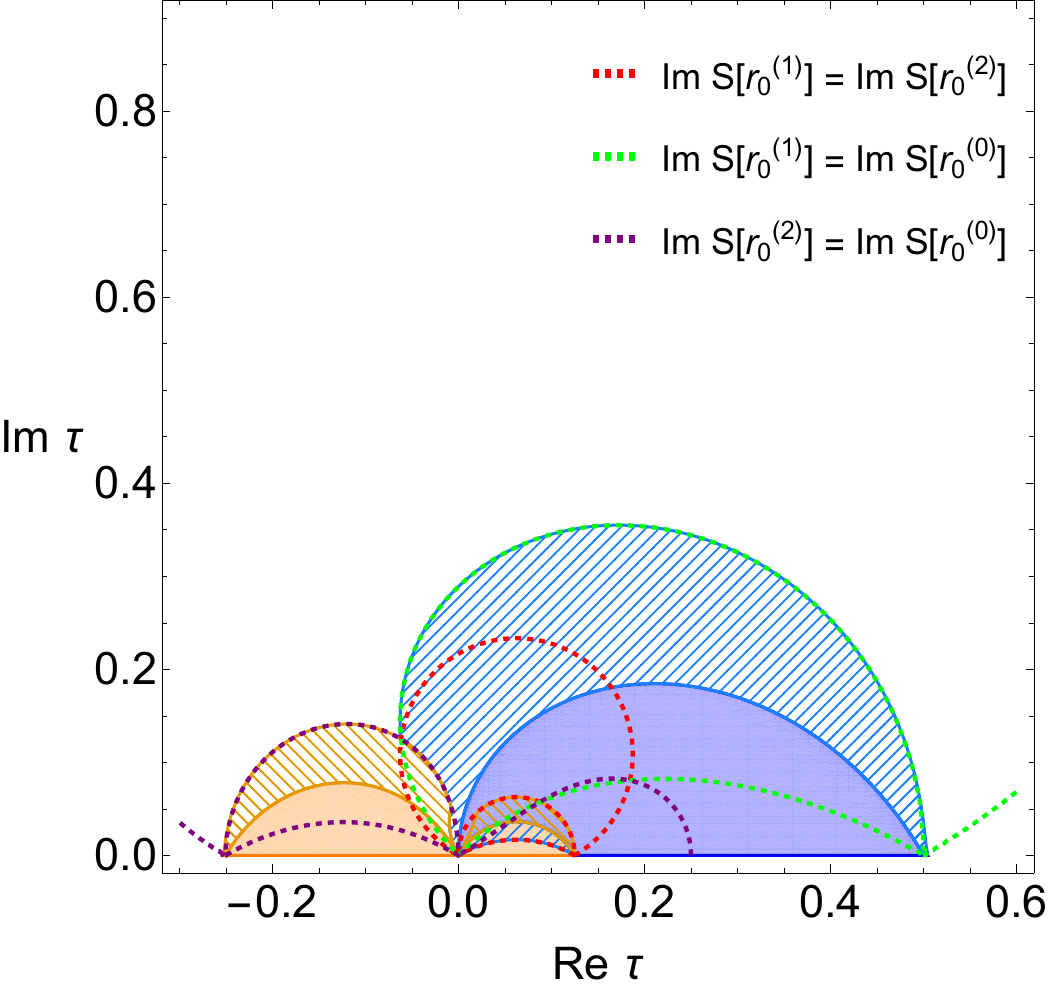} }}
    
   \caption{
   \textbf{Left:} The light blue and orange shadings show the regions in the complex $\tau$-plane for which the $r_0^{(1)}$ and $r_0^{(2)}$ saddles contribute to \eqref{eq:Zgrav_rplus}; i.e., where 
   ascent contours (dual thimbles) cross the positive real axis. The red, green, and purple dashed lines mark the loci on which the imaginary parts coincide for the indicated pair of saddle-point Euclidean actions.  These are thus locations where Stokes' phenomenon may potentially occur, though the actual Stokes' phenomena happen only where the shading changes. \textbf{Right:}  The blue and orange shaded regions show the domains in the complex $\tau$-plane where the $r_0^{(1)}$ and $r_0^{(2)}$ saddles are the dominant contributions. The blue and orange hatched regions show domains where the thermal AdS saddle $r_0^{(0)}$ dominates but where the $r_0^{(1)}$ and $r_0^{(2)}$ saddles are respectively the second most dominant contributions.
   }
    \label{fig:contributingsaddles}
\end{figure}

Numerically performing a Picard-Lefschetz thimble  analysis for the other two saddles\footnote{As discussed in section \ref{sec:PLO}, Stokes phenomena can occur only when the imaginary part of the Euclidean action for one saddles coincides with those of another.  Such loci can be found numerically with high accuracy.  It then suffices to numerically construct thimbles with low  density near and along such loci to determine the regions in which each saddle contributes.}  yields the results shown in figure \ref{fig:contributingsaddles}.
The left panel indicates the regions in which the various contribute to the integral \eqref{eq:Zgrav_rplus}. The black hole saddles contribute only in a compact region and, in particular, only when the imaginary part of $\tau$ is sufficiently small. As a result, Stokes' phenomena catalyzed by the vacuum AdS endpoint occur at the largest values of ${\rm Im}\, \tau$ where the black holes contribute.  This fits the picture of thermal AdS dominance at large ${\rm Im}\, \tau$.     The right panel adds further information by showing the dominant contributing saddle in each region as well as the relative magnitudes of subdominant contributing saddles.

\begin{figure}[h!]
    \centering
    {{\includegraphics[width=0.48\linewidth]{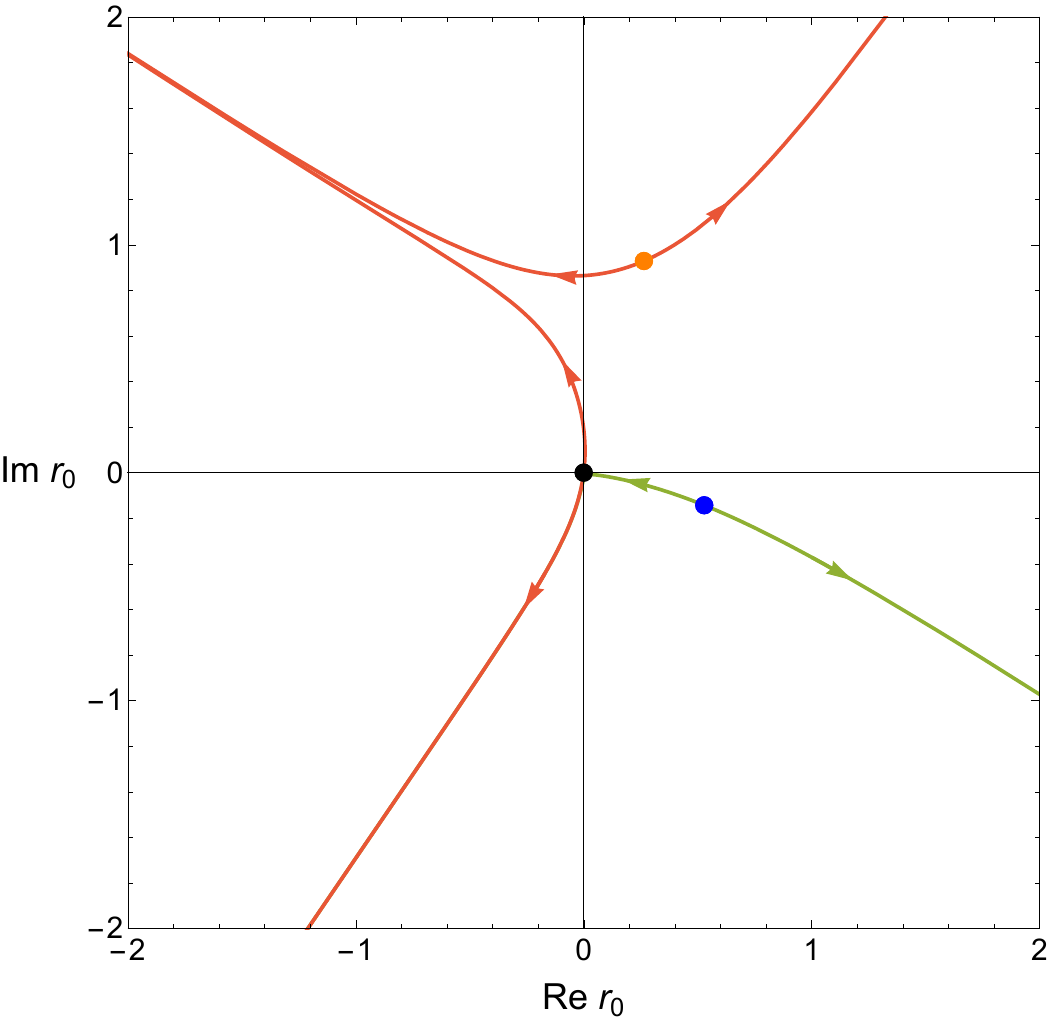}}}
    {{\includegraphics[width=0.48\linewidth]{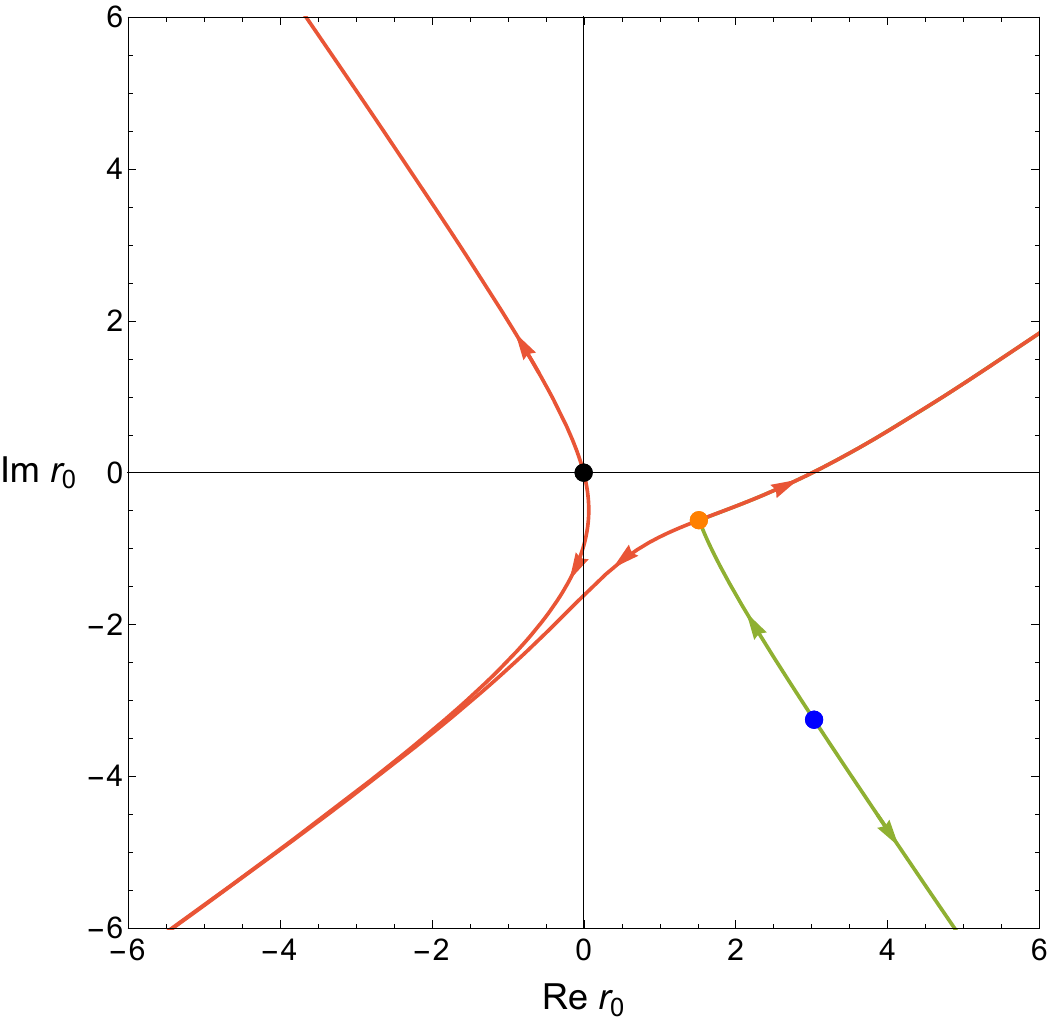} }}
    {{\includegraphics[width=0.48\linewidth]{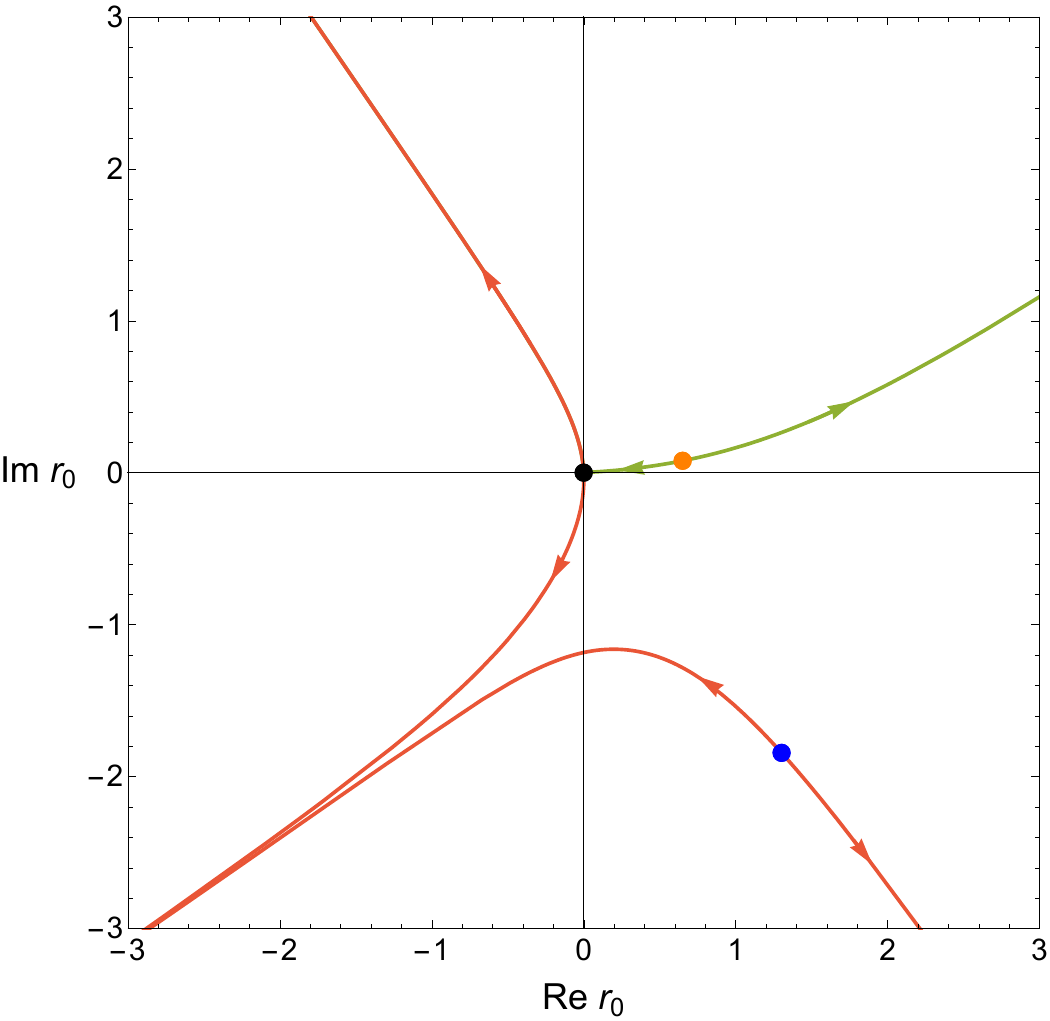} }}
    {{\includegraphics[width=0.48\linewidth]{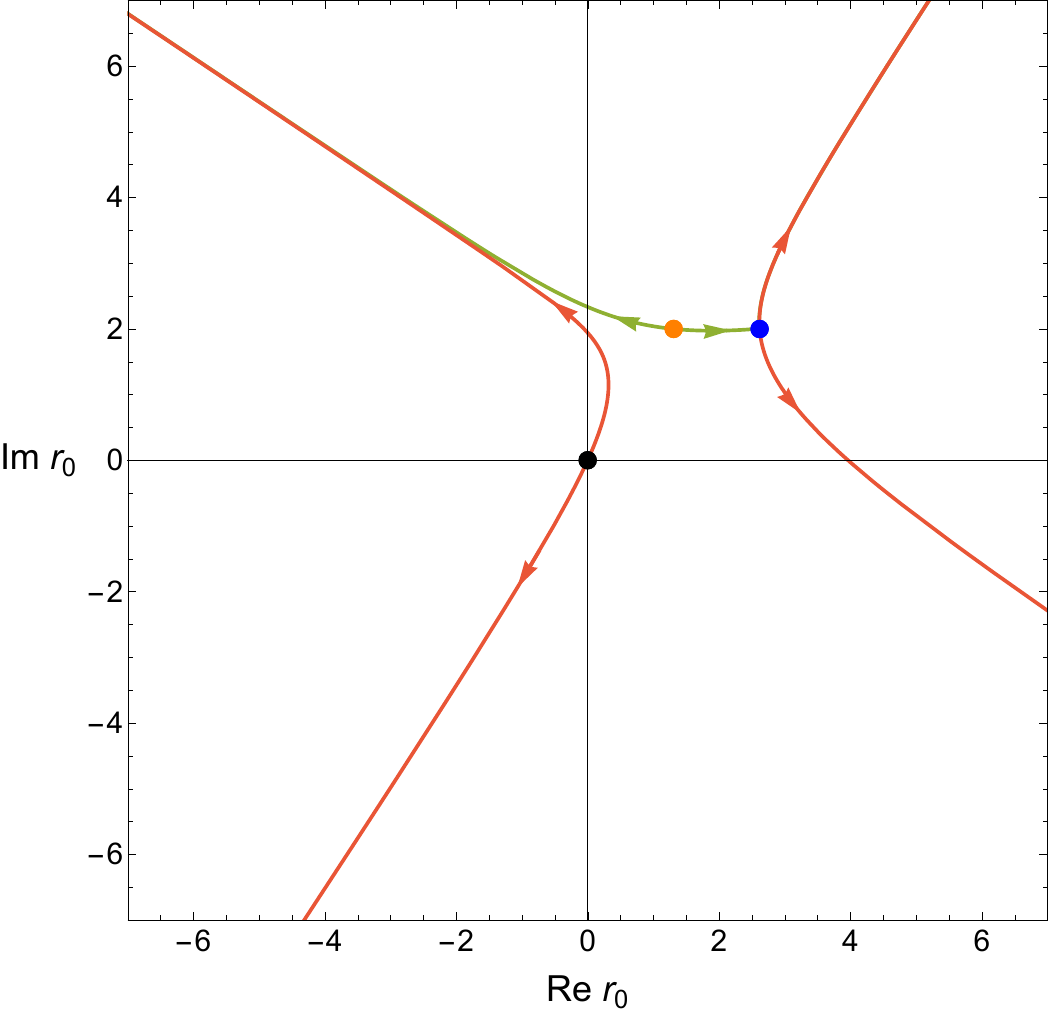}}}
    
   \caption{
   Examples of our Stokes' phenomena for selected values of $\tau$.  Ascent contours are  shown for the saddles  $r_0^{(0)}$ (Black), $r_0^{(1)}$ (Blue), and $r_0^{(2)}$ (Orange) in the complex $r_0$-plane. The ascent contour for the saddle experiencing the Stokes' discontinuity is shown in green, while those for other saddles are shown in red.
   Arrows indicate the direction in which $-\text{Re}S_{\rm E}$ increases.  \textbf{Top Left:} $\tau=0.315+0.318i$.   The blue saddle ($r_0^{(1)}$) experiences a Stokes' transition  catalyzed by the black saddle ($r_0^{(0)}$). \textbf{Top Right:} $\tau=-0.0543+0.0637i$.  The blue ($r_0^{(1)}$) truly-BPS saddle experiences a Stokes' transition catalyzed by the only-asymptotically-BPS orange saddle ($r_0^{(2)}$); see again footnote \ref{foot:OABPS}. \textbf{Bottom Left:} $\tau=-0.127+0.141i$. The orange saddle ($r_0^{(2)}$) experiences a Stokes' transition catalyzed by the black saddle ($r_0^{(0)}$). \textbf{Bottom Right:} $\tau=0.0637+0.0625i$. The orange saddle ($r_0^{(2)}$) experiences a Stokes' transition catalyzed by the blue saddle ($r_0^{(1)}$).  In all 4 cases the relevant $n_\sigma$ changes from $\pm 1$ to zero as can be seen by following flows on either side of the green contour.
    }
    \label{fig:StokesEx}
\end{figure}

Figure \ref{fig:StokesEx} then presents a few examples illustrating our Stokes phenomena. When $\tau$ reaches a critical line at which the imaginary parts of the actions of two saddles coincide, the steepest-ascent contour (upward-flow cycle) from one saddle $\sigma$ can flow to another saddle $\sigma'$.  When this occurs, the upward-flow cycle from $\sigma$ changes discontinuously as $\tau$ crosses the relevant line.  

As a result, the intersection number $n_\sigma$ with the original integration contour may change discontinuously as well.  The most interesting case shown is perhaps that in the upper right panel, in which the only-asymptotically-BPS saddle\footnote{\label{foot:OABPS} Recall from section \ref{subsec:NonBPSSUSY} that any non-BPS saddle with $2\tau_n-3\Delta_{n'}=\pm 1$ is continuously connected to a large $\beta$ limit that converges to a saddle satisfying the BPS condition $E-2J-\frac{3}{2}Q=0$.} $r_0^{(2)}$  catalyzes a transition for the BPS 
 $(r_0^{(1)})$.  This transition is responsible for the fact that the lower left portion of the blue shaded region in figure \ref{fig:contributingsaddles} ends on a dashed red line.
 
\subsection{Including rotation}
\label{subsec:gen}

When we include both charge and rotation (though still with $J_1=J_2$
 and $Q_1=Q_2=Q_3$),  the non-BPS ansatz is defined by the multi-dimensional integral in~\eqref{eq:Zgrav} over the variables $(a,r_0,q)$. Let us refer to the term associated with integers $n,m$ as ${\mathcal I}_{n,m}(\beta, \Omega, \Phi)$. Since the dependence on $q$ is Gaussian at fixed $a,r_0$,  it is convenient to first integrate over $q$ to obtain an effective two-dimensional integral
\begin{equation}
{\mathcal I}_{n,m}(\beta, \Omega, \Phi)
=\int_{{r_0 >0}\atop {1> a> -1}} dr_0 da\; f(a,r_0)\, e^{-S_{\rm eff}(a,r_0)} \, ,
\label{eq:Z-eff}
\end{equation}
where the effective action of course also depends on $(\beta, \Omega, \Phi)$. 
The prefactor $f(a,r_0)$ collects the Jacobian from the change of variables to $a,r_0,q$ together with the determinant produced by the Gaussian integral over $q$. In particular, $f(a,r_0)$ is in general a meromorphic function and introduces additional poles in the complexified $(a,r_0)$ plane.

The presence of multiple extra poles from the measure factor $f(a,r_0)$ substantially complicates a direct Lefschetz-thimble decomposition of~\eqref{eq:Z-eff}.
Nevertheless, as has been used already many times in the main text, one may study a simple condition that is \emph{necessary} for the supersymmetric black-hole saddle to contribute: 
If the steepest-descent thimble $\mathcal J_{\rm BH}$ attached to ths saddle has non-vanishing intersection number with the original integration cycle  $\mathcal C$, then $\mathcal C$ must contain at least one point $p$ at which  i) the phase of the integrand at $p$ agrees with the phase at the saddle, and also ii) the magnitude of the integrand  at $p$ is larger than at the saddle.  Equivalently, there must exist a point $p=(a,r_0)$ on $\mathcal C$ such that
\begin{equation}
{\rm Im}\, S_{\rm eff}(p)={\rm Im}\, S_{\rm BH}.
\qquad {\rm and} \qquad
{\rm Re}\, S_{\rm eff}(p)<{\rm Re}\, S_{\rm BH}.
\label{eq:necessary-condition}
\end{equation}

To study this condition, we again specialize to potentials satisfying \eqref{eq:susy-constraint2} for some choice of sign, thinking of $\Phi$ as the corresponding function $\Phi^{(+)}(\beta, \Omega)$ or $\Phi^{(-)}(\beta, \Omega)$ as in section \ref{sec:SUSY}.  
Using $(a_\ast,r_{0,\ast})$ to denote the corresponding supersymmetric black-hole solution for the given choice of potentials, we again use the notation of section \ref{sec:SUSY} to define 
\begin{equation}
S^{(\pm)}_{\rm BH}(\tau):=S^{(\pm)}_{\beta, \Omega}(a_\ast,r_{0,\ast}) \, ,
\label{eq:SBH}
\end{equation}
as representing the action of the $\pm$ supersymmetric saddle. 
Here the argument $\tau$ reminds us
that, as we saw in section \ref{sec:SUSY}, the action of the supersymmetric black hole is independent of $\beta$ when the parameter $\tau$ is held fixed which, in the present context, also holds $\Delta$ fixed by \eqref{eq:chem_constraint}.
By construction, 
the pair $(a_\ast,r_{0,\ast})$ is also a saddle for \eqref{eq:Z-eff} at which $S_{eff}(a_\ast,r_{0,\ast})$ is again given by \eqref{eq:SBH}. Corresponding statements also hold for the non-BPS saddles discussed in section \ref{subsec:NonBPSSUSY}.

Indeed, this condition is especially interesting  for potentials satisfying \eqref{eq:SUSYPot}, where we may hope that our ansatz approximates the supersymmetric index.  Recall that, for fixed values of $\sigma, \tau, \vec \Delta$, the true index is known to be independent of $\beta$.  As a result, any given term in the trans-series expansion of ${\mathcal I}$ must also be independent of $\beta$ at fixed $\sigma, \tau, \vec \Delta$,  So if the index is computed by a gravitational path integral, and if there is a saddle of Euclidean action $S_E$ that contributes at some value of $\beta$, then at all other values of $\beta$ (with the same $\tau, \sigma, \vec \Delta$) there must also be a saddle that contributes and which has identical Euclidean action\footnote{Though other aspects of the bulk physics might potentially depend on $\beta$; e.g., it might be possible for the saddle to be just a complex black hole at some $\beta$, while at other $\beta$ the saddle might be a dual dressed black hole of the sort described in \cite{Choi:2024xnv}.} $S_E$. The necessary condition \eqref{eq:necessary-condition} would then need to be satisfied at all values of $\beta$ for the given $\tau,\sigma, \vec \Delta$.

\begin{figure}[h]
    \centering
    {{\includegraphics[width=0.5\linewidth]{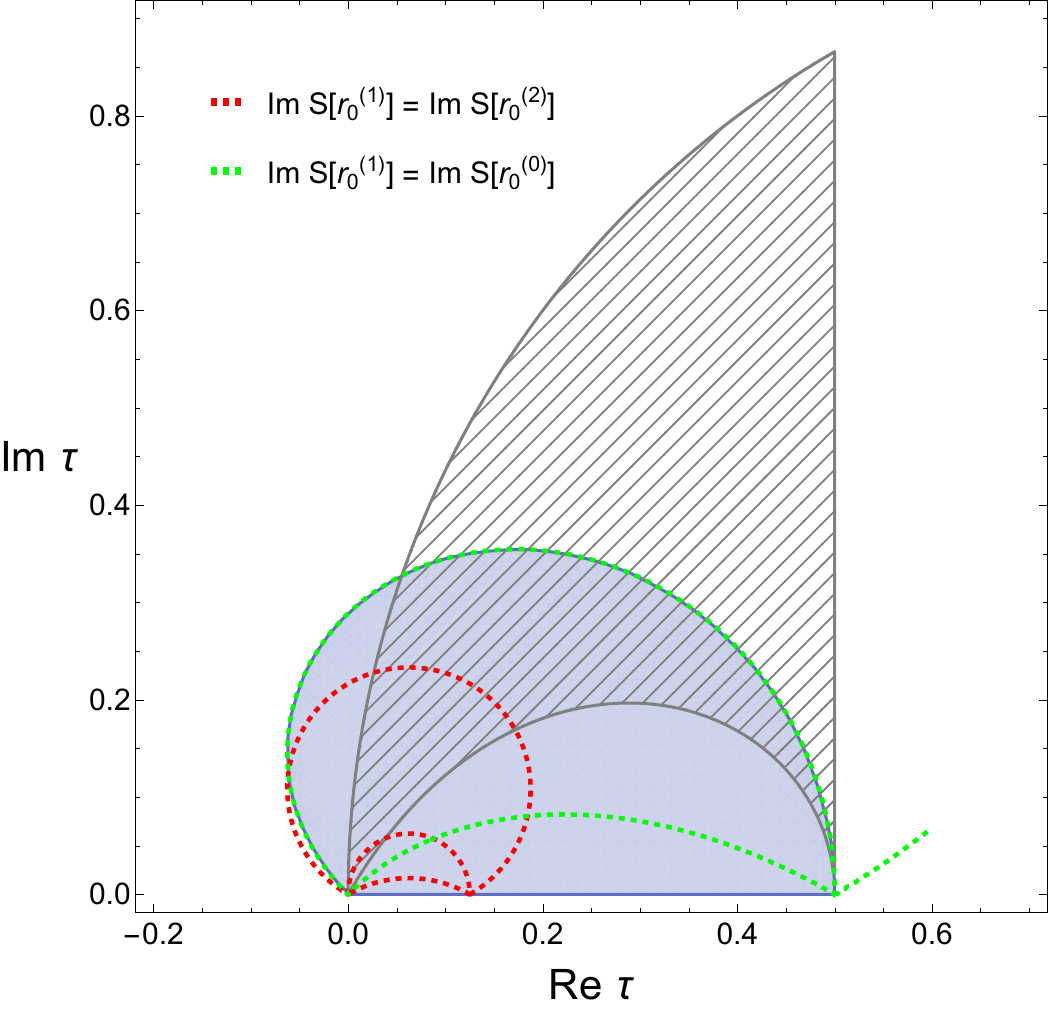}}}
   \caption{ In the lavender region the necessary condition \eqref{eq:necessary-condition} is satisfied for all $\beta$ with $\text{Re }\beta > \pi\text{Im }\tau > 0$. At such $\beta, \tau$ the saddle {\it may} contribute.  In the gray hatched region there exists a $\beta$ with $\text{Re }\beta > \pi\text{Im }\tau > 0$ such that the supersymmetric black hole saddle can have real $r_0>0$ and $-1<a<1$, so that this saddle lies directly on the integration contour. Near such $\beta, \tau$ this saddle {\it must} contribute. The green dashed lines indicate the $\beta$-independent locus where the imaginary part of the action vanishes for the supersymmetric saddle and thus where it may experience Stokes' phenomena catalyzed by the thermal AdS saddle. Since they coincide with the green dashed lines of figure \ref{fig:contributing_tau}, they show that the shaded region here also coincides with that of \ref{fig:contributing_tau}. However,  the red dashed lines show that the lavender region differs slightly from the blue region in the left panel of Figure \ref{fig:contributingsaddles}.}
    \label{fig:NC}
\end{figure}

The results of our numerical investigations are shown in figure \ref{fig:NC}.  As shown there,  we do indeed find a region of the $\tau$-plane in which the necessary condition is satisfied for all complex $\beta$ with $\text{Re }\beta >0$. This region coincides with the region shown in figure \ref{fig:contributing_tau} where the supersymmetric saddle contributed to the one-dimensional BPS-only ansatz of section \ref{subsec:one-dimensional-BPS-integral}.  

\begin{figure}[h!]
    \centering
    {{\includegraphics[width=0.5\linewidth]{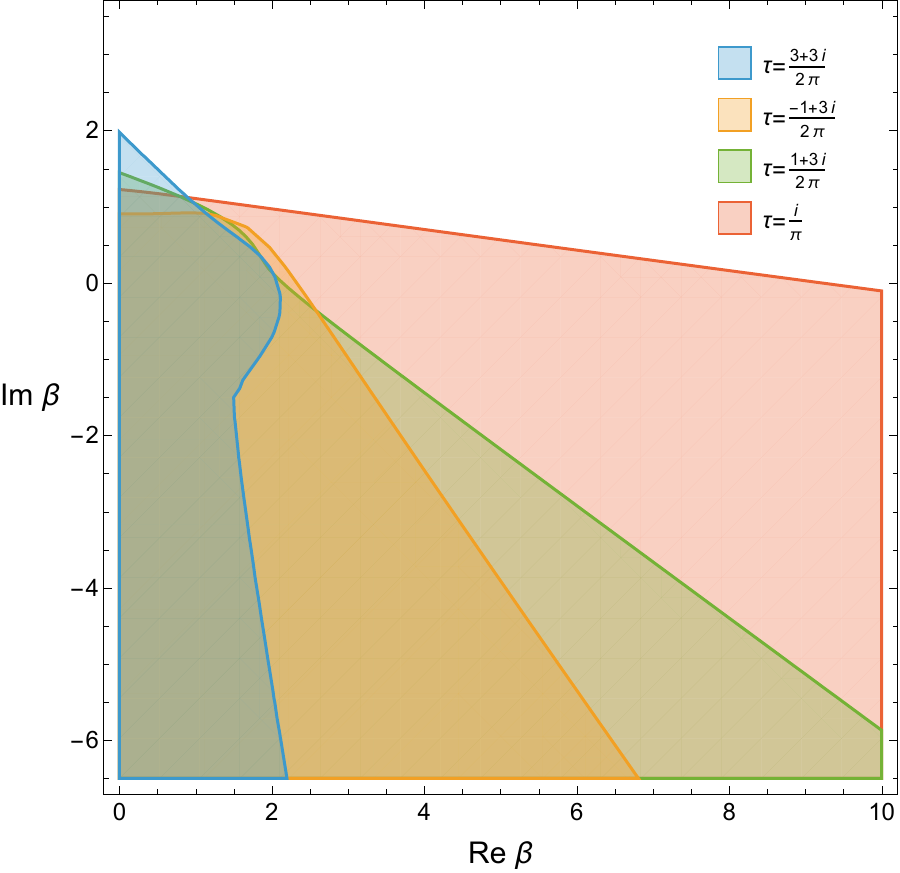}}}
   \caption{The regions in the $\beta$-plane where the necessary condition \eqref{eq:necessary-condition} is satisfied, for fixed $\tau$.
    }\label{fig:NC2}
\end{figure}

The reason for this agreement can be seen from figure \ref{fig:NC2}.  If the saddle contributes at some $\beta$, then it always continues to contribute when the real part is decreased; i.e., failures occur only at large ${\rm Re}\, \beta$.  So requiring the condition to be fulfilled for all $\beta$ is equivalent to requiring it to hold in the limit of large ${\rm Re} \, \beta$.  But in direct parallel with the discussion of thimbles in section \ref{subsec:consistency}, this is then equivalent to requiring the condition to hold for the one-dimensional BPS-only integral studied in section \ref{subsec:one-dimensional-BPS-integral}.  It then simply turns out for this case that applying the necessary condition to the BPS-only integral correctly predicts the results of the full Picard-Lefschetz analysis for that integral.

However, in the unshaded region of figure \ref{fig:NC}, we find that there is nevertheless {\it some} region of $\beta$ for which the necessary condition holds, though there is also some region where it fails.  There are then three possibilities:  i) Although the necessary condition is satisfied at some $\beta$, the ascent contour never actually intersects the integration contour and the saddle never contributes. 
ii) A new BPS saddle with precisely the same action (and precisely the same quantum corrections at each order) becomes relevant at some $\beta$, though it involves new degrees of freedom
(e.g., dual giant gravitons) that we have not studied explicitly.  Thus the relevant point $p$ does not lie on the particular three-dimensional integration contour used in our ansatz \eqref{eq:ZgravAJQ}.
 Nevertheless, the two saddles conspire so that the necessary condition is satisfied at all $\beta$ in the full path integral.  
iii) Fermion zero modes make quantum corrections large, and  dropping such corrections as we have done has introduced spurious $\beta$-dependence. 

Option (i) can be ruled out as follows: At fixed $\tau, \sigma, \vec \Delta$, the saddles move in the complex $r_0,a$ planes as a function of the complex parameter $\beta$.  Since the integration contour is codimension-2, for any given saddle there should be codimension-zero regions of the $(\tau, \sigma, \vec \Delta)$ parameter space where we can move  that saddle onto the integration contour by tuning ${\rm Re} \, \beta$ and ${\rm Im} \beta$.  Furthermore, when the saddle lies on the integration contour, it will contribute unless its ascent contour is tangent to the integration contour.  
The set of $\tau$ where this is possible is shown as the hatched region of figure \ref{fig:NC}.  The fact that it extends well outside the shaded region then rules out option (i) for the relevant values of $\tau$.  (It should not be a surprise that the hatched region does not include the entire shaded region, as saddles can be relevant even if they never lie on the contour of integration.) 

This establishes that the ansatz \eqref{eq:ansatz} is in fact $\beta$-dependent. As discussed in the main text, ignoring quantum corrections as we have done can certainly lead to such a result.

\section{The large \texorpdfstring{$\beta$}{beta} limit}
\label{subsec:large-beta-limit}

This appendix confirms the statement in section \ref{sec:BPSansatz} that the semiclassical expansion of the BPS-only ansatz \eqref{eq:indexansatz} agrees with what is obtained by first performing the semiclassical expansion of the non-BPS \eqref{eq:ansatz} and then taking the limit $\beta \rightarrow \infty$.  In particular, we begin by showing in section \ref{subsec:largebpert} that saddles and ascent flows for \eqref{eq:ansatz} at finite $\beta$ can be constructed perturbatively about the restricted BPS-only flows for \eqref{eq:indexansatz}.  This statement is equally true for both only-asymptotically-BPS saddles and truly-BPS saddles.  We then provide some brief numerics illustrating this agreement in section \ref{subsec:largebnumerics}. 

\subsection{Large $\beta$ flows as perturbations of BPS-only flows}
\label{subsec:largebpert}

We will now try to understand in what sense the only-BPS ansatz is approximated by the path integral over non-BPS configurations  in $\beta \to \infty$ limit. Let us start by rewriting the bosonic action as
\begin{subequations}
\begin{equation}
    -S = \frac{A}{4} - \beta \tilde E + 4\pi i\tau J + 3 \pi i Q \Delta ,
\end{equation}
where
\begin{equation}
    \tilde E = E - 2J - \frac{3}{2}Q.
\end{equation}
\end{subequations}
For simplicity, we will restrict to the case
\begin{equation}
    2\tau - 3\Delta =1.
\end{equation}
We should express everything in terms of the actual variables over which we will integrate.   It will be convenient to introduce the following parametrization:
\begin{subequations}
   \begin{equation}
        X = q+a^2-(1+2a)r_0^2
   \end{equation}
   \begin{equation}
       Y = r_0^2 - 2a-a^2,
   \end{equation}
   and, as in previous sections
   \begin{equation}
       a = -1 + \frac{1}{2}\left(t+t^{-1} \right).
   \end{equation}
\end{subequations}
In terms of this parametrization, we may write our action as
\begin{equation}
    \tilde{E} = f(t,Y) (4 X^2 t^2 + Y^2 ( (t^2-1)^2 + 4t^2 Y)) \approx f(t) (4X^2 t^2 +Y^2 (t^2-1)^2)
\end{equation}
\begin{equation}
    \frac{A}{4}+4\pi i \tau J + 3\pi i \Delta Q = \tilde{S}(t,X,Y) \approx S_0(t) + S_X(t) X + S_Y(t) Y,
\end{equation}
where we have set $G=1$ and used the fact that at large $\beta$ we expect $X$ and $Y$ to both be of order $O(\beta^{-1})$ (as will be justified below). We have also denoted $f(t) = f(t,0)$. In the following, the exact form of $f, \tilde{S}, S_0, S_X, S_Y$ is not important (although it is important that $S_0$ is given by the action of the BPS-only ansatz \eqref{eq:indexansatz}, which means that it agrees with \eqref{eq:BPS_action_t}). What is important is that they depend only on $t$ (i.e., they are independent of $X,Y$) and that they are rational functions with poles at $t=\pm i, \pm 1,2\pm \sqrt{3}$. We thus must be especially careful around these points. The idea we will follow is that, as $\beta \to \infty$, the saddles and thimble flows will localize at the $X=Y=0$ surface.  We then want to solve perturbatively in $\beta^{-1}$ for the thimbles. To this end, it is convenient to write the action as a sum of squares:
\begin{eqnarray}
    -S &\approx& -\beta f \left( \left(
2X t - \frac{1}{4t \beta f} S_X
    \right)^2  + \left(  
    Y (t^2-1) - \frac{1}{2 \beta f (t^2-1)} S_Y
    \right)^2\right) + S_0 \nonumber \\&+& \frac{1}{16t^2 \beta f} S_X^2 + \frac{1}{4\beta f (t^2-1)^2} S_Y^2,
    \label{eq:SXY}
\end{eqnarray}
where we have again dropped higher order corrections in $X,Y$, and $1/\beta$.

Let us start by looking at the saddles.  We will denote the location of the saddles by $t_\star, X_\star, Y_\star$. In regions where  the full action $S$ is smooth, we expect the saddles at large $\beta$ to be perturbatively close to points $(X_\star, Y_\star, t_\star) = (0,0,t_0)$ where $t_0$ is a saddle of $S_0$. The smoothness requirement holds at saddles of $S_0$ where $t_0$ is $t_+$, $t_-$ or $t = \frac{4+3i}{5}$. For these cases we write $t_\star = t_0 + \beta^{-1} t_1 + O(\beta^{-2})$.  We will return later to the more singular cases $t=\pm i$ which, despite the fact that they are poles of $S_X,S_Y$, can in fact be studied using a  slightly different choice of coordinates on $\mathbb{C}^3$.

For the smooth saddles $t=t_\pm, \frac{4+3i}{5}$, at order $\beta^{-1}$ the above expansion immediately yields expressions for $X,Y$ in terms of $t_0$:
\begin{subequations}
    \begin{equation}
        X = \frac{S_X(t_0)}{8\beta f(t_0) t_0^2},
    \end{equation}
    \begin{equation}
        Y = \frac{S_Y(t_0)}{2\beta f(t_0) (t_0^2-1)^2}.
    \end{equation}
\end{subequations}
We thus see that our expectation that $X,Y \sim \beta^{-1}$ is justified. 
However, we still need to solve for $t$. Let us write
\begin{equation}
    h(t) = \frac{S_X^2}{16t^2 f} + \frac{S_Y^2}{4(t^2-1)^2 f},
\end{equation}
so the equation for $t$ reads
\begin{equation}
    S_0' +\frac{h'}{\beta} = 0.
\end{equation}
Expanding around $t_0$, we find:
\begin{equation}
    0 = S_0'(t_0) + \frac{1}{\beta} \left(S_0''(t_0) t_1 + h'(t_0) \right). 
\end{equation}
Since $t_0$ was the saddle of $S_0$, 
we have $S_0'(t_0)=0$ and thus
\begin{equation}
    t_1 = - \frac{h'(t_0)}{S_0''(t_0)}.
\end{equation}
These results remain valid as long as the saddle at $\beta = \infty$ is non-degenerate and $f(t_0)t_0^2 (t_0^2-1) \neq 0$ (which is the case for generic $\tau$).

We have thus showed that saddles will remain parametrically close to those of $S_0$ as we turn on the temperature. We want to show that the same can be said about thimbles. These are going to be 3-(real) dimensional surfaces. The problem in $X,Y$ variables is almost Gaussian. We may simplify it further if we write
\begin{subequations}
\begin{equation}
    x = \sqrt{\beta f} \left(2Xt - \frac{S_X}{4tf \beta} \right),
\end{equation}
\begin{equation}
    y = \sqrt{\beta f} \left(Y (t^2-1) - \frac{S_Y}{2\beta f (t^2-1)} \right).
\end{equation}
\end{subequations}
Since these definitions involve $\sqrt{f}$, one should be careful in keeping track of relevant complex sheets, though in the end they turn out only to be relevant  for the flow in $t$.
Note that $x,y \sim \beta^{-1/2}$.

In this way, we have reduced the three-dimensional problem to three one-dimensional ones.
In these variables (assuming a flat metric in $x,y,t$), the flow equations for $x,y$ that follow from the explicit terms in \eqref{eq:SXY} take the form:
\begin{subequations}
  \begin{equation}
  \label{eq:xflow}
        \dot{x} = -2\overline{ x}
  \end{equation}
  and
  \begin{equation}
        \dot{y} = -2 \overline{ y},
  \label{eq:yflow}  
  \end{equation}
  \end{subequations}
  where a dot denotes the derivative with respect to the flow parameter $s$ and the bars denote complex conjugation.  These can be easily solved upon a choice of the right boundary conditions. To impose them, we should remember that we are looking for steepest ascent contours starting from the saddle\footnote{At higher orders in $1/\beta$ we may correct the definitions of $x,y$ to keep the saddle at $x=y=0.$} at $x=y=0$.  That implies that as $s \to -\infty$, we must approach the origin from the side with imaginary $x,y$. Using $\beta$-independent real constants $\gamma_x,\gamma_y$ to parametrize the possible initial directions of the flow, the solutions to \eqref{eq:xflow} and \eqref{eq:yflow} are simply
  $x = {|\beta|}^{-1/2} i \gamma_x e^{2s}$ and $y = |\beta|^{-1/2} i \gamma_y e^{2s}$.   

  We are thus left with the task of solving for the $t$ flow. Since this is a one-dimensional flow, instead of solving a differential equation for the flow we will simply find constant phase curves. Moreover, we will do it perturbatively in $\beta^{-1}$. Let us write $t(s) = t_\infty (s) + \beta^{-1} T (s)$. Since we are not trying to solve flow equations, we have the freedom to make arbitrary reparameterizations $\tilde s(s)$, which we will use to set $T =i \frac{\beta}{|\beta|} \frac{d t_\infty(\tilde s)}{d\tilde s} j(\tilde s)$ for some purely real $j(\tilde s)$ to be determined. This gauge is chosen to make certain expressions below manifestly imaginary and thus simplify the taking of their imaginary parts.
The equation for a constant phase curve is given by
\begin{equation}
    {\rm Im} \, S_0(t(\tilde s)) + {\rm Im} \, \frac{ h(t(\tilde s))}{\beta} = {\rm Im} \, S_0(t_\star) +{\rm Im} \, \frac{ h(t_\star)}{\beta}.
\end{equation}
Expanding to leading order, we find
\begin{eqnarray}
    {\rm Im} \, S_0(t_\infty(\tilde s)) &+&{\rm Im} \, \left(\frac{1}{\beta} S'_0(t_\infty(\tilde s)) T(\tilde s)
    \right) + {\rm Im} \, \frac{h(t_\infty(\tilde s))}{\beta} \\ &=& {\rm Im} \, S_0(t_0) + {\rm Im} \, \left( S'_0(t_0) \frac{t_1}{\beta} \right) + {\rm Im} \, \frac{h(t_0)}{\beta}.
\end{eqnarray}
Since $t_0$ was an (unperturbed) saddle, we have $S'_0(t_0) = 0$. Note also that 
\begin{equation}
    \frac{1}{\beta} S_0'(t_\infty(\tilde s)) T(\tilde s) = \frac{i}{|\beta|} S'_0(t_\infty(\tilde s)) t'_\infty(\tilde s) j(\tilde s),
\end{equation}
which is purely imaginary along the original contour (since it was a constant phase curve). For the same reason we have ${\rm Im} \, S_0(t_\infty(\tilde s)) = {\rm Im} \, S_0(t_0)$. Thus, we may solve for $j(\tilde s)$:
\begin{equation}
    j(\tilde s) = \frac{{\rm Im} \, \left(\frac{|\beta|}{\beta} \left( h(t_0) - h(t_\infty(\tilde s)) \right) \right)}{S'_0(t_\infty(\tilde s)) t'_\infty(\tilde s)}.
\end{equation}
Note that $j(\tilde s)$ is a real function (in accordance with our gauge choice) and does not depend on the magnitude of $\beta$. Going back to our original parametrization, we have\footnote{Since the difference between $s$ and $\tilde{s}$ is subleading at large $\beta$, we simply put $s=\tilde{s}$ in expressions for $X,Y$}
\begin{subequations}
    \begin{equation}
        t(\tilde s) = t_\infty(\tilde s) + i \frac{{\rm Im} \, \left(\frac{1}{\beta} \left( h(t_0) - h(t_\infty(\tilde s)) \right) \right)}{S'_0(t_\infty(\tilde s))}
    \end{equation}
    \begin{equation}
        X(\gamma_x,\tilde s) = \frac{i \gamma_x}{2\sqrt{\beta |\beta|} t\sqrt{f(t)}} e^{2\tilde s} + \frac{S_X(t)}{8t^2 f(t) \beta}
    \end{equation}
    and
    \begin{equation}
        Y(\gamma_y,\tilde s) = \frac{i \gamma_y}{\sqrt{\beta  |\beta|}\sqrt{f(t)} (t^2-1)} e^{2\tilde s} + \frac{S_Y(t)}{2 \beta f(t) (t^2-1)^2},
    \end{equation}
\end{subequations}
where $t = t(\tilde s)$ and $\tilde s,\gamma_x, \gamma_y \in \mathbb{R}^3$.
The only remaining question is what happens to a given intersection with the real plane when we turn on a small temperature. At zero temperature, the intersection will be $(t,X,Y) = (t_\heartsuit, 0,0)$ (and a value of a flow parameter $\tilde{s}_\heartsuit)$, let us solve for intersections perturbatively in $\beta^{-1}$ around these points. It is clear that we can always find $\gamma_x, \gamma_y$ such that the $X,Y$ will be real, provided that $t_\heartsuit \sqrt{\beta f(t_\heartsuit)}$ and $ \sqrt{\beta f(t_\heartsuit)} (t_\heartsuit^2-1)$ are not purely imaginary.  This is the case for generic values of $\tau$ and $\beta$.  These points are of course going to be at the distance of order $\beta^{-1}$ away from the origin. The $t$-intersection changes as
\begin{equation}
    t \left(\tilde s_\heartsuit+\frac{1}{\beta} \delta \tilde s \right) =t_\infty(\tilde s_\heartsuit) + \frac{1}{\beta} t_\infty'(\tilde s_\heartsuit) \delta \tilde s + i \frac{{\rm Im} \, \left(\frac{1}{\beta} \left( h(t_0) - h(t_\heartsuit) \right) \right)}{S'_0(t_\heartsuit)} \in \mathbb{R}.
\end{equation}
Since $\delta \tilde s \in \mathbb{R}$, the above relation has (a unique) solution provided that $\frac{1}{\beta} t'_\infty(\tilde s_\heartsuit) \notin \mathbb{R}$ which again is true for generic $\tau, \beta$.  Thus, we established that the perturbed thimble will still cross the real plane. Furthermore, the original intersection point $(t_\heartsuit, 0,0)$ was in the interior of our contour, not on the boundary, since due to our inclusion of both inner and outer horizons in the ansatz \eqref{eq:ansatz}, $X$ and $Y$ can have generic signs in the original contour and $t_\heartsuit$ was a generic value of $t$. The perturbed thimble will thus have the same intersection number with our integration contour as the unperturbed one. In particular, if the saddle contributed to the BPS-only integral \eqref{eq:indexansatz}, it will also contribute at sufficiently large finite $\beta$. 
Using continuity in $\tau, \beta$, we may also conclude that the same result follows at the non-generic values of $\tau,\beta$ where the above first-order argument fails. 

Let us now return to the remaining saddles of $S_0$ at $t=\pm i$. These can be studied in an entirely analogous manner, except we need to use different variables defined as
\begin{subequations}
    \begin{equation}
        X=2g(t) \left(a^2-2 a r_0^2+q-r_0^2\right)
    \end{equation}
    \begin{equation}
        Y=g \left(-a^2 -2 a +  r_0^2\right),
    \end{equation}
    where
    \begin{equation}
        g(t) = \sqrt{\frac{2\pi}{(1+t^2)^3}}
    \end{equation}
\end{subequations}
Note that the map between $(X,Y,t)$ and $(r_0,q,t)$ is not smooth at $t= \pm i$ which is exactly the reason why it can be used to remove the poles from the action\footnote{In these variables, a few new poles appear at $t=-1$ and $t=4+\sqrt{15}$. Since the flow of interest stays far away from these points, we may simply ignore these issues.}. This fact does not change the leading saddle point analysis but could potentially modify one-loop determinants around these points. The rest of analysis follows mutatis mutandis in the same ways as for other saddles. In particular, using an expansion in powers of $\beta^{-1}$, we find that perturbed thimbles stay close to the unperturbed ones and that the (vanishing) intersection numbers do not change.

\subsection{Numerics for large-$\beta$ flows}
\label{subsec:largebnumerics}


This section provides some brief numerics illustrating the correspondence between large-$\beta$ flows of \eqref{eq:ansatz} and the BPS-only flows of \eqref{eq:indexansatz} for saddles of \eqref{eq:ansatz} that become BPS at large $\beta$.
We now study the full equal-charge ansatz \eqref{eq:ansatz3}, including both rotation and charge,  directly at large $\beta$. At fixed $a$ and $r_0$, the dependence of the action in \eqref{eq:Zgrav} on $q$ is quadratic, so the Gaussian $q$ integral can be performed exactly. Writing $r\equiv r_0$, we obtain a two-real-dimensional integral over the contour $C=\left\{ -1<a<1,\ r>0\right\}$. After complexifying $a$ and $r$, the corresponding gradient-flow problem is two-complex-dimensional.

Let $z_\sigma=(a_\sigma,r_\sigma)$ be a critical point of $S_{\rm eff}$, and let $K_\sigma$ denote its upward-flow cycle, or dual thimble. We wish to determine the intersection number $n_\sigma=\langle C,K_\sigma\rangle$. Since both $C$ and $K_\sigma$ are real two-dimensional manifolds in the four-real-dimensional space $\mathbb C^2$, their intersections are generically isolated points. Any point $p\in C\cap K_\sigma$ must satisfy the necessary conditions
\begin{equation}
\operatorname{Im}S_{\rm eff}(p)=\operatorname{Im}S_{\rm eff}(z_\sigma),\qquad
\operatorname{Re}S_{\rm eff}(p)<\operatorname{Re}S_{\rm eff}(z_\sigma).
\label{eq:2D-thimble-necessary-condition}
\end{equation}

On the real contour $C$, the first condition generically defines a one-real-dimensional curve, while the second restricts this curve to a segment on which an intersection remains possible.
To test whether a candidate point on this segment belongs to $K_\sigma$, we will evolve it along the {\it downward} flow. If this flow approaches $z_\sigma$, then reversing the trajectory gives an upward flow from the saddle to the original contour.

We now specialize to the integral with $2\tau_n-3\Delta_{n'} =1$. For the parameters studied, generic starting points on the candidate segment fall into one of two basins: their flows terminate at the singularity $a=-1$ or at the singularity $a=1$. Along this segment, the two basins are separated by a single transition point. We locate this point numerically by repeatedly bisecting a pair of starting points whose flows terminate at opposite singularities.

We first consider $\tau=0.2+0.1i$. As shown in figure~\ref{fig:AdS5_phase_diagram}, at this point the $t_+$ saddle contributes to the one-dimensional BPS-only ansatz \eqref{eq:indexansatz2}, whereas the only-asymptotically-BPS saddle at $t=\frac{4+3i}{5}$, for which $a=-\frac{1}{5}$ in the large-$\beta$ limit, does not. We will see that the two-complex-dimensional flows at $\beta=10$ reproduce both conclusions.

\begin{figure}[h]
    \centering
    \includegraphics[width=0.92\linewidth]{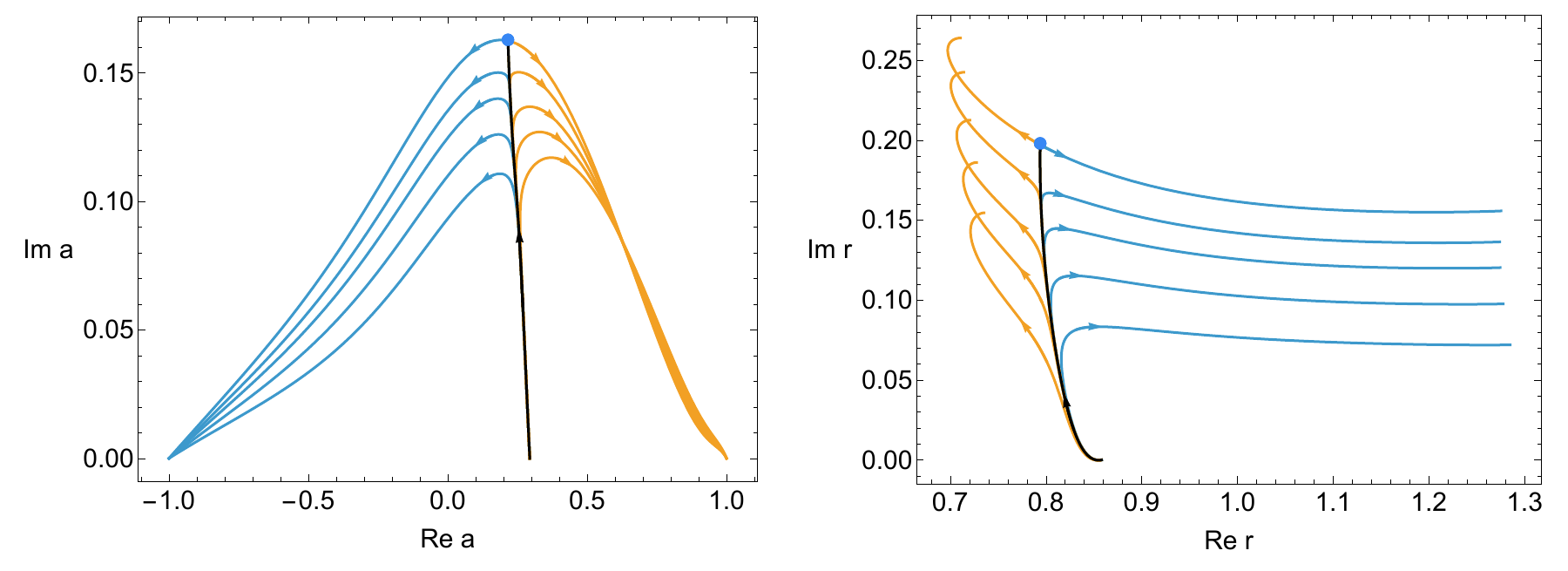}
    \caption{Downward flows for $\tau=0.2+0.1i$ and $\beta=10$ that start on the segment $N_C$ of the real contour $C$ selected by the necessary conditions \eqref{eq:2D-thimble-necessary-condition} for the BPS saddle. The left and right panels show projections onto the $a$- and $r$-planes respectively. The blue dot marks the saddle, and the arrows indicate the downward-flow direction. The points at which the blue and orange trajectories begin define basins that flow respectively to $a=-1$ or $a=+1$ and which appear to divide $N_C$ into two intervals.   The black curve is the limiting trajectory obtained by tuning the initial conditions to the boundary between these two intervals. Nearby trajectories initially follow it toward the saddle and then peel away along the blue or orange branches. For each near-transition trajectory, the closest approach to the saddle occurs at the same value of the flow parameter in the $a$-plane and $r$-plane projections, confirming that the limiting trajectory approaches the full critical point in $\mathbb C^2$. }
\label{fig:2D_thimble_BPS}
\end{figure}

Figure~\ref{fig:2D_thimble_BPS} shows the result for the finite-$\beta$ BPS saddle. As the initial point is tuned toward the boundary between the two basins, the downward trajectory passes increasingly close to the saddle and remains near it for an increasingly long flow time. Within numerical accuracy, the limiting black trajectory ends at the saddle. Reversing this trajectory therefore gives an upward flow from the saddle to $C$, showing that its intersection number is non-zero. The near-transition flows also remain close, up to corrections suppressed at large $\beta$, to the complexified BPS locus $r^2=a^2+2a$.

\begin{figure}[h]
    \centering
    \includegraphics[width=0.92\linewidth]{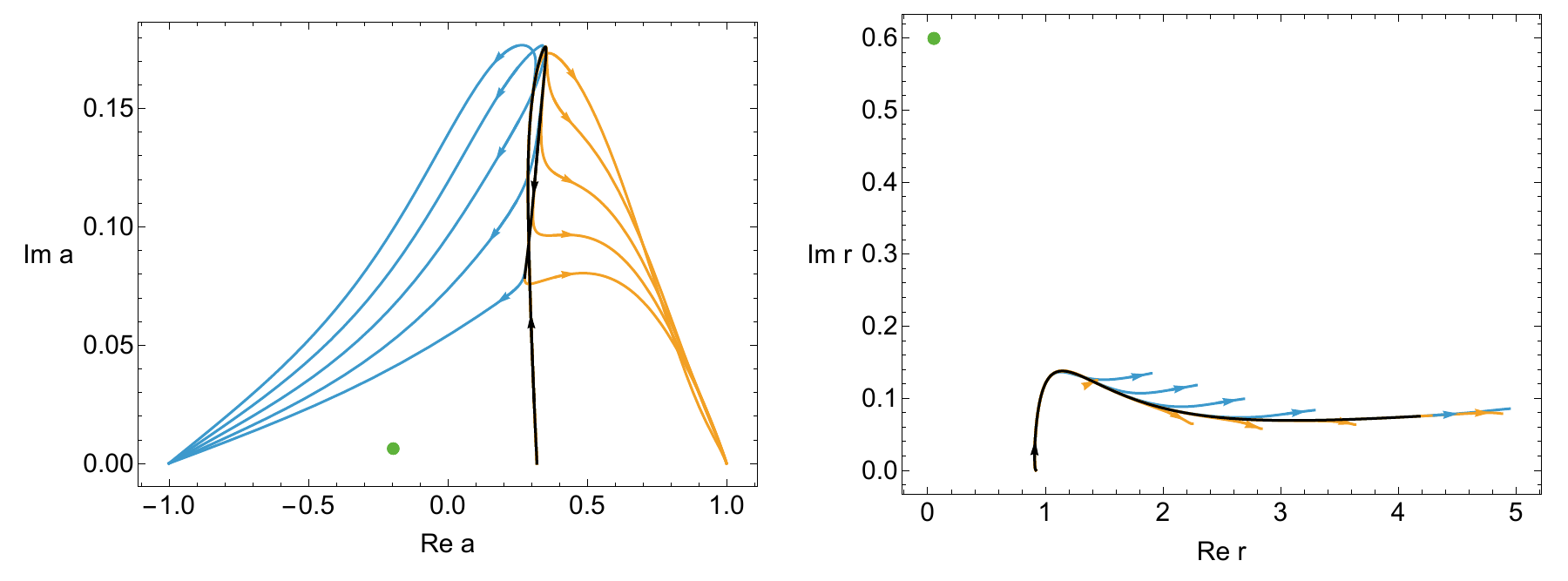}
    \caption{Downward flows $\tau=0.2+0.1i$ and $\beta=10$ that start on the segment $N'{}_C$ of $C$ where the necessary condition is satisfied for the finite-$\beta$ only-asymptotically-BPS saddle. This $\tau$ lies outside the red region of figure \ref{fig:AdS5_phase_diagram}, so the $\beta \rightarrow \infty$ limit does not contribute to the BPS-only ansatz \eqref{eq:indexansatz2}. The setup and conventions are as in figure~\ref{fig:2D_thimble_BPS}, with the projections of the saddle now marked by green dots. The black transition trajectory now runs toward large $\operatorname{Re} r$ rather than toward the saddle, so no downward flow is found to run from $C$ to the saddle.}
    \label{fig:2D_thimble_nonBPS}
\end{figure}

The same procedure gives a different result for the finite-$\beta$ saddle whose large-$\beta$ behavior is $(a,r)=\left(-\frac{1}{5},\frac{3i}{5}\right)+\mathcal O(\beta^{-1})$. As shown in figure~\ref{fig:2D_thimble_nonBPS}, the limiting transition trajectory does not approach this saddle but instead runs toward large $\operatorname{Re} r$. No point found in the scan of the candidate segment flows to the saddle. This numerically confirms, in the full two-complex-dimensional problem, the conclusion from the reduced one-dimensional analysis that the saddle does not contribute at $\tau=0.2+0.1i$.

\begin{figure}[h]
    \centering
    \includegraphics[width=0.92\linewidth]{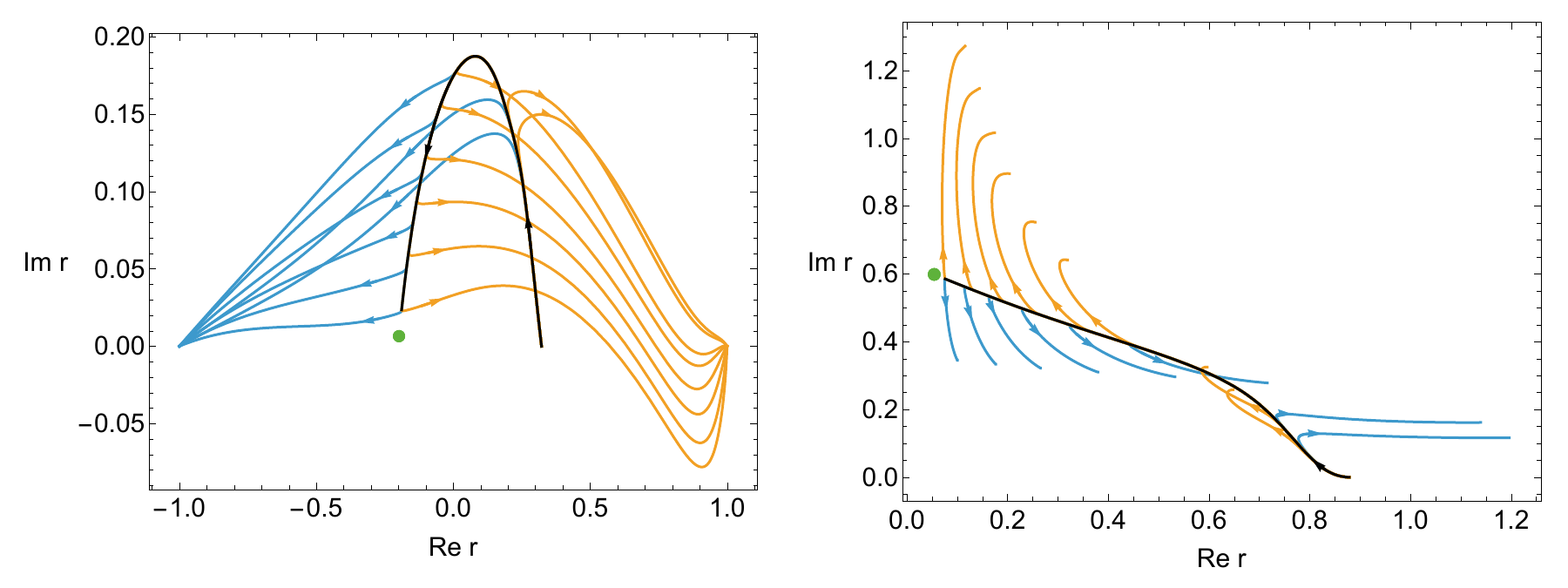}
    \caption{Downward flows $\tau=0.2+0.05i$ and $\beta=10$ that start on the segment $N'{}_C$ of $C$ where the necessary condition is satisfied for the finite-$\beta$ only-asymptotically-BPS saddle. This value of $\tau$ lies inside the red region of figure \ref{fig:AdS5_phase_diagram}, so the $\beta \rightarrow \infty$ limit of this saddle contributes to the BPS-only ansatz \eqref{eq:indexansatz2}. The setup and conventions are as in figure~\ref{fig:2D_thimble_BPS}, with the saddle marked by green dots. Here the black transition trajectory approaches the saddle at the same value of the flow parameter in both projections, showing that it reaches the full critical point in $\mathbb C^2$ and connects $C$ to the saddle.}
\label{fig:2D_thimble_nonBPS_contributing}
\end{figure}

We next take $\tau=0.2+0.05i$, which lies in the red subregion of figure~\ref{fig:AdS5_phase_diagram}. Here the one-dimensional BPS-only analysis predicts that the $t=\frac{4+3i}{5}$ saddle contributes. Repeating the same bisection procedure at $\beta=10$, we find that the limiting transition trajectory now approaches the corresponding finite-$\beta$ non-BPS saddle, as shown in figure~\ref{fig:2D_thimble_nonBPS_contributing}. Within numerical accuracy, the approach occurs at the same value of the flow parameter in both projections, so the limiting trajectory reaches the full critical point in $\mathbb C^2$. Reversing this trajectory gives an upward flow from the saddle to $C$ and hence a non-zero intersection number, again in agreement with the BPS-only analysis.

Taken together, these examples illustrate how the one-complex-dimensional BPS reduction captures the relevant thimble geometry and intersection numbers of the full two-complex-dimensional problem at large $\beta$, for both the truly-BPS saddle and the only-asymptotically-BPS saddle considered here.

\bibliographystyle{JHEP}
\bibliography{reference}
\end{document}